\documentclass[prd,twocolumn,showpacs,superscriptaddress,nofootinbib,floatfix,showkeys,10pt]{revtex4-2}

\usepackage{titlesec}
\titlespacing*{\section}{0pt}{3ex plus 1ex minus .2ex}{2ex}
\usepackage{graphicx}
\usepackage{amsmath}
\usepackage{bm}
\usepackage{yhmath}
\usepackage{mathtools}
\usepackage{wasysym}
\usepackage[colorlinks,citecolor=blue,urlcolor=blue,linkcolor=blue]{hyperref}
\usepackage{subfigure}
\usepackage{color}
\usepackage{cases}
\usepackage{subfigure}
\usepackage{times}
\usepackage{dcolumn,booktabs,bm}
\usepackage{slashed}
\usepackage{amsfonts,amssymb,stmaryrd,latexsym,amsmath}
\usepackage{textcomp}
\usepackage{multirow}
\usepackage{cancel}
\usepackage{array}
\usepackage{orcidlink}
\usepackage{enumitem}
\usepackage{booktabs}
\usepackage{tabularx}
\usepackage{caption}
\usepackage{dashrule}
\usepackage{float}

\newcommand{\change}[1]{#1}
\renewcommand{\arraystretch}{1.8}
\begin{document}

% %%%%%%%%%%%%%%open the reply mode%%%%%%%%%%%%%%%%%%
% \onecolumngrid
% \input{reply2.tex}
% \newpage
% \setcounter{page}{0}
% %%%%%%%%%%%%%%open the reply mode%%%%%%%%%%%%%%%%%%

\title{Investigation of S-wave tetraquark bound and resonant states with all Jacobi coordinates}

\author{Xin-He Zheng\,\orcidlink{0009-0002-2550-331X}}\email{zhengxh@stu.pku.edu.cn}
\affiliation{School of Physics, Peking University, Beijing 100871, China}

\author{Yao Ma\,\orcidlink{0000-0002-5868-1166}}\email{yao.ma@tum.de}
\affiliation{Technical University of Munich, TUM School of Natural Sciences, Physics Department, James-Franck-Str. 1, 85748 Garching, Germany.}

\author{Liang-Zhen Wen\,\orcidlink{0009-0006-8266-5840}}\email{ wenlzh hep-th@stu.pku.edu.cn}
\affiliation{School of Physics and Center of High Energy Physics, Peking University, Beijing 100871, China}

\author{Shi-Lin Zhu\,\orcidlink{0000-0002-4055-6906}}\email{zhusl@pku.edu.cn}
\affiliation{School of Physics and Center of High Energy Physics, Peking University, Beijing 100871, China}

\begin{abstract}
		We systematically explore the $S$-wave tetraquark systems $Qs\bar{n}\bar{n}$, $QQ\bar{n}\bar{n}$, $QQ\bar{Q}\bar{Q}$, and $ss\bar{s}\bar{s}$ ($Q=c,b$; $n=u,d$) within the constituent quark potential model. We incorporate all K-type Jacobi coordinates in addition to the conventional H-type configurations, optimize the basis expansion via a stochastic parameter generation strategy, and apply the complex scaling method to identify bound and resonant states. Our calculations demonstrate that while conventional H-type configurations suffice for low-lying states such as the $T_{cc}(3875)^+$ molecular candidate, the inclusion of K-type configurations becomes important for extracting highly excited resonances, allowing higher-energy resonances absent in H-only calculations to be identified. Furthermore, we identify resonance candidates for the $T_{cs0}(2900)$, $X(6900)$, and $X(7200)$, whereas the absence of fully-strange compact poles below 2.6 GeV challenges the interpretation of $\phi(2170)$ and $X(2370)$ as $S$-wave compact $s s \bar{s} \bar{s}$ tetraquarks. 
	\end{abstract}

\maketitle

\section{Introduction}
\label{sec:intro}

Since the discovery of the $X(3872)$~\cite{Belle:2003nnu}, numerous exotic hadron candidates have been observed experimentally. Compared to conventional hadrons, multiquark states, hybrids, and glueballs exhibit more intricate color structures~\cite{Jaffe:1976ig,Jaffe:1975fd,Fritzsch:1973pi}. Investigating these exotic states provides a unique window into the nonperturbative nature of quantum chromodynamics (QCD), which has continuously stimulated extensive theoretical and experimental efforts (see Refs.~\cite{Chen:2016qju,Hosaka:2016pey,Esposito:2016noz,Ali:2017jda,Lebed:2016hpi,Guo:2017jvc,Liu:2019zoy,Brambilla:2019esw,Meng:2022ozq,Chen:2022asf,Mai:2022eur} for recent reviews).

In 2020, the LHCb Collaboration observed two tetraquarks $T_{cs}(2900)$ in the $D^-K^+$ invariant mass distribution~\cite{LHCb:2020bls,LHCb:2020pxc}, representing the first tetraquark candidates with four different flavors ($cs\bar{u}\bar{d}$). Various interpretations have been proposed to elucidate their mass spectra and line shapes, including $D^*K^*$ hadronic molecules~\cite{Molina:2010tx,Chen:2020aos,He:2020btl,Liu:2020nil,Hu:2020mxp,Agaev:2020nrc,Wang:2021lwy,Wang:2023hpp},
compact tetraquarks~\cite{Karliner:2020vsi,He:2020jna,Wang:2020xyc,Zhang:2020oze,Wang:2020prk,Lu:2020qmp,Tan:2020cpu,Albuquerque:2020ugi,Yang:2021izl,Agaev:2022eeh,Liu:2022hbk}, 
and kinematic effects arising from triangle singularities~\cite{Liu:2020orv,Burns:2020epm}. Among these, a unified calculation incorporating both molecular and compact configurations has recently accommodated the $T_{cs0}(2900)$ as a $D^*\bar{K}^*$ molecular resonance~\cite{Chen:2023syh}.

Shortly after the discovery of the $T_{cs}$ states, in 2021, the LHCb Collaboration discovered the first doubly charmed tetraquark state $T_{cc}(3875)^+$ in the $D^0D^0\pi^+$ mass spectrum~\cite{LHCb:2021vvq,LHCb:2021auc}. It is a narrow state with a mass extremely close to the $D^{*+}D^0$ threshold, favoring its interpretation as a $D^*D$ molecular state. Theoretical investigations on the possible existence of doubly heavy tetraquark states have been conducted for decades~\cite{Pepin:1996id,Gelman:2002wf,Vijande:2003ki}, while subsequent mass predictions were highly inconsistent across different frameworks~\cite{Ebert:2007rn,Du:2012wp,Karliner:2017qhf,Eichten:2017ffp,Yang:2019itm}. The precise measurement of the near-threshold $T_{cc}(3875)^+$ stimulated extensive investigations into its doubly heavy analogues, expanding the focus to various $bb\bar{n}\bar{n}$ and $bc\bar{n}\bar{n}$ bound and resonant states across different frameworks~\cite{Albaladejo:2021vln,Meng:2024yhu,Whyte:2024ihh}. Recently, a unified calculation~\cite{Wu:2024zbx} successfully accommodated the $T_{cc}$ and its bottom analog as molecular states while identifying various compact configurations, further enriching the dynamics of open-flavor heavy tetraquarks.

Beyond these systems containing light quarks, the fully heavy tetraquark states $QQ\bar{Q}\bar{Q}$ stand out as relatively pristine environments, unaffected by the unquenched dynamics such as the creation and annihilation of light $n\bar{n}$ pairs. The LHCb Collaboration initially discovered a fully charmed tetraquark candidate $X(6900)$~\cite{LHCb:2020bwg}, and subsequently, the CMS and ATLAS collaborations reported additional structures such as $X(6400)$, $X(6600)$, and $X(7200)$~\cite{CMS:2023owd,ATLAS:2023bft}. These discoveries not only provide insights into the $cc\bar{c}\bar{c}$ system but also stimulate extensive theoretical interest in fully bottomed tetraquarks. Furthermore, the underlying symmetries imply the possible existence of their strange analogues, the $ss\bar{s}\bar{s}$ states. In experiments, there are several promising fully strange candidates, such as the $\phi(2170)$ and $X(2370)$~\cite{BaBar:2006gsq,BaBar:2007ptr,BaBar:2007ceh,BaBar:2011btv,BESIII:2010gmv,BESIII:2019wkp}. However, their inner dynamics and whether they are genuine compact tetraquarks remain controversial. Various interpretations have been proposed in earlier theoretical studies~\cite{liu:2020eha,Lu:2020cns,Dong:2020okt,Giron:2020wpx,Wang:2020wrp,Dong:2020nwy,Wang:2006ri,Liu:2020lpw}, 
ranging from diquark-antidiquark compact structures to meson-meson molecular configurations. To clarify these ambiguities, comprehensive benchmark calculations incorporating both spatial configurations have been performed to systematically investigate the resonant states of the fully heavy~\cite{Wu:2024euj} and fully strange~\cite{Ma:2024vsi} tetraquarks. These studies successfully identified multiple compact resonance candidates corresponding to the $X(6900)$ and $X(7200)$, while casting doubt on the compact $S$-wave assignments for the low-lying $\phi(2170)$ and $X(2370)$.

However, these previous constituent quark model calculations typically restricted their spatial basis to H-type configurations, namely the dimeson and diquark-antidiquark structures~\cite{Chen:2023syh,Wu:2024zbx,Wu:2024euj,Ma:2024vsi}. While adequate for low-lying states, this limited basis lacks the spatial degrees of freedom required to fully capture complex multiquark correlations. To overcome this structural limitation and conduct a more rigorous evaluation of these tetraquark candidates, it is crucial to perform comprehensive four-body calculations that incorporate all Jacobi configurations, particularly the K-type coordinates.

In the present work, we systematically investigate the $S$-wave $Qs\bar{n}\bar{n}$, $QQ\bar{n}\bar{n}$, $QQ\bar{Q}\bar{Q}$, and $ss\bar{s}\bar{s}$ tetraquark systems using the AL1 constituent quark potential model~\cite{Semay:1994ht,Silvestre-Brac:1996myf}. To mitigate the rapid growth of the basis dimension in the Gaussian expansion method (GEM)~\cite{Hiyama:2003cu}, we implement a stochastic parameter generation strategy combined with parallel computing. Furthermore, we adopt the complex scaling method (CSM)~\cite{Aguilar:1971ve,Balslev:1971vb,Aoyama:2006hrz} to identify genuine resonant states and bound states from the scattering continua simultaneously.

The paper is arranged as follows. In Sec.~\ref{sec:framework}, we introduce the theoretical framework. In Sec.~\ref{sec:results}, we present the numerical results and discuss the structures of the tetraquark states. Finally, a summary is given in Sec.~\ref{sec:summary}.

\section{Theoretical framework}~\label{sec:framework}

In this work, we employ the nonrelativistic constituent quark potential model to systematically investigate the fully $S$-wave $Qs\bar{n}\bar{n}$, $QQ\bar{n}\bar{n}$, $QQ\bar{Q}\bar{Q}$, and $ss\bar{s}\bar{s}$ tetraquark systems. The four-body Schr\"odinger equation is solved using the Gaussian expansion method, while bound states and resonances are identified via the complex scaling method. We introduce the K-type Jacobi coordinates in the construction of spatial wave functions and optimize the Gaussian basis expansion via a stochastic parameter generation strategy to achieve a significant improvement in computational performance.

\subsection{Hamiltonian}~\label{subsec:Hamiltonian}

The dynamics of the tetraquark system in the center-of-mass frame are governed by the following nonrelativistic Hamiltonian:
\begin{align}\label{eq:Hamiltonian}
	H=\sum_{i=1}^4\left(m_i+\frac{\boldsymbol{p}_i^2}{2m_i}\right)+\sum_{i<j=1}^4 V_{i j}\,,
\end{align}
where $m_i$ and $\boldsymbol{p}_i$ denote the constituent mass and the momentum of the $i$-th (anti)quark, respectively. The pairwise interaction $V_{ij}$ between quarks is described by the AL1 potential model~\cite{Semay:1994ht,Silvestre-Brac:1996myf}, which combines one-gluon exchange and linear confinement:
\begin{align}\label{eq:AL1}
	V_{i j}=-\frac{3}{16}&\lambda_i^c \cdot \lambda_j^c\Big(-\frac{\kappa}{r_{ij}}+\lambda r_{i j}-\Lambda\nonumber\\
	&+\frac{8\pi\kappa'}{3m_{i}m_{j}}\frac{\exp(-r_{ij}^{2}/r_{0}^{2})}{\pi^{3/2}r_{0}^{3}}\boldsymbol{s}_{i}\cdot\boldsymbol{s}_{j}
	\Big).
\end{align}
Here, the smearing parameter is defined as $r_0=A\left(\frac{2m_im_j}{m_i+m_j}\right)^{-B}$. The operators $\lambda^c$ and $\boldsymbol{s}_i$ stand for the SU(3) color Gell-Mann matrices and the spin operator, respectively. The adopted AL1 potential parameters, which were originally calibrated to reproduce the meson spectra~\cite{Silvestre-Brac:1996myf}, alongside the corresponding constituent quark masses, are explicitly listed in Table~\ref{tab:paraAL1}. Furthermore, the theoretical masses and root-mean-square (rms) radii of the relevant mesons calculated within this model are summarized in Table~\ref{tab:meson}.

\begin{table}[htbp]
	\centering
	\caption{The parameters in the AL1 quark potential model and the constituent quark masses.}
	\label{tab:paraAL1}
	\begin{tabular*}{\hsize}{@{}@{\extracolsep{\fill}}cccccc@{}}
		\hline\hline
		$ \kappa $ & $ \lambda { [\mathrm{GeV}^{2}]}$ & $ \Lambda {\rm [GeV]} $ & 
		$ \kappa^\prime $ & $ A { [\mathrm{GeV}^{B-1}]}$ & $ B $ \\
		\hline
		0.5069 & 0.1653 & 0.8321 & 1.8609 & 1.6553 & 0.2204 \\
		\hline
		$ m_b \,[\mathrm{GeV}] $ & $ m_c \,[\mathrm{GeV}] $ & 
		$ m_s \,[\mathrm{GeV}] $ & $ m_q \,[\mathrm{GeV}] $ & & \\
		\hline
		5.227 & 1.836 & 0.577 & 0.315 & & \\
		\hline\hline
	\end{tabular*}
\end{table}

\begin{table}[htbp]
	\renewcommand{\arraystretch}{1.4}
	\centering
	\caption{\label{tab:meson} Theoretical masses (MeV) and rms radii (fm) of mesons in the AL1 model.\footnote{Note that the constituent quark model typically has a systematic uncertainty on the order of tens of MeV.} Experimental values are taken from Ref.~\cite{ParticleDataGroup:2022pth}.}
	\begin{tabular*}{\hsize}{@{\extracolsep{\fill}}cccc|ccc@{}}
		\hline\hline
		Meson & $m_{\mathrm{Exp.}}$ & $m_{\mathrm{Theo.}}$ & $r^{\mathrm{rms}}_{\mathrm{Theo.}}$ 
		& Meson & $m_{\mathrm{Theo.}}$ & $r^{\mathrm{rms}}_{\mathrm{Theo.}}$ \\
		\hline 
		$K$ & 496 & 491 & 0.59 & $K(2S)$ & 1465 & 1.31 \\
		$K^*$ & 894 & 904 & 0.81 & $K^*(2S)$ & 1643 & 1.42 \\
		$\ensuremath{\eta^{\prime}}$\footnote{No mixing between $I=0$ $\eta(n\bar{n})$ and $\eta'(s\bar{s})$ is assumed.} 
		& -- & 713 & 0.54 
		& $\ensuremath{\eta^{\prime}(2S)}$ & 1565 & 1.17 \\ 
		$\ensuremath{\phi}$ & 1020 & 1021 & 0.70 
		& $\ensuremath{\phi(2S)}$ & 1695 & 1.25 \\
		$D$ & 1867 & 1862 & 0.61 & $D(2S)$ & 2643 & 1.23 \\
		$D^*$ & 2009 & 2016 & 0.70 & $D^*(2S)$ & 2715 & 1.27 \\
		$B$ & 5279 & 5294 & 0.63 & $B(2S)$ & 6013 & 1.20 \\
		$B^*$ & 5325 & 5351 & 0.66 & $B^*(2S)$ & 6041 & 1.22 \\
		$\eta_c$ & 2984 & 2983 & 0.35 & $\eta_c(2S)$ & 3605 & 0.78 \\
		$J/\psi$ & 3097 & 3103 & 0.40 & $\psi(2S)$ & 3646 & 0.81 \\
		$\eta_b$ & 9399 & 9424 & 0.20 & $\eta_b(2S)$ & 10003 & 0.49 \\
		$\Upsilon$ & 9460 & 9462 & 0.21 & $\Upsilon(2S)$ & 10013 & 0.49 \\
		\hline \hline
	\end{tabular*}
\end{table}

\subsection{Wave function construction}~\label{subsec:wavefunction}

The total wave function $\psi$ for the fully $S$-wave tetraquark systems is constructed as a global antisymmetric product of the spatial wave function $\phi$ and the color-spin counterpart $\chi$:
\begin{equation}\label{eq:Abasis}
	\psi=\mathcal{A}\left(\chi \otimes \phi\right).
\end{equation}

\begin{figure}[H]
	\centering
	\includegraphics[width=0.48\textwidth]{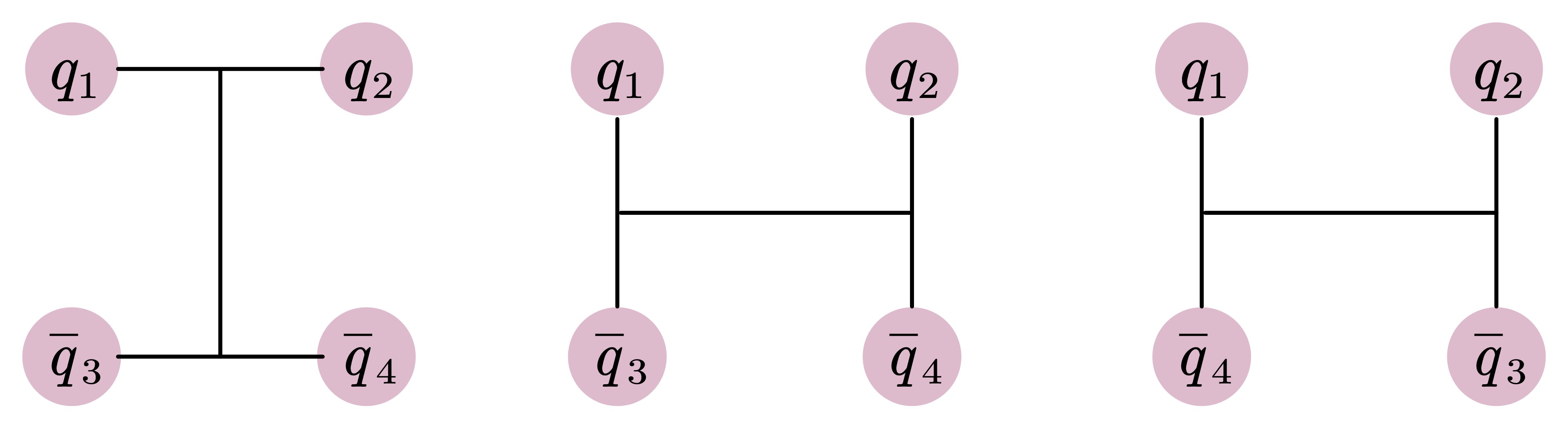}
	\caption{\label{fig:H_type}The traditional H-type Jacobi coordinate structures of the tetraquark system, including the diquark-antidiquark and dimeson configurations.}
\end{figure}

The spatial wave functions are constructed using the Gaussian expansion method~\cite{Hiyama:2003cu}. In traditional quark model studies of tetraquark states, the basis space is frequently restricted to the H-type subset of Jacobi coordinates~\cite{Chen:2023syh,Wu:2024zbx,Yang:2025wqo,Wu:2024euj,Wu:2024hrv,Ma:2024vsi,Wu:2024ocq,Zheng:2025uzy}, namely the dimeson and diquark-antidiquark configurations (as illustrated in Fig.~\ref{fig:H_type}). Historically, adopting this H-only framework was not merely a pragmatic compromise to circumvent computational limitations. It was also driven by physical intuition, as such structures naturally mirror the anticipated quark clustering behaviors dictated by the color confinement mechanism. Nevertheless, while this restricted basis is adequate for describing low-lying states, highly excited resonances generally have more extended spatial wave functions and may involve components associated with a different particle clustering. To go beyond this limitation and ensure a reliable extraction of highly excited resonance poles, we expand the basis space by incorporating the 12 distinct K-type Jacobi configurations (as illustrated in Fig.~\ref{fig:K_type}), which, together with the H-type structures, constitute all possible Jacobi coordinate arrangements for a four-body system~\cite{Kamada:2001tv,Hiyama:2001zt,Hiyama:2003cu}.

\begin{widetext}
\noindent
\makebox[\textwidth][c]{%
    \includegraphics[width=0.85\textwidth]{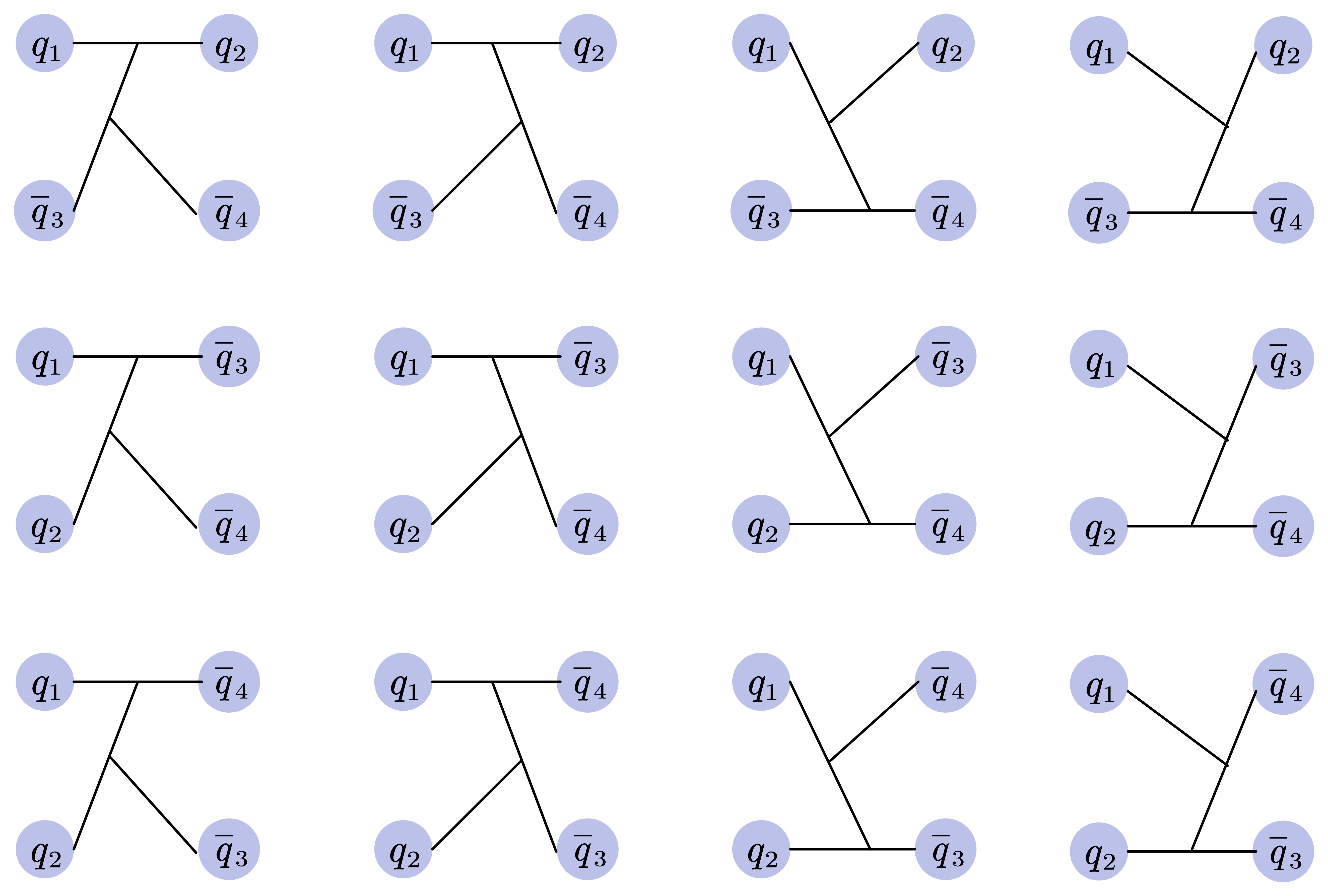}%
}

\refstepcounter{figure}
\begin{center}
\small
FIG.~\thefigure. The 12 distinct K-type Jacobi coordinate structures for the tetraquark system.
\label{fig:K_type}
\end{center}
\end{widetext}

% \begin{figure*}[htbp]
% 	\centering
% 	\includegraphics[width=0.85\textwidth]{K.jpg}
% 	\caption{\label{fig:K_type}The 12 distinct K-type Jacobi coordinate structures for the tetraquark system.}
% \end{figure*}

Furthermore, for the $QQ\bar{Q}\bar{Q}$ and $ss\bar{s}\bar{s}$ systems investigated in our study, the physical states must possess definite charge conjugation parity. This is accomplished by applying the charge conjugation transformation to the basis components. By forming symmetric or antisymmetric linear combinations of the initial basis functions and their charge-conjugated counterparts, we can project the total wave function into independent subspaces featuring positive ($C=+1$) and negative ($C=-1$) C-parities.

\subsection{Complex scaling method}~\label{subsec:method}

Since the wave functions of resonant states diverge at spatial infinity, they are not square-integrable and cannot be directly treated as standard Hermitian eigenvalue problems. To circumvent this difficulty, we apply the complex scaling method~\cite{Aguilar:1971ve,Balslev:1971vb,Aoyama:2006hrz}. This method analytically rotates the spatial coordinates and their conjugate momenta by a scaling angle $\theta$:
\begin{align}\label{eq:complexRotation}
	U(\theta) \boldsymbol{r}=\boldsymbol{r} e^{i \theta}, \quad U(\theta) \boldsymbol{p}=\boldsymbol{p} e^{-i \theta}.
\end{align}
Applying this transformation yields a non-Hermitian Hamiltonian $H(\theta)$. According to the ABC theorem~\cite{Aguilar:1971ve,Balslev:1971vb}, the continuous scattering spectra in the complex energy plane rotate downward by an angle of $2\theta$ starting from their corresponding physical thresholds. In contrast, the discrete eigenvalues, including the bound states on the negative real axis and the complex resonance poles at $E_R = M_R - i\Gamma_R/2$, are independent of the scaling angle $\theta$.

In addition, the inclusion of K-type Jacobi coordinates enlarges the basis space and will improve the numerical convergence of the resonance pole positions with respect to the complex scaling angle $\theta$ and reduces the risk of artificial distortions or missing highly excited resonances, thereby allowing for a more stable extraction of resonance poles in the higher-energy region.

\subsection{Spatial Structure}

The root-mean-square radius acts as a vital metric for probing the internal spatial configurations of multiquark systems, particularly for distinguishing compact tetraquarks from hadronic molecules. In the conventional approach, the rms radius is calculated using the fully antisymmetrized wave function. However, for tetraquark systems containing identical fermions, such as the $Qs\bar{n}\bar{n}$, $QQ\bar{n}\bar{n}$, $QQ\bar{Q}\bar{Q}$, and $ss\bar{s}\bar{s}$ systems studied here, the antisymmetrization required by the Pauli principle introduces exchange terms. However, these exchange terms inevitably mix different clustering structures when evaluating spatial properties, thereby obscuring the distinction of internal structures. 

To eliminate this structural ambiguity, we adopt an alternative scheme that evaluates the rms radii based on the non-antisymmetrized part of the wave function. Specifically, color configurations like $\bar{3}_c \otimes 3_c$ or $6_c \otimes \bar{6}_c$ can be decomposed into linear combinations of color-singlet subsets, e.g., $[(q_1\bar{q}_3)_{1_c}(q_2\bar{q}_4)_{1_c}]$ and $[(q_1\bar{q}_4)_{1_c}(q_2\bar{q}_3)_{1_c}]$, which form a complete, albeit non-orthogonal, basis for the color space. By gathering all direct terms associated with a specified meson-meson clustering, the total wave function $\Psi(\theta)$ can be expressed as an antisymmetrization over a non-antisymmetrized component $\Psi_{nA}(\theta)$:
\begin{align}
	\Psi(\theta) &= \mathcal{A} \left[ \sum_{s_1,s_2} \left[ (q_1 \bar{q}_3)_{1_c}^{s_1} (q_2 \bar{q}_4)_{1_c}^{s_2} \right]_{1_c}^{J} \otimes \psi(\boldsymbol{r}_1, \boldsymbol{r}_2, \boldsymbol{r}_3, \boldsymbol{r}_4; \theta) \right] \nonumber \\
	&\equiv \mathcal{A} \Psi_{nA}(\theta),
\end{align}
where $s_1$ and $s_2$ denote the spin of each subcluster coupled to the total angular momentum $J$, and $\mathcal{A}$ is the global antisymmetrization operator. It is essential to note that while only the fully antisymmetrized $\Psi(\theta)$ is a valid eigenfunction of the Hamiltonian, the core component $\Psi_{nA}(\theta)$ successfully filters out the exchange interference and serves as a reliable probe for the underlying cluster sizes.

Consequently, the effective rms radii are determined solely through this non-antisymmetrized component $\Psi_{nA}(\theta)$:
\begin{equation}
	r_{ij}^{\mathrm{rms}} \equiv \mathrm{Re} \left[ \sqrt{ \frac{\langle \Psi_{nA}(\theta) | r_{ij}^2 e^{2i\theta} | \Psi_{nA}(\theta) \rangle}{\langle \Psi_{nA}(\theta) | \Psi_{nA}(\theta) \rangle} } \right].
\end{equation}
Here, the inner product is computed using the c-product rule common in CSM frameworks, which omits the complex conjugation of the bra state.  By employing this criterion, we can quantitatively identify whether a state possesses a molecular nature based on a comparison between its inter-cluster and intra-cluster distances.

\subsection{Computational optimization}~\label{subsec:optimization}
	
	Including all H- and K-type configurations greatly increases the basis dimension. In the standard Gaussian expansion method (GEM), the scale parameters $\nu_{n}$ are generated using geometric sequences. For a given Jacobi configuration $\beta$, the spatial wave function is usually expanded on a tensor-product grid:
	\begin{equation}
		\Phi_{\beta} = \sum_{n_1=1}^{N_1} \sum_{n_2=1}^{N_2} \sum_{n_3=1}^{N_3} C_{n_1, n_2, n_3}^{(\beta)} \phi_{n_1}(\mathbf{r}_1) \phi_{n_2}(\mathbf{r}_2) \phi_{n_3}(\mathbf{r}_3),
	\end{equation}
	with the parameters defined as:
	\begin{equation}
		\nu_{n} = \nu_{1} \left( \frac{\nu_{N}}{\nu_{1}} \right)^{\frac{n-1}{N-1}}, \quad (n = 1, 2, \dots, N).
	\end{equation}
	For a four-body system with $J$ Jacobi configurations, the total spatial dimension is $J \times N^3$. For example, in our previous work using only H-type configurations on an $N=12$ grid, we generated $3 \times 12^3 = 5184$ spatial bases~\cite{Zheng:2025uzy}. For the $J^P=1^+$ state, multiplying by the 2 color and 3 spin discrete channel numbers expands this to $5184 \times 2 \times 3 = 31104$. If we applied this tensor-product method to all 15 Jacobi configurations with $N=12$ precision of spatial wavefuction, the basis dimension would grow to $15 \times 12^3 \times 2 \times 3 = 155520$. Furthermore, the fixed tensor-product grid inevitably produces many highly extended basis states that yield almost no contribution due to color confinement, thus merely wasting computer memory and calculation time.
	
\begin{figure*}[htbp]
		\centering
		\includegraphics[width=0.85\textwidth]{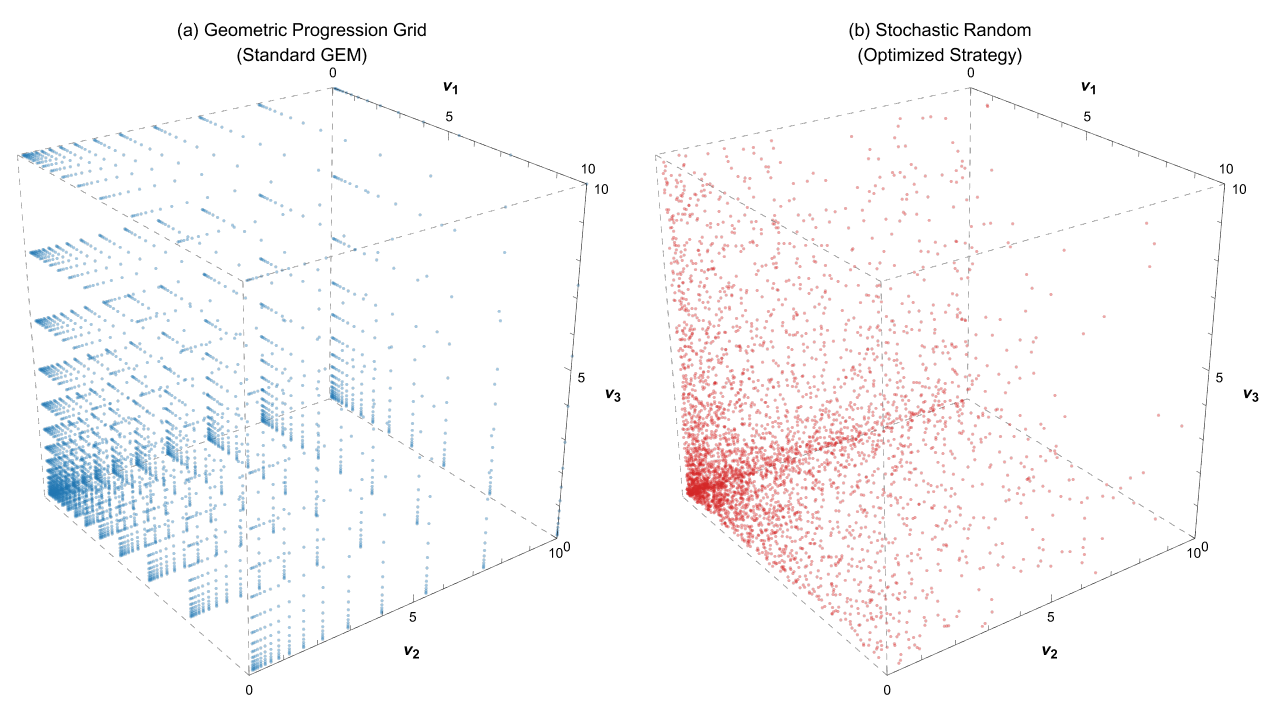}
		\caption{\label{fig:parameter_space} Schematic comparison of the parameter generation strategies.}
	\end{figure*}

	To solve this problem, we use a stochastic method to generate the parameters. Instead of using a fixed grid, we write the spatial wave function as a single sum over randomly generated bases:
	\begin{equation}
		\Phi_{\beta} = \sum_{m=1}^{N'} C_{m}^{(\beta)} \Big[ \phi_{\nu_{m,1}}(\mathbf{r}_1) \phi_{\nu_{m,2}}(\mathbf{r}_2) \phi_{\nu_{m,3}}(\mathbf{r}_3) \Big].
	\end{equation}
	Here, the set of parameters $\{\nu_{m,1}, \nu_{m,2}, \nu_{m,3}\}$ is sampled together randomly. We use a log-normal distribution for this sampling:
	\begin{equation}
		P(\nu) = \frac{1}{\nu \sigma \sqrt{2\pi}} \exp\left[ -\frac{(\ln \nu - \mu)^2}{2\sigma^2} \right],
	\end{equation}
	where $\mu$ and $\sigma$ are the mean and standard deviation. As intuitively illustrated in Fig.~\ref{fig:parameter_space}, this stochastic sampling approach mitigates the rapid expansion of basis states inherent to the rigid tensor grid. Moreover, it naturally concentrates the generated parameters within the physically efficient regions, effectively eliminating redundant, overly extended bases.
	
	In practice, we find that generating $N^{\prime}=500$ random bases per Jacobi configuration is enough to achieve comparable numerical accuracy to the full tensor-product grid for these four-body systems. This reduces the total spatial dimension from $J \times 12^3$ to $J \times 500$, which is less than $30\%$ of the original size, while keeping the same accuracy, thereby reducing the computational cost of the $\mathcal{O}(N^2)$ matrix evaluation by approximately an order of magnitude. In addition, we use parallel computing with OpenMP~\cite{Dagum:1998rbq} and AVX instructions via BLAS~\cite{Wang:2013pot} to speed up the matrix evaluation, which yields a further practical speedup by dozens of times through lock-free memory access and optimized cache locality. The diagonalization is expedited by the Implicitly Restarted Arnoldi Method (ARPACK)~\cite{Lehoucq:1989aug}, which drops the algorithmic complexity from $\mathcal{O}(N^3)$ to $\mathcal{O}(N^2)$, providing an acceleration by a factor of over $20$ in typical scenarios. This efficient approach allows us to study more complex multiquark systems, such as pentaquarks and hexaquarks, in future studies.

\section{Numerical results}~\label{sec:results}
	
In the following results, the mass uncertainties are on the order of tens of MeV, representing the typical precision scale of the constituent quark model. Furthermore, since the finite widths of the constituent mesons are neglected and only two-body strong decays are considered, the theoretical calculations are expected to underestimate the actual decay widths.

\change{We compare the pole positions obtained with and without the K-type configurations to estimate their effect. For the low-lying states below the $1S2S$ thresholds, the pole positions change only slightly, showing that the K-type configurations have little effect on these states. However, for highly excited states, the K-type configurations become essential, as evidenced by substantial pole shifts and the production of new resonances.}
	
\subsection{$cs\bar{n}\bar{n}$}~\label{subsec:csnn}

\begin{figure*}[htbp]
	\centering
	\includegraphics[width=0.95\textwidth]{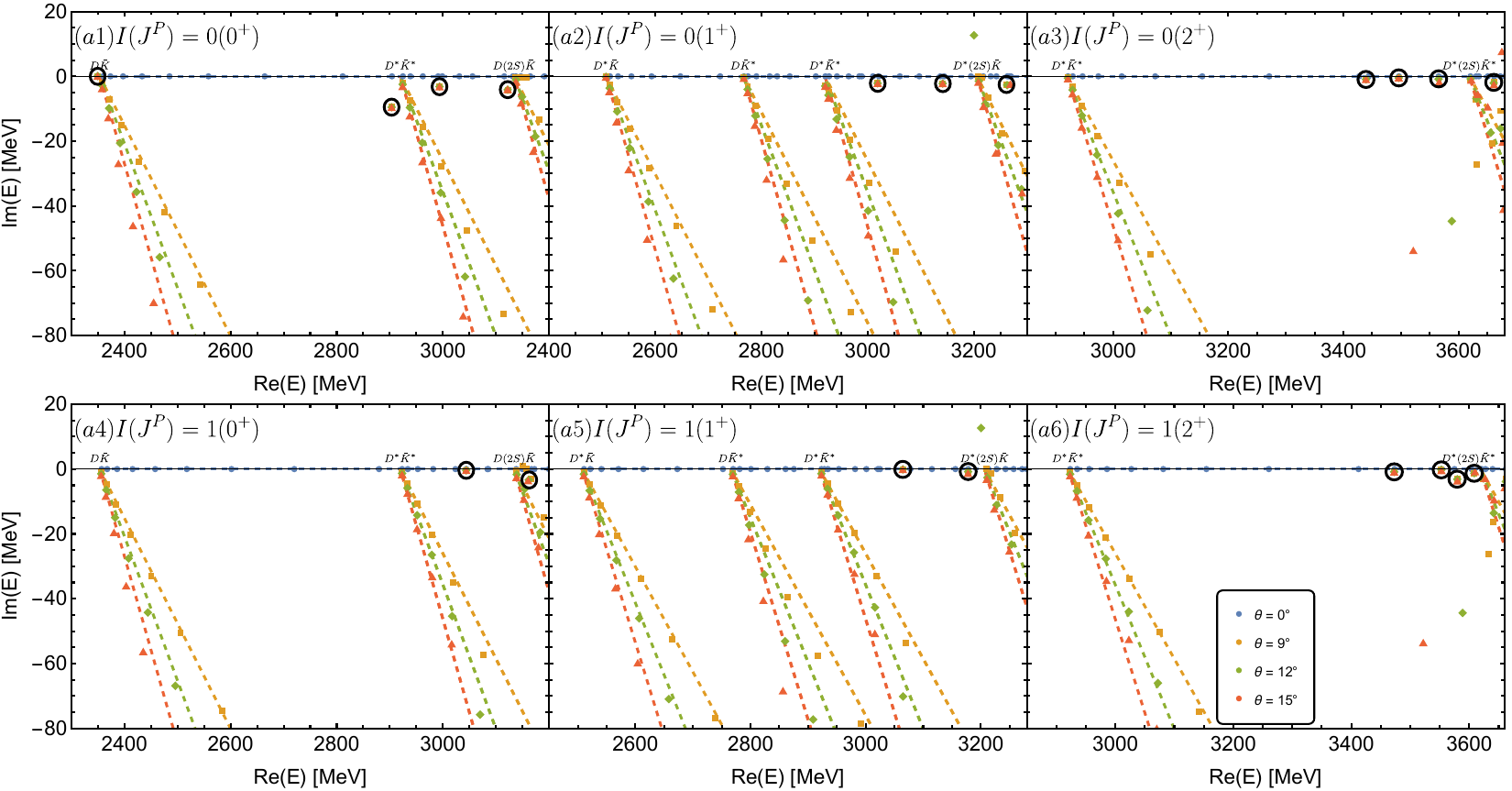}
	\caption{\label{fig:csnn_PlotGrid} Complex energy eigenvalues of the $cs\bar{n}\bar{n}$ states for different $I(J^P)$ quantum numbers in the AL1 potential with varying $\theta$ in the CSM. The dashed lines represent the continuum lines rotating along $\mathrm{Arg}(E)=-2\theta$. The bound and resonant states do not shift as $\theta$ changes and are marked out by the black circles.}
\end{figure*}

In this section, we systematically investigate the $S$-wave $cs\bar{n}\bar{n}$ tetraquark system with quantum numbers $I=0, 1$ and $J^P=0^+, 1^+, 2^+$. The complex energy spectra calculated with the CSM are presented in Fig.~\ref{fig:csnn_PlotGrid}. The numerical results for the extracted resonance poles and bound states, including their complex energies $E = M - i\Gamma/2$, color configurations, and rms radii, are summarized in Table~\ref{tab:csnn_poles}.

\begin{table*}[htbp]
	\caption{Numerical results for the bound state and resonance poles of the $cs\bar{n}\bar{n}$ system. The complex energies $E_{\rm H}$ (with H-only configurations~\cite{Chen:2023syh}) and $E_{\rm All}$ (with all Jacobi coordinates) are in units of MeV. The binding energy $\Delta E$ is in units of MeV. The proportions of color configurations ($\chi_{\bar{3}_c\otimes 3_c}$ and $\chi_{6_c\otimes\bar{6}_c}$) are given in percentages. The root-mean-square radii of the relative distances between quarks are in units of fm. The quarks are denoted as $c$, $s$, and $\bar{n}_3, \bar{n}_4 = \bar{u}/\bar{d}$. The last column denotes the spatial configuration of the states, where M. and C. represent molecular and compact configurations, respectively, and M.$^*$ denotes a molecular candidate whose intra-cluster sizes differ from those of the corresponding 1S or 2S mesons.}
	\label{tab:csnn_poles}
	\begin{ruledtabular}
		\begin{tabular}{ccccccccccccc}
			$I(J^P)$ & $E_{\rm H}$ & $E_{\rm All}$ & $\Delta E$ & $\chi_{\bar{3}_c\otimes3_c}$ & $\chi_{6_c\otimes\bar{6}_c}$ & $r_{c\bar{n}_3}$ & $r_{s\bar{n}_4}$ & $r_{cs}$ & $r_{\bar{n}_3\bar{n}_4}$ & $r_{c\bar{n}_4}$ & $r_{s\bar{n}_3}$ & Configuration \\
			\midrule
			\multirow{4}{*}{$0(0^+)$} 
			& $2350$ & $2350$ & $-3.8$ & 32\% & 68\% & 0.61 & 0.59 & 2.21 & 2.29 & 2.23 & 2.27 & M. \\
			& $2906-10i$ & $2903-10i$ & & 59\% & 41\% & 0.76 & 0.85 & 1.09 & 1.23 & 1.19 & 1.24 & C. \\
			& -- & $2994-3i$ & & 29\% & 71\% & 1.05 & 0.59 & 1.52 & 1.53 & 1.55 & 1.51 & C. \\
			& -- & $3122-4i$ & & 33\% & 67\% & 1.16 & 0.60 & 2.16 & 2.10 & 2.19 & 2.08 & M.$^*$ \\
			\midrule
			\multirow{3}{*}{$0(1^+)$}
			& -- & $3019-2i$ & & 31\% & 69\% & 1.08 & 0.59 & 1.57 & 1.61 & 1.59 & 1.58 & C. \\
			& -- & $3142-2i$ & & 33\% & 67\% & 1.17 & 0.60 & 2.23 & 2.03 & 2.25 & 2.01 & M.$^*$ \\
			& -- & $3262-3i$ & & 34\% & 66\% & 1.27 & 0.61 & 2.67 & 2.15 & 2.69 & 2.13 & M. \\
			\midrule
			\multirow{5}{*}{$0(2^+)$}
			& -- & $3439-1i$ & & 26\% & 74\% & 1.07 & 0.83 & 1.71 & 1.83 & 1.77 & 1.75 & C. \\
			& -- & $3496-0.5i$ & & 24\% & 76\% & 0.73 & 1.25 & 1.34 & 1.54 & 1.42 & 1.47 & C. \\
			& -- & $3565-0.8i$ & & 29\% & 71\% & 1.18 & 0.83 & 2.30 & 2.12 & 2.33 & 2.08 & M.$^*$ \\
			& $3607-2i$ & -- & & & & & & & & & & \\
			& -- & $3661-2i$ & & 34\% & 66\% & 0.73 & 1.41 & 2.60 & 2.75 & 2.69 & 2.67 & M. \\
			\midrule
			\multirow{2}{*}{$1(0^+)$}
			& -- & $3044-0.4i$ & & 36\% & 64\% & 1.07 & 0.60 & 1.79 & 1.85 & 1.82 & 1.82 & C. \\
			& -- & $3163-3i$ & & 36\% & 64\% & 1.15 & 0.62 & 2.39 & 2.61 & 2.41 & 2.59 & M.$^*$ \\
			\midrule
			\multirow{2}{*}{$1(1^+)$}
			& -- & $3064-0.1i$ & & 36\% & 64\% & 1.10 & 0.60 & 1.84 & 1.86 & 1.88 & 1.83 & C. \\
			& -- & $3178-0.1i$ & & 36\% & 64\% & 1.19 & 0.61 & 2.30 & 2.07 & 2.33 & 2.05 & M.$^*$ \\
			\midrule
			\multirow{5}{*}{$1(2^+)$}
			& -- & $3473-0.8i$ & & 44\% & 56\% & 1.10 & 0.86 & 1.72 & 1.79 & 1.82 & 1.75 & C. \\
			& -- & $3553-0.2i$ & & 53\% & 47\% & 0.95 & 1.15 & 1.54 & 1.68 & 1.73 & 1.60 & C. \\
			& -- & $3580-3i$ & & 60\% & 40\% & 1.08 & 1.07 & 1.62 & 1.64 & 1.70 & 1.64 & C. \\
			& $3605-3i$ & -- & & & & & & & & & & \\
			& -- & $3608-1i$ & & 57\% & 43\% & 1.25 & 1.03 & 1.70 & 1.72 & 1.86 & 1.77 & C. \\
		\end{tabular}
	\end{ruledtabular}
\end{table*}

For the isoscalar sector, we obtain a shallow bound state located at 2350 MeV in the $J^P=0^+$ channel, which lies 3.8 MeV below the $D\bar{K}$ theoretical threshold. Its calculated root-mean-square radii suggest a molecular spatial structure, consistent with previous calculations~\cite{Chen:2023syh}. Since this state exists below the lowest physical threshold, it can only decay weakly~\cite{Yu:2017pmn,Chen:2020eyu}. Above the threshold, we identify a resonance pole at $E = 2903 - 10i$ MeV. Theoretically, this result is in good agreement with both previous calculations~\cite{Chen:2023syh} and a recent comprehensive coupled-channel analysis within the meson exchange model~\cite{Wang:2024ukc}. Furthermore, considering the intrinsic uncertainties of the constituent quark model, the mass, width, and quantum numbers of this pole are generally consistent with the exotic $T_{cs0}(2900)$ state discovered experimentally by the LHCb collaboration.

\begin{figure*}[htbp]
	\centering
	\includegraphics[width=0.95\textwidth]{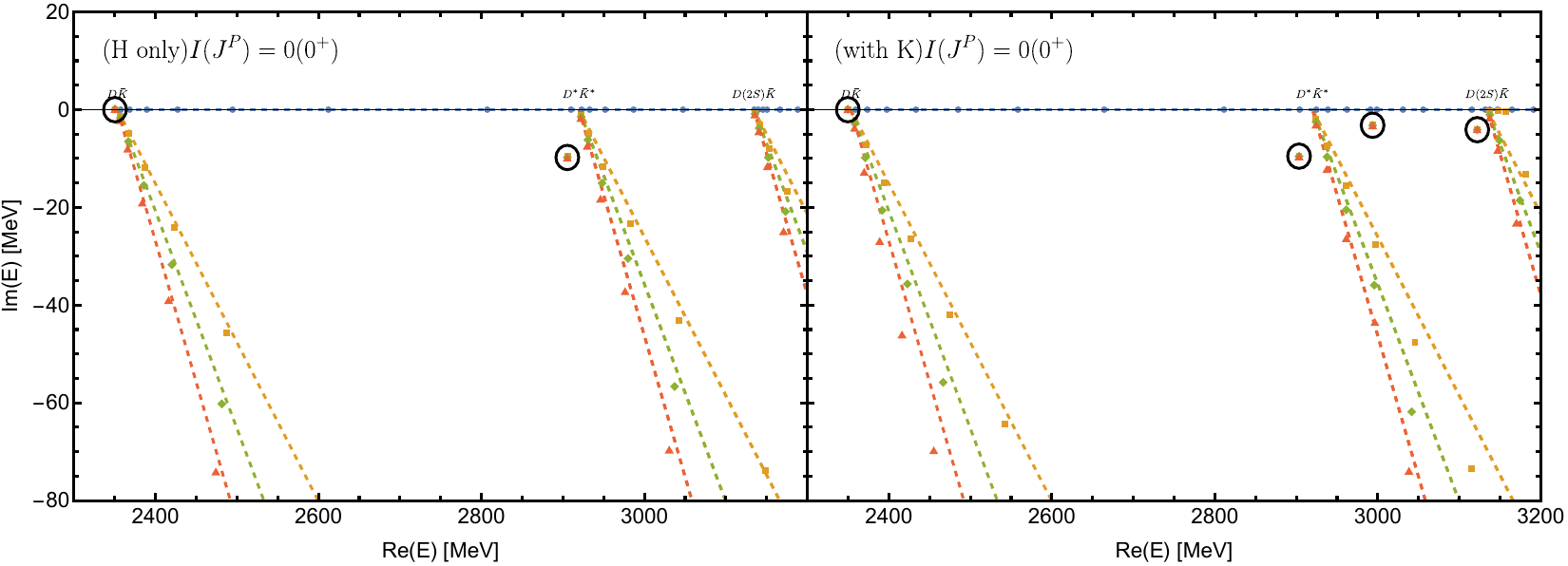}
	\caption{\label{fig:csnn_compare} Comparison of the complex energy spectra for the $cs\bar{n}\bar{n}$ $I(J^P)=0(0^+)$ system. The left panel shows the results obtained using exclusively H-type configurations~\cite{Chen:2023syh}, while the right panel incorporates all H- and K-type configurations. The stationary poles are marked by black circles.}
\end{figure*}

We now turn to the effects of the expanded K-type spatial basis. For brevity, we focus on several representative cases, including systems that contain bound states. To evaluate the dynamical effects of the K-type Jacobi coordinates, we compare our all-Jacobi results for the $S$-wave $cs\bar{n}\bar{n}$ $I(J^P)=0(0^+)$ system with those from the H-only calculation in Ref.~\cite{Chen:2023syh}. As shown in Fig.~\ref{fig:csnn_compare} and Table~\ref{tab:csnn_compare}, the effects of the K-type configurations differ considerably between low-lying and highly excited states.

For low-lying states, the conventional H-type configurations are generally adequate. For instance, upon expanding the basis, the lowest resonance pole experiences a negligible shift from $E = 2906 - 10i$~MeV to $E = 2903 - 10i$~MeV, indicating that resonance signals below the $1S2S$ thresholds are relatively reliable.

\begin{table*}[htbp]
	\caption{A comparative summary of the calculated resonance poles and spatial root-mean-square radii for the $cs\bar{n}\bar{n}$ state with $I(J^P)=0(0^+)$. We contrast the results obtained using the traditional H-only Jacobi coordinates with those derived from the all-Jacobi configurations. The binding energy $\Delta E$ and complex eigenenergy $E = M - i\Gamma/2$ are given in units of MeV, while the spatial sizes are expressed in fm. The constituent quarks are labeled as $c$, $s$, and $\bar{n}_{3,4}$.}
	\label{tab:csnn_compare}
	\setlength{\tabcolsep}{3pt}
	\footnotesize
	\begin{tabular}{cccccccccc|cccccccccc}
		\hline\hline
		\multicolumn{10}{c|}{\textbf{H-only framework}} & \multicolumn{10}{c}{\textbf{All-Jacobi framework}} \\
		\hline
		$E$ & $\Delta E$ & $\chi_{\bar{3}\otimes3}$ & $\chi_{6\otimes\bar{6}}$ & $r_{c\bar{n}_3}$ & $r_{s\bar{n}_4}$ & $r_{cs}$ & $r_{\bar{n}\bar{n}}$ & $r_{c\bar{n}_4}$ & $r_{s\bar{n}_3}$ & $E$ & $\Delta E$ & $\chi_{\bar{3}\otimes3}$ & $\chi_{6\otimes\bar{6}}$ & $r_{c\bar{n}_3}$ & $r_{s\bar{n}_4}$ & $r_{cs}$ & $r_{\bar{n}\bar{n}}$ & $r_{c\bar{n}_4}$ & $r_{s\bar{n}_3}$ \\
		\hline
		$2350$ & $-3.1$ & 32\% & 68\% & 0.61 & 0.59 & 2.45 & 2.52 & 2.47 & 2.50 & $2350$ & $-3.8$ & 32\% & 68\% & 0.61 & 0.59 & 2.21 & 2.29 & 2.23 & 2.27 \\
		$2906-10i$ & & 57\% & 43\% & 0.75 & 0.86 & 1.11 & 1.25 & 1.20 & 1.26 & $2903-10i$ & & 59\% & 41\% & 0.76 & 0.85 & 1.09 & 1.23 & 1.19 & 1.24 \\
		& &      &      &      &      &      &      &      &      & $2994-3i$  & & 29\% & 71\% & 1.05 & 0.59 & 1.52 & 1.53 & 1.55 & 1.51 \\
		& &      &      &      &      &      &      &      &      & $3122-4i$  & & 33\% & 67\% & 1.16 & 0.60 & 2.16 & 2.10 & 2.19 & 2.08 \\
		\hline\hline
	\end{tabular}
\end{table*}

In contrast, for highly excited states, a strong dependence on the spatial degrees of freedom provided by the K-type structures is observed. At higher excitation energies, the complex multi-quark spatial structures cannot be adequately captured by simple dimeson or diquark-antidiquark clustering. By incorporating all K-type configurations, we uncover two distinct resonance states located at $E = 2994 - 3i$~MeV and $E = 3122 - 4i$~MeV. Notably, these poles were entirely unresolved in the H-only calculation within the same energy window. This comparison clearly demonstrates that while a limited basis suffices for low-lying states, employing the comprehensive spatial configurations is indispensable for authentically mapping the highly excited multi-quark dynamics.

A broader comparison of the complex energies between the H-only and all-Jacobi calculations across all channels can be found in the aforementioned Table~\ref{tab:csnn_poles}. 

For the $I(J^P)=0(1^+)$ channel, the relevant physical thresholds include the ground-state $D^*\bar{K}$ and $D\bar{K}^*$ channels, as well as their higher excitation thresholds. As shown in Fig.~\ref{fig:csnn_PlotGrid} (a2), the distribution of the complex energy eigenvalues reveals three distinct resonance poles at $3019$~MeV, $3142$~MeV, and $3262$~MeV.

For the highest spin configuration, $I(J^P)=0(2^+)$, the lowest allowed $S$-wave threshold is $D^*\bar{K}^*$. As depicted in Fig.~\ref{fig:csnn_PlotGrid} (a3), we extract several relatively narrow resonances in this channel. These include the lowest pole located at $E = 3439 - 1i$~MeV, along with two higher-mass states at $E = 3496 - 0.5i$~MeV and $E = 3565 - 0.8i$~MeV.

For the isovector sector of the $cs\bar{n}\bar{n}$ system, resonances are extracted across the $J^P=0^+, 1^+$, and $2^+$ channels. For the $1(0^+)$ states, two poles are identified at $E = 3044 - 0.4i$~MeV and $E = 3163 - 3i$~MeV. For the $1(1^+)$ channel, two narrow states are located at $E = 3064 - 0.1i$~MeV and $E = 3178 - 0.1i$~MeV.

For the highest spin channel, $1(2^+)$, four resonances are extracted within the considered range. Their decay widths are consistently small, with the broadest being only $\Gamma = 6$~MeV for the state $E = 3580 - 3i$~MeV. Notably, the internal color structure undergoes a transition in these higher spin states, where the $\bar{3}_c \otimes 3_c$ configuration gradually becomes favored, reaching $60\%$ for the resonance poles. 

\subsection{$bs\bar{n}\bar{n}$}~\label{subsec:bsnn}
	
\begin{figure*}[htbp]
		\centering
		\includegraphics[width=0.95\textwidth]{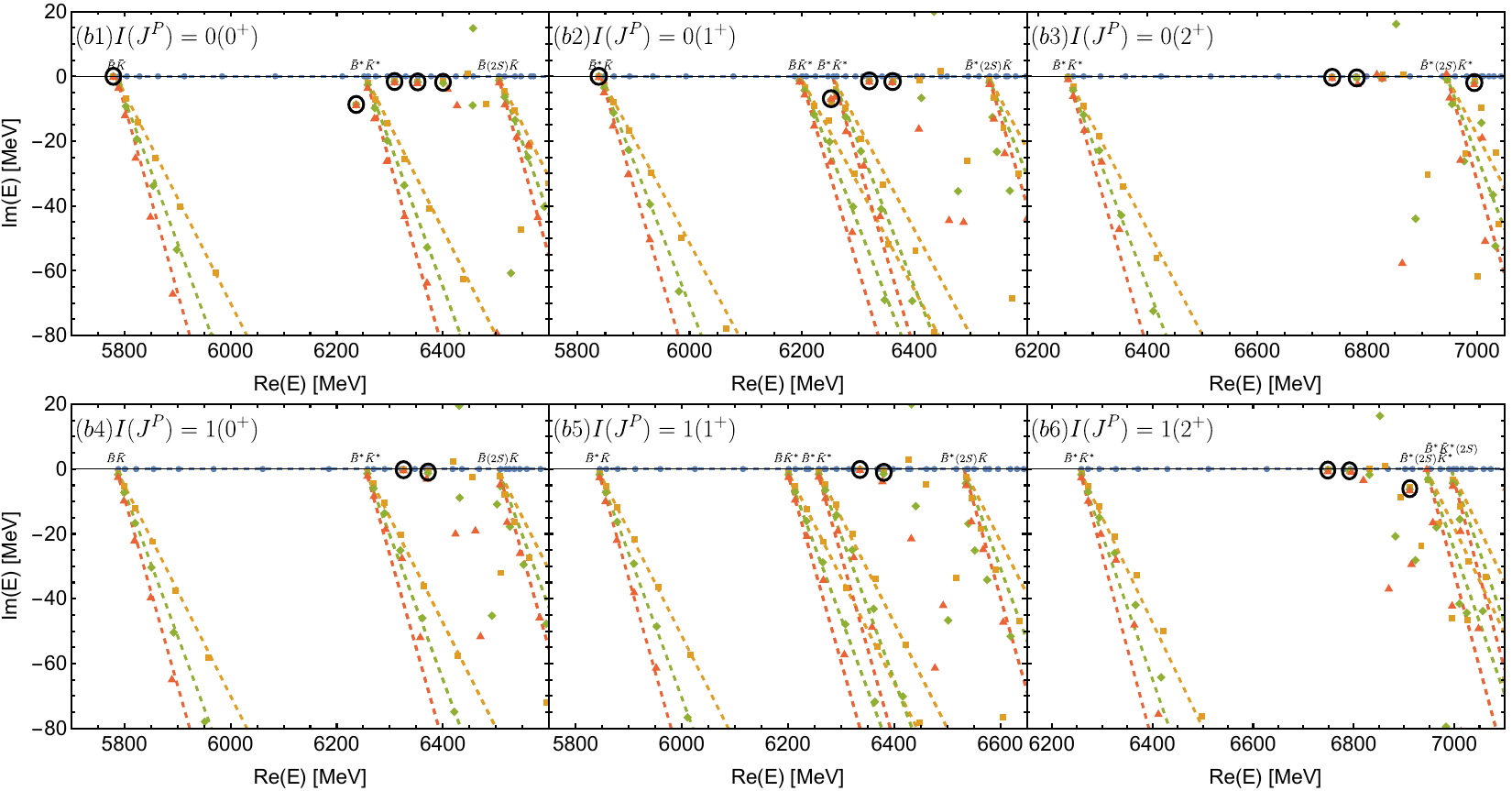}
		\caption{\label{fig:bsnn_PlotGrid} Complex energy eigenvalues of the $bs\bar{n}\bar{n}$ states for different $I(J^P)$ quantum numbers in the AL1 potential with varying $\theta$ in the CSM. The dashed lines represent the continuum lines rotating along $\mathrm{Arg}(E)=-2\theta$. The bound and resonant states do not shift as $\theta$ changes and are marked out by the black circles.}
	\end{figure*}

	For the $S$-wave $bs\bar{n}\bar{n}$ tetraquark system, the complex energy spectra for both the isoscalar and isovector sectors are presented in Fig.~\ref{fig:bsnn_PlotGrid}. The extracted bound states and resonance poles, including their masses, decay widths, proportions of color configurations, and root-mean-square radii, are summarized in Table~\ref{tab:bsnn_poles}.
	
\begin{table*}[htbp]
	\caption{Numerical results for the bound state and resonance poles of the $bs\bar{n}\bar{n}$ system. The complex energies $E_{\rm H}$ (with H-only configurations~\cite{Chen:2023syh}) and $E_{\rm All}$ (with all Jacobi coordinates) are in units of MeV. The binding energy $\Delta E$ is in units of MeV. The proportions of color configurations ($\chi_{\bar{3}_c\otimes 3_c}$ and $\chi_{6_c\otimes\bar{6}_c}$) are given in percentages. The root-mean-square radii of the relative distances between quarks are in units of fm. The ``?" indicates that the rms radii results are numerically unstable. The quarks are denoted as $b$, $s$, and $\bar{n}_3, \bar{n}_4 = \bar{u}/\bar{d}$. The last column denotes the spatial configuration of the states, where M. and C. represent molecular and compact configurations, respectively, and M.$^*$ denotes a molecular candidate whose intra-cluster sizes differ from those of the corresponding 1S or 2S mesons.}
	\label{tab:bsnn_poles}
	\begin{ruledtabular}
		\begin{tabular}{ccccccccccccc}
			$I(J^P)$ & $E_{\rm H}$ & $E_{\rm All}$ & $\Delta E$ & $\chi_{\bar{3}_c\otimes3_c}$ & $\chi_{6_c\otimes\bar{6}_c}$ & $r_{b\bar{n}_3}$ & $r_{s\bar{n}_4}$ & $r_{bs}$ & $r_{\bar{n}_3\bar{n}_4}$ & $r_{b\bar{n}_4}$ & $r_{s\bar{n}_3}$ & Configuration \\
			\midrule
			\multirow{5}{*}{$0(0^+)$} 
			& $5781$ & $5779$ & $-5.3$ & 32\% & 68\% & 0.63 & 0.59 & 1.86 & 1.96 & 1.88 & 1.94 & M. \\
			& $6240-9i$ & $6237-9i$ & & 58\% & 42\% & 0.71 & 0.85 & 1.03 & 1.19 & 1.12 & 1.19 & C. \\
			& -- & $6309-2i$ & & 31\% & 69\% & 0.99 & 0.59 & 2.00 & 2.05 & 2.03 & 2.03 & M.$^*$ \\
			& -- & $6353-2i$ & & 32\% & 68\% & 1.02 & 0.59 & 3.08 & 2.94 & 3.09 & 2.92 & M.$^*$ \\
			& -- & $6400-2i$ & & 33\% & 67\% & 1.06 & 0.59 & 3.76 & 3.51 & 3.77 & 3.49 & M.$^*$ \\
			\midrule
			\multirow{4}{*}{$0(1^+)$}
			& $5840$ & $5838$ & $-3.0$ & 32\% & 68\% & 0.66 & 0.59 & 2.18 & 2.28 & 2.20 & 2.26 & M. \\
			& $6253-6i$ & $6252-7i$ & & 32\% & 68\% & 0.68 & 0.83 & ? & ? & ? & ? & ? \\
			& -- & $6320-1i$ & & 32\% & 68\% & 1.00 & 0.59 & 2.06 & 2.11 & 2.09 & 2.09 & M.$^*$ \\
			& -- & $6361-2i$ & & 32\% & 68\% & 1.03 & 0.59 & 3.11 & 2.97 & 3.13 & 2.95 & M.$^*$ \\
			\midrule
			\multirow{3}{*}{$0(2^+)$}
			& -- & $6736-0.3i$ & & 30\% & 70\% & 1.00 & 0.82 & 2.28 & 2.33 & 2.32 & 2.28 & M.$^*$ \\
			& -- & $6781-0.5i$ & & 31\% & 69\% & 1.03 & 0.82 & 3.27 & 3.13 & 3.30 & 3.10 & M.$^*$ \\
			& -- & $6995-2i$ & & 35\% & 75\% & 0.69 & 1.41 & 2.86 & 3.01 & 2.94 & 2.93 & M. \\
			\midrule
			\multirow{2}{*}{$1(0^+)$}
			& -- & $6326-0.2i$ & & 34\% & 66\% & 1.00 & 0.59 & 2.61 & 2.57 & 2.63 & 2.55 & M.$^*$ \\
			& -- & $6372-0.9i$ & & 34\% & 66\% & 1.03 & 0.59 & 3.32 & 3.16 & 3.34 & 3.14 & M.$^*$ \\
			\midrule
			\multirow{2}{*}{$1(1^+)$}
			& -- & $6336-0.1i$ & & 34\% & 66\% & 1.01 & 0.59 & 2.65 & 2.61 & 2.67 & 2.59 & M.$^*$ \\
			& -- & $6381-1i$ & & 34\% & 66\% & 1.04 & 0.60 & 3.35 & 3.19 & 3.37 & 3.17 & M.$^*$ \\
			\midrule
			\multirow{4}{*}{$1(2^+)$}
			& -- & $6748-0.3i$ & & 36\% & 64\% & 1.01 & 0.82 & 2.63 & 2.61 & 2.67 & 2.57 & M.$^*$ \\
			& -- & $6791-0.5i$ & & 40\% & 60\% & 1.04 & 0.84 & 3.17 & 3.03 & 3.21 & 3.01 & M.$^*$ \\
			& -- & $6912-6i$ & & 54\% & 46\% & 0.87 & 1.20 & 1.22 & 1.37 & 1.21 & 1.45 & C. \\
			& $6916-2i$ & -- & & & & & & & & & & \\
		\end{tabular}
	\end{ruledtabular}
\end{table*}
	
	For the isoscalar sector, we obtain two bound states below the lowest $S$-wave two-body thresholds. In the $0(0^+)$ channel, a bound state is found at $M = 5779$~MeV with a binding energy of $-5.3$~MeV relative to the theoretical $\bar{B}\bar{K}$ threshold. In the $0(1^+)$ channel, another bound state is obtained at $M = 5838$~MeV with $\Delta E = -3.0$~MeV below the $\bar{B}^*\bar{K}$ threshold. As listed in Table~\ref{tab:bsnn_poles}, the $r_{b\bar{n}_3}$ ($0.63$~fm and $0.66$~fm) and $r_{s\bar{n}_4}$ ($0.59$~fm) of these two bound states are highly comparable to the theoretical rms radii of the individual $\bar{B}$ ($0.63$~fm), $\bar{B}^*$ ($0.66$~fm), and $\bar{K}$ ($0.59$~fm) mesons, respectively. Furthermore, these intra-cluster sizes are significantly smaller than the other relative distances in the tetraquark system, which extend from $1.86$~fm to $2.28$~fm, indicating a molecular spatial structure. Since these states exist below the physical thresholds, they can only decay weakly or electromagnetically~\cite{Yu:2017pmn,Qin:2020zlg}.
	
\begin{figure*}[htbp]
		\centering
		\includegraphics[width=0.95\textwidth]{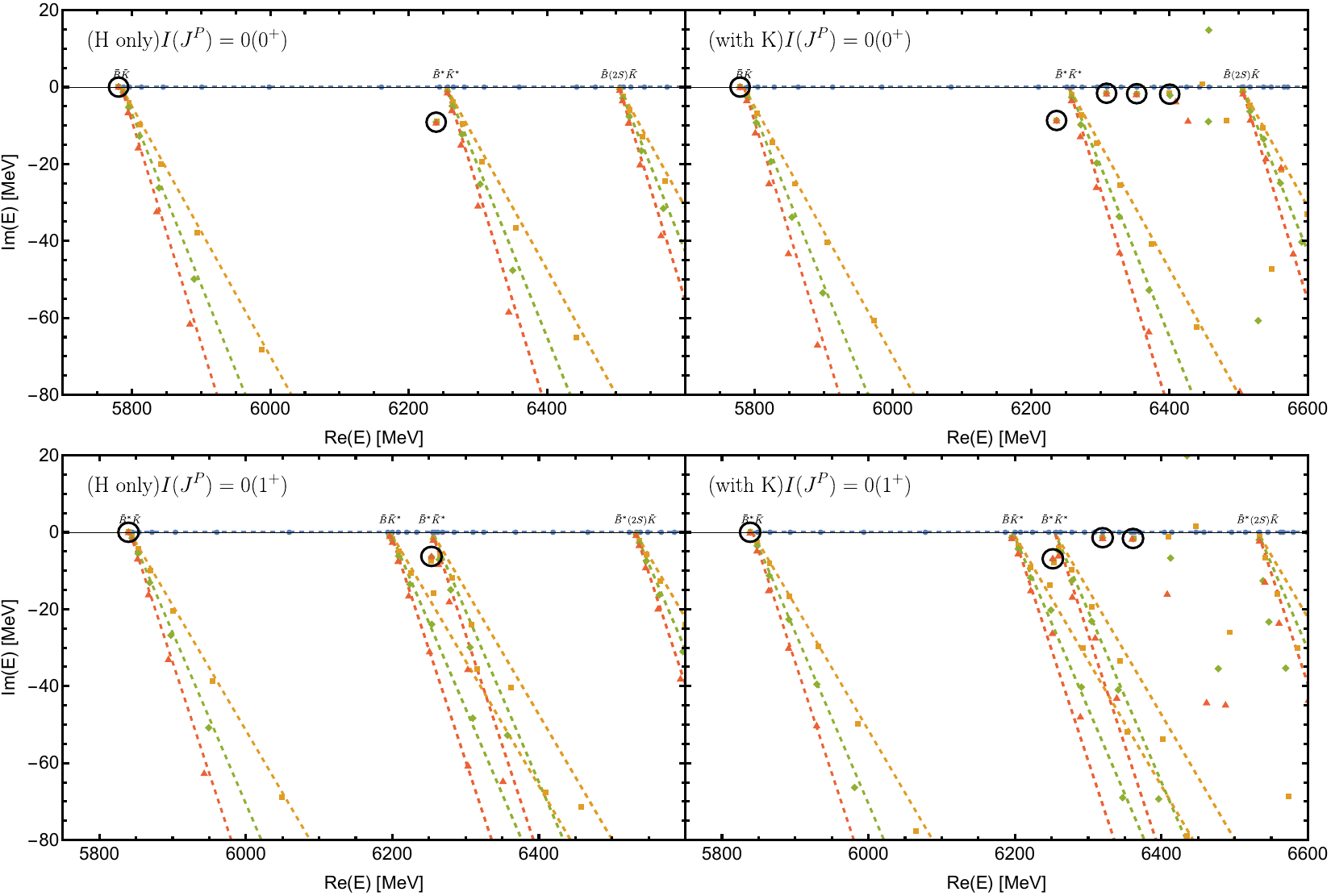}
		\caption{\label{fig:bsnn_compare} Comparison of the complex energy spectra for the $bs\bar{n}\bar{n}$ $I(J^P)=0(0^+)$ and $0(1^+)$ systems. The left panel shows the results obtained using exclusively H-type configurations, while the right panel incorporates all H- and K-type configurations. The stationary poles are marked by black circles.}
	\end{figure*}

	To demonstrate the impact of the K-type Jacobi coordinates in the $bs\bar{n}\bar{n}$ system, we compare the results obtained using the restricted H-only basis with those from the comprehensive all-Jacobi framework for the isoscalar $0^+$ and $1^+$ channels, as illustrated in Fig.~\ref{fig:bsnn_compare} and detailed in Table~\ref{tab:bsnn_compare}. A broader comparison of the complex energies between the H-only and all-Jacobi calculations across all channels can be found in the aforementioned Table~\ref{tab:bsnn_poles}.
	
	\begin{table*}[htbp]
		\caption{A comparative summary of the calculated resonance poles and spatial root-mean-square radii for the $bs\bar{n}\bar{n}$ states with $I(J^P)=0(0^+)$ and $0(1^+)$. We contrast the results obtained using the traditional H-only Jacobi coordinates with those derived from the all-Jacobi configurations. The binding energy $\Delta E$ and complex eigenenergy $E = M - i\Gamma/2$ are given in units of MeV, while the spatial sizes are expressed in fm. The ``?" indicates that the rms radii results are numerically unstable. The constituent quarks are labeled as $b$, $s$, and $\bar{n}_{3,4}$.}
		\label{tab:bsnn_compare}
		\resizebox{\textwidth}{!}{
			\begin{tabular}{ccccccccccc|cccccccccc}
				\hline\hline
				\multicolumn{11}{c|}{\textbf{H-only framework}} & \multicolumn{10}{c}{\textbf{All-Jacobi framework}} \\
				\hline
				$J^P$ & $E$ & $\Delta E$ & $\chi_{\bar{3}\otimes3}$ & $\chi_{6\otimes\bar{6}}$ & $r_{b\bar{n}_3}$ & $r_{s\bar{n}_4}$ & $r_{bs}$ & $r_{\bar{n}\bar{n}}$ & $r_{b\bar{n}_4}$ & $r_{s\bar{n}_3}$ & $E$ & $\Delta E$ & $\chi_{\bar{3}\otimes3}$ & $\chi_{6\otimes\bar{6}}$ & $r_{b\bar{n}_3}$ & $r_{s\bar{n}_4}$ & $r_{bs}$ & $r_{\bar{n}\bar{n}}$ & $r_{b\bar{n}_4}$ & $r_{s\bar{n}_3}$ \\
				\hline
				\multirow{5}{*}{$0^+$} & $5781$ & $-3.9$ & 32\% & 68\% & 0.63 & 0.59 & 2.12 & 2.22 & 2.15 & 2.20 & $5779$ & $-5.3$ & 32\% & 68\% & 0.63 & 0.59 & 1.86 & 1.96 & 1.88 & 1.94 \\
				& $6240-9i$ & & 57\% & 43\% & 0.70 & 0.86 & 1.05 & 1.22 & 1.13 & 1.22 & $6237-9i$ & & 58\% & 42\% & 0.71 & 0.85 & 1.03 & 1.19 & 1.12 & 1.19 \\
				& & & & & & & & & & & $6309-2i$ & & 31\% & 69\% & 0.99 & 0.59 & 2.00 & 2.05 & 2.03 & 2.03 \\
				& & & & & & & & & & & $6353-2i$ & & 32\% & 68\% & 1.02 & 0.59 & 3.08 & 2.94 & 3.09 & 2.92 \\
				& & & & & & & & & & & $6400-2i$ & & 33\% & 67\% & 1.06 & 0.59 & 3.76 & 3.51 & 3.77 & 3.49 \\
				\hline
				\multirow{4}{*}{$1^+$} & $5840$ & $-1.8$ & 32\% & 68\% & 0.66 & 0.59 & 2.77 & 2.85 & 2.78 & 2.83 & $5838$ & $-3.0$ & 32\% & 68\% & 0.66 & 0.59 & 2.18 & 2.28 & 2.20 & 2.26 \\
				& $6253-6i$ & & 32\% & 68\% & 0.67 & 0.83 & ? & ? & ? & ? & $6252-7i$ & & 32\% & 68\% & 0.68 & 0.83 & ? & ? & ? & ? \\
				& & & & & & & & & & & $6320-1i$ & & 32\% & 68\% & 1.00 & 0.59 & 2.06 & 2.11 & 2.09 & 2.09 \\
				& & & & & & & & & & & $6361-2i$ & & 32\% & 68\% & 1.03 & 0.59 & 3.11 & 2.97 & 3.13 & 2.95 \\
				\hline\hline
			\end{tabular}
		}
	\end{table*}
	
	Above the physical thresholds, the expanded spatial basis becomes crucial. For the $0(0^+)$ channel, both frameworks identify a resonance near $6240$~MeV, which can be regarded as the bottom partner of the $T_{cs0}(2900)$ state due to its color configurations and rms radii. However, the all-Jacobi calculation successfully extracts three additional narrow resonances at $6309$~MeV, $6353$~MeV, and $6400$~MeV. The states at $6353$~MeV and $6400$~MeV exhibit significantly large inter-cluster distances up to $3.7$~fm, representing extended multi-quark configurations. These highly excited resonances are entirely unresolved in the H-only continuum, demonstrating the necessity of the K-type structures for accurately capturing the complex spatial correlations at higher energies.
	
	For the $0(1^+)$ channel, a similar pattern emerges. We identify a resonance pole near $6252$~MeV in both frameworks. Nevertheless, the all-Jacobi approach successfully resolves two higher-mass compact resonances at $E = 6320 - 1i$~MeV and $E = 6361 - 2i$~MeV, highlighting once again the indispensability of the complete spatial degrees of freedom.
	
	For the $0(2^+)$ channel, three relatively narrow resonances are extracted from $6736$~MeV to $6995$~MeV, with widths ranging from $0.6$~MeV to $4$~MeV. The state at $6781 - 0.5i$~MeV is identified as a molecular resonance, while the others are compact ones.
	
	In the isovector sector, our framework extracts several narrow resonance poles at elevated energies for the $0^+, 1^+$, and $2^+$ channels. Examples include the states located at $E = 6326 - 0.2i$~MeV and $E = 6336 - 0.1i$~MeV in the $1(0^+)$ and $1(1^+)$ quantum numbers, respectively. According to the numerical results in Table~\ref{tab:bsnn_poles}, these $I=1$ resonances generally possess large rms radii exceeding $2.5$~fm, with the maximum extension reaching approximately $3.3$~fm. Despite such significant spatial sizes, the comparable relative distances between the constituent quarks in several states denoted by ``C." indicate that they are extended multiquark resonances where all four (anti)quarks play similar roles, rather than typical hadronic molecules.

\subsection{$cc\bar{n}\bar{n}$}~\label{subsec:ccnn}
	
\begin{figure*}[htbp]
		\centering
		\includegraphics[width=0.95\textwidth]{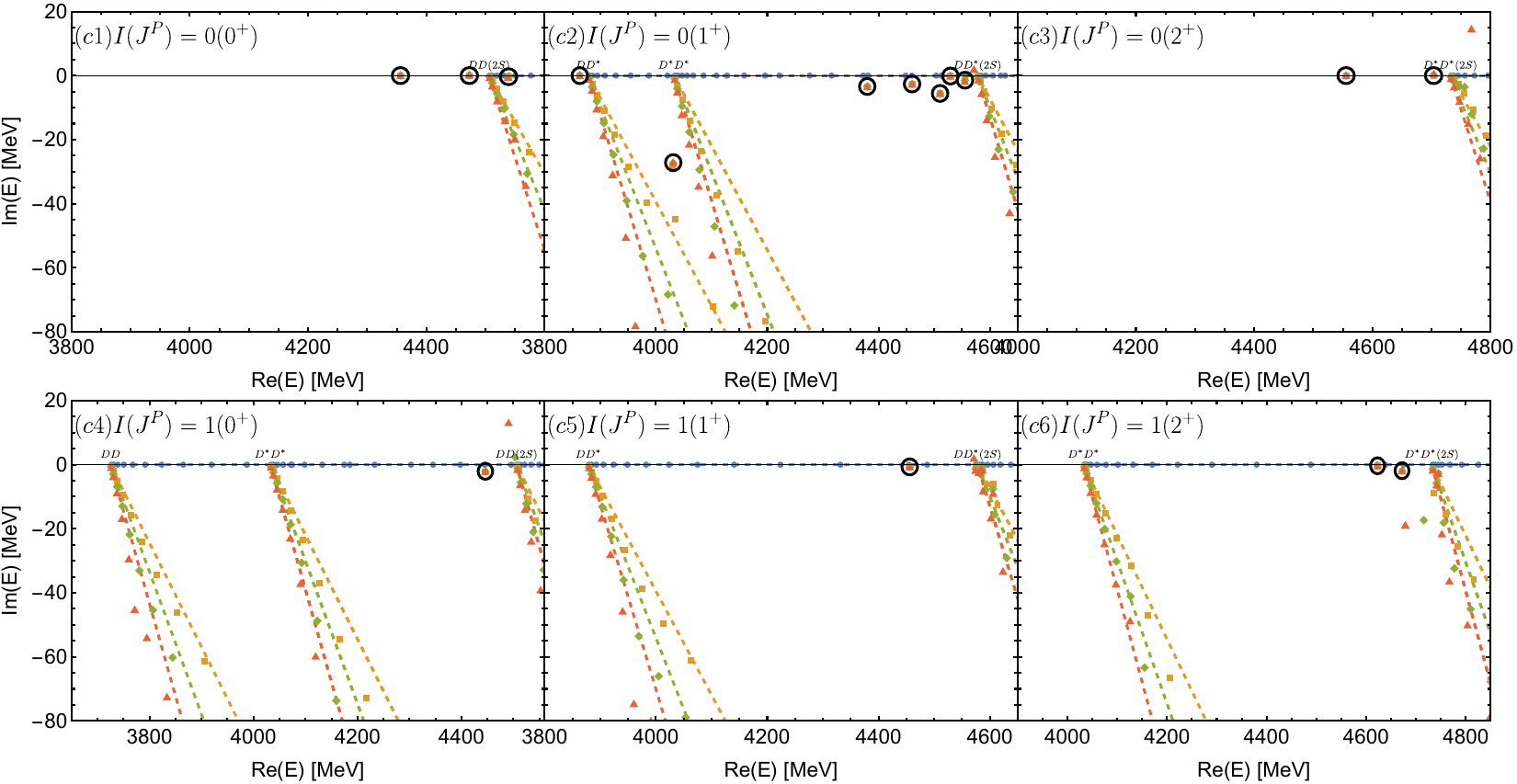}
		\caption{\label{fig:ccnn_PlotGrid} Complex energy eigenvalues of the $cc\bar{n}\bar{n}$ states for different $I(J^P)$ quantum numbers in the AL1 potential with varying $\theta$ in the CSM. The dashed lines represent the continuum lines rotating along $\mathrm{Arg}(E)=-2\theta$. The bound and resonant states do not shift as $\theta$ changes and are marked out by the black circles.}
	\end{figure*}

\begin{figure*}[htbp]
		\centering
		\includegraphics[width=0.95\textwidth]{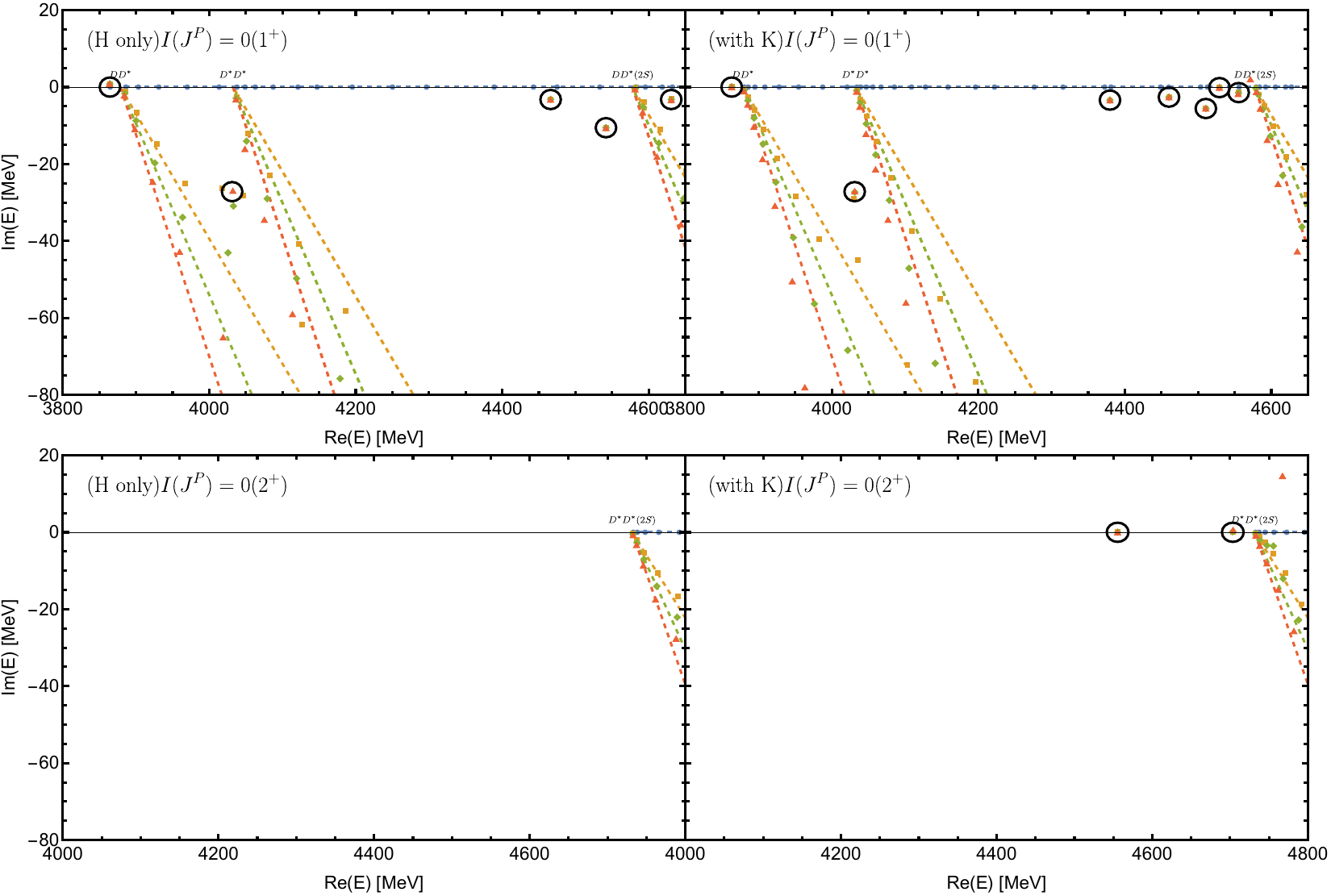}
		\caption{\label{fig:ccnn_compare} Comparison of the complex energy spectra for the $cc\bar{n}\bar{n}$ $I(J^P)=0(1^+)$ and $0(2^+)$ systems. The left panels show the results obtained using exclusively H-type configurations, while the right panels incorporate all H- and K-type configurations. The stationary poles are marked by black circles.}
	\end{figure*}

	In this section, we investigate the $S$-wave $cc\bar{n}\bar{n}$ tetraquark system. The complex energy spectra for different isospin and spin-parity channels are shown in Fig.~\ref{fig:ccnn_PlotGrid}, and the numerical results for the extracted states are summarized in Table~\ref{tab:ccnn_poles}. To demonstrate the effects of K-type Jacobi coordinates, we take the $I(J^P)=0(1^+)$ and $0(2^+)$ channels as examples to compare the results using the traditional H-only basis and the all-Jacobi basis. The comparative results are detailed in Table~\ref{tab:ccnn_compare} and Fig.~\ref{fig:ccnn_compare}. A broader comparison of the complex energies between the H-only and all-Jacobi calculations across all channels can be found in Table~\ref{tab:ccnn_poles}.
	
	\begin{table*}[htbp]
		\caption{Numerical results for the bound state and resonance poles of the $cc\bar{n}\bar{n}$ system. The complex energies $E_{\rm H}$ (with H-only configurations~\cite{Wu:2024zbx}) and $E_{\rm All}$ (with all Jacobi coordinates) are in units of MeV. The binding energy $\Delta E$ is in units of MeV. The proportions of color configurations ($\chi_{\bar{3}_c\otimes 3_c}$ and $\chi_{6_c\otimes\bar{6}_c}$) are given in percentages. The root-mean-square radii of the relative distances between quarks are in units of fm. The ``?" indicates that the rms radii results are numerically unstable. The quarks are denoted as $c_1$, $c_2$, and $\bar{n}_3, \bar{n}_4 = \bar{u}/\bar{d}$. The last column denotes the spatial configuration of the states, where M. and C. represent molecular and compact configurations, respectively, and M.$^*$ denotes a molecular candidate whose intra-cluster sizes differ from those of the corresponding 1S or 2S mesons.}
		\label{tab:ccnn_poles}
		\begin{ruledtabular}
			\begin{tabular}{ccccccccccccc}
				$I(J^P)$ & $E_{\rm H}$ & $E_{\rm All}$ & $\Delta E$ & $\chi_{\bar{3}_c\otimes3_c}$ & $\chi_{6_c\otimes\bar{6}_c}$ & $r_{c_1\bar{n}_3}$ & $r_{c_2\bar{n}_4}$ & $r_{c_1c_2}$ & $r_{\bar{n}_3\bar{n}_4}$ & $r_{c_1\bar{n}_4}$ & $r_{c_2\bar{n}_3}$ & Configuration \\
				\midrule
				\multirow{3}{*}{$0(0^+)$} 
				& -- & $4356$ & $-149$ & 40\% & 60\% & 0.88 & 0.88 & 1.07 & 1.14 & 1.17 & 1.17 & C. \\
				& -- & $4473$ & $-33$  & 66\% & 34\% & 0.95 & 0.95 & 1.07 & 1.14 & 1.23 & 1.23 & C. \\
				& -- & $4539-0.3i$&        & 32\% & 68\% & 0.95 & 0.95 & 1.44 & 1.41 & 1.49 & 1.49 & C. \\
				\midrule
				\multirow{9}{*}{$0(1^+)$}
				& $3864$ & $3863$ & $-15$  & 58\% & 42\% & 0.71 & 0.64 & 1.01 & 1.21 & 1.13 & 1.15 & M. \\
				& $4031-27i$ & $4031-27i$ &        & 1\%  & 99\% & 0.71 & 0.76 & ? & ? & ? & ? & ? \\
				& -- & $4380-3i$ &        & 37\% & 63\% & 1.07 & 0.65 & 1.16 & 1.36 & 1.26 & 1.27 & C. \\
				& $4466-3i$ & $4460-3i$  &        & 82\% & 18\% & 1.11 & 1.06 & 0.74 & 0.98 & 1.14 & 1.21 & C. \\
				& -- & $4511-6i$ &        & 55\% & 45\% & 0.97 & 0.94 & 1.10 & 1.31 & 1.29 & 1.26 & C. \\
				& -- & $4529-0.1i$&        & 22\% & 78\% & 0.90 & 0.94 & 1.36 & 1.52 & 1.44 & 1.42 & C. \\
				& $4542-10i$ & -- &        & & & & & & & & & \\
				& -- & $4555-1i$ &        & 37\% & 63\% & 1.12 & 0.77 & 1.47 & 1.54 & 1.57 & 1.43 & C. \\
				& $4631-3i$ & -- &        & & & & & & & & & \\
				\midrule
				\multirow{2}{*}{$0(2^+)$}
				& -- & $4555$ & $-176$  & 22\% & 78\% & 0.91 & 0.91 & 1.26 & 1.42 & 1.39 & 1.39 & C. \\
				& -- & $4703$ & $-28$  & 28\% & 72\% & 1.00 & 1.00 & 1.56 & 1.52 & 1.58 & 1.58 & C. \\
				\midrule
				\multirow{1}{*}{$1(0^+)$}
				& -- & $4446-2i$ &        & 40\% & 60\% & 0.89 & 0.89 & 1.27 & 1.65 & 1.45 & 1.45 & C. \\
				\midrule
				\multirow{1}{*}{$1(1^+)$}
				& -- & $4456-0.7i$&        & 40\% & 60\% & 1.10 & 0.65 & 1.26 & 1.51 & 1.40 & 1.43 & C. \\
				\midrule
				\multirow{3}{*}{$1(2^+)$}
				& -- & $4623-0.3i$&        & 62\% & 38\% & 0.96 & 0.96 & 1.16 & 1.45 & 1.31 & 1.31 & C. \\
				& -- & $4672-2i$ &        & 83\% & 17\% & 1.14 & 1.14 & 0.76 & 1.29 & 1.31 & 1.31 & C. \\
				& $4673-2i$ & -- &        & & & & & & & & & \\
			\end{tabular}
		\end{ruledtabular}
	\end{table*}
	
	\begin{table*}[htbp]
		\caption{A comparative summary of the calculated resonance poles and spatial root-mean-square radii for the $cc\bar{n}\bar{n}$ states with $I(J^P)=0(1^+)$ and $0(2^+)$. We contrast the results obtained using the traditional H-only Jacobi coordinates with those derived from the all-Jacobi configurations. The binding energy $\Delta E$ and complex eigenenergy $E = M - i\Gamma/2$ are given in units of MeV, while the spatial sizes are expressed in fm. The ``?" indicates that the rms radii results are numerically unstable. The constituent quarks are labeled as $c_{1,2}$ and $\bar{n}_{3,4}$.}
		\label{tab:ccnn_compare}
		\resizebox{\textwidth}{!}{
			\begin{tabular}{ccccccccccc|cccccccccc}
				\hline\hline
				\multicolumn{11}{c|}{\textbf{H-only framework}} & \multicolumn{10}{c}{\textbf{All-Jacobi framework}} \\
				\hline
				$J^P$ & $E$ & $\Delta E$ & $\chi_{\bar{3}\otimes3}$ & $\chi_{6\otimes\bar{6}}$ & $r_{c_1\bar{n}_3}$ & $r_{c_2\bar{n}_4}$ & $r_{c_1c_2}$ & $r_{\bar{n}_3\bar{n}_4}$ & $r_{c_1\bar{n}_4}$ & $r_{c_2\bar{n}_3}$ & $E$ & $\Delta E$ & $\chi_{\bar{3}\otimes3}$ & $\chi_{6\otimes\bar{6}}$ & $r_{c_1\bar{n}_3}$ & $r_{c_2\bar{n}_4}$ & $r_{c_1c_2}$ & $r_{\bar{n}_3\bar{n}_4}$ & $r_{c_1\bar{n}_4}$ & $r_{c_2\bar{n}_3}$ \\
				\hline
				\multirow{9}{*}{$1^+$} & $3864$ & $-14$ & 58\% & 42\% & 0.71 & 0.64 & 1.01 & 1.22 & 1.13 & 1.16 & $3863$ & $-15$ & 58\% & 42\% & 0.71 & 0.64 & 1.01 & 1.21 & 1.13 & 1.15 \\
				& $4031-27i$ & & 0\% & 100\% & 0.71 & 0.75 & ? & ? & ? & ? & $4031-27i$ & & 1\% & 99\% & 0.71 & 0.76 & ? & ? & ? & ? \\
				& & & & & & & & & & & $4380-3i$ & & 37\% & 63\% & 1.07 & 0.65 & 1.16 & 1.36 & 1.26 & 1.27 \\
				& $4466-3i$ & & 91\% & 9\% & 1.13 & 1.09 & 0.63 & 0.86 & 1.12 & 1.13 & $4460-3i$ & & 82\% & 18\% & 1.11 & 1.06 & 0.74 & 0.98 & 1.14 & 1.21 \\
				& & & & & & & & & & & $4511-6i$ & & 55\% & 45\% & 0.97 & 0.94 & 1.10 & 1.31 & 1.29 & 1.26 \\
				& & & & & & & & & & & $4529-0.1i$ & & 22\% & 78\% & 0.90 & 0.94 & 1.36 & 1.52 & 1.44 & 1.42 \\
				& $4542-10i$ & & 61\% & 39\% & 0.95 & 0.91 & 0.88 & 1.18 & 1.01 & 1.07 & & & & & & & & & & \\
				& & & & & & & & & & & $4555-1i$ & & 37\% & 63\% & 1.12 & 0.77 & 1.47 & 1.54 & 1.57 & 1.43 \\
				& $4631-3i$ & & 33\% & 67\% & 0.98 & 0.99 & 1.04 & 1.76 & 1.38 & 1.52 & & & & & & & & & & \\
				\hline
				\multirow{2}{*}{$2^+$} & \multicolumn{10}{c|}{\multirow{2}{*}{--}} & $4555$ & $-176$ & 22\% & 78\% & 0.91 & 0.91 & 1.26 & 1.42 & 1.39 & 1.39 \\
				& \multicolumn{10}{c|}{} & $4703$ & $-28$ & 28\% & 72\% & 1.00 & 1.00 & 1.56 & 1.52 & 1.58 & 1.58 \\
				\hline\hline
			\end{tabular}
		}
	\end{table*}
	
	For the $I(J^P)=0(1^+)$ channel, a bound state below the $D^*D$ threshold is obtained in both calculations. In the H-only framework, this state is located at $3864$ MeV with a binding energy of $14$ MeV. With all Jacobi configurations, its mass slightly shifts to $3863$ MeV ($\Delta E = -15$ MeV). The proportions of color configurations and the rms radii remain almost unchanged. The rms radii $r_{c_1\bar{n}_3}^{\mathrm{rms}}$ ($0.71$~fm) and $r_{c_2\bar{n}_4}^{\mathrm{rms}}$ ($0.64$~fm) are close to the sizes of the $D^*$ ($0.70$~fm) and $D$ ($0.61$~fm) mesons, respectively, indicating a $D^*D$ molecular configuration. Considering the typical systematic uncertainty of tens of MeV in the constituent quark model, this state serves as a good qualitative candidate for the experimental $T_{cc}(3875)^+$. The consistent results suggest that the H-only coordinates are sufficient to provide a qualitative description for near-threshold bound states.
	
	Above the physical thresholds, we identify a resonant state located at $E = 4031 - 27i$ MeV. In the H-only calculation, the extraction of this pole position is highly unstable. However, incorporating all Jacobi coordinates effectively resolves this instability, leading to a robust extraction. However, its rms radii remain numerically unstable and change dramatically as the complex scaling angle $\theta$ varies. This instability may result from the state being sandwiched between the $D^*D$ and $D^*D^*$ thresholds, making it strongly coupled to the adjacent scattering states. Higher numerical precision is needed to determine its exact spatial structure.
	
	Moreover, the expanded basis incorporating all Jacobi configurations profoundly alters the highly excited spectrum. In the restricted H-only calculation, we find three resonant states at $4466-3i$, $4542-10i$, and $4631-3i$ MeV. However, when all K-type configurations are incorporated, only the pole near $4460$~MeV remains relatively stable. The states at $4542-10i$ MeV and $4631-3i$ MeV do not possess robust counterparts in the comprehensive framework, suggesting they might be heavily distorted or artificially produced by the incomplete spatial basis. Conversely, the all-Jacobi calculation successfully extracts four entirely new compact resonant states in this energy region, located at $4380-3i$, $4511-6i$, $4529-0.1i$, and $4555-1i$ MeV. This demonstrates that the conventional H-only configurations are insufficient to fully capture the complex spatial correlations at higher excitation energies.
	
	As depicted in Fig.~\ref{fig:ccnn_compare}, a more pronounced contrast emerges in the $0(2^+)$ channel. Our calculations reveal deeply bound states within the isoscalar $0(0^+)$ and $0(2^+)$ configurations. These states are strictly prohibited from undergoing $S$-wave decays into the respective ground-state $DD$ and $D^*D^*$ channels. This dynamical suppression is a direct consequence of Bose-Einstein statistics: the two identical bosonic mesons in the final state require a completely symmetric total wave function. However, an isoscalar $S$-wave system possesses an antisymmetric isospin wave function, which inherently violates this requisite symmetry. Their lowest allowed $S$-wave dimeson thresholds are shifted to higher radial excitations, such as the $D^{(*)}(1S)D^{(*)}(2S)$ channels, where the two mesons are no longer identical. Because our current framework is restricted to $S$-wave configurations, couplings to the lower-lying $P$-wave channels are explicitly excluded. Lacking these corresponding decay thresholds, they manifest as ``pseudo-bound states'', which are intrinsically physical $P$-wave scattering states. Notably, these ``pseudo-bound states'' cannot be obtained within the H-only framework. The K-type configurations are indispensable, as they partially encapsulate the effects of higher partial waves. The explicit inclusion of $P$-wave dynamics in future studies could alter their positions and reveal their true physical nature.
	
	For the isovector sector, no bound states are found below the lowest thresholds due to the absence of the strongly attractive $[\bar{n}\bar{n}]_{I=0}^{3_c}$ diquark configuration. Nevertheless, several narrow resonances are identified at higher energies. In the $J^P=0^+$ and $1^+$ channels, we obtain two compact resonances at $E = 4446 - 2i$ MeV and $E = 4456 - 0.7i$ MeV, respectively.
	
	In the $I(J^P)=1(2^+)$ channel, two compact resonant poles are extracted at $E = 4623 - 0.3i$ MeV and $E = 4672 - 2i$ MeV. The state at $4672$ MeV possesses a $17\%$ $6_c \otimes \bar{6}_c$ color component. To satisfy the Pauli exclusion principle for the identical light antiquarks, the spatial wave function within this $[\bar{n}\bar{n}]_{I=1}^{\bar{6}_c}$ component must be antisymmetric. Such an antisymmetric spatial wave function is naturally accommodated in our $S$-wave calculation due to the projections between all Jacobi configurations. Due to angular momentum and parity conservation, these $2^+$ states decay exclusively into the $D^*D^*$ channel, providing targets for future experimental searches.

\subsection{$bb\bar{n}\bar{n}$}
	
\begin{figure*}[htbp]
		\centering
		\includegraphics[width=0.95\textwidth]{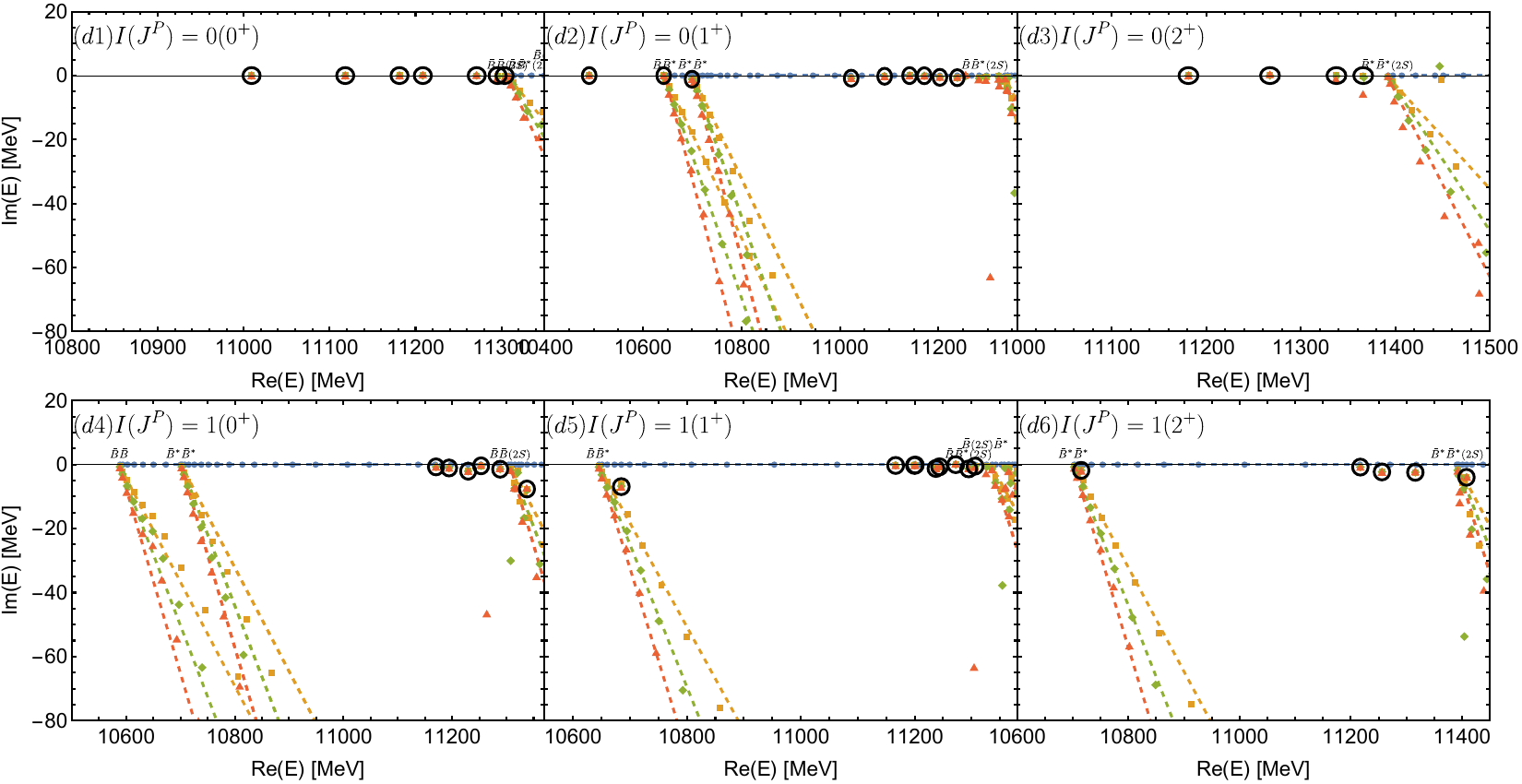}
		\caption{\label{fig:bbnn_PlotGrid} Complex energy eigenvalues of the $bb\bar{n}\bar{n}$ states for different $I(J^P)$ quantum numbers in the AL1 potential with varying $\theta$ in the CSM. The dashed lines represent the continuum lines rotating along $\mathrm{Arg}(E)=-2\theta$. The bound and resonant states do not shift as $\theta$ changes and are marked out by the black circles.}
	\end{figure*}

	For the $S$-wave $bb\bar{n}\bar{n}$ tetraquark system, the complex energy spectra for both the isoscalar and isovector sectors are presented in Fig.~\ref{fig:bbnn_PlotGrid}. The extracted bound states and resonance poles are summarized in Tables~\ref{tab:bbnn_poles_I0} and \ref{tab:bbnn_poles_I1}.
	
	\begin{table*}[htbp]
		\caption{Numerical results for the bound state and resonance poles of the $I=0$ $bb\bar{n}\bar{n}$ system. The complex energies $E_{\rm H}$ (with H-only configurations~\cite{Wu:2024zbx}) and $E_{\rm All}$ (with all Jacobi coordinates) are in units of MeV. The binding energy $\Delta E$ is in units of MeV. The proportions of color configurations ($\chi_{\bar{3}_c\otimes 3_c}$ and $\chi_{6_c\otimes\bar{6}_c}$) are given in percentages. The root-mean-square radii of the relative distances between quarks are in units of fm. The quarks are denoted as $b_1$, $b_2$, and $\bar{n}_3, \bar{n}_4 = \bar{u}/\bar{d}$. The last column denotes the spatial configuration of the states, where M. and C. represent molecular and compact configurations, respectively, and M.$^*$ denotes a molecular candidate whose intra-cluster sizes differ from those of the corresponding 1S or 2S mesons.}
		\label{tab:bbnn_poles_I0}
		\begin{ruledtabular}
			\begin{tabular}{ccccccccccccc}
				$I(J^P)$ & $E_{\rm H}$ & $E_{\rm All}$ & $\Delta E$ & $\chi_{\bar{3}_c\otimes3_c}$ & $\chi_{6_c\otimes\bar{6}_c}$ & $r_{b_1\bar{n}_3}$ & $r_{b_2\bar{n}_4}$ & $r_{b_1b_2}$ & $r_{\bar{n}_3\bar{n}_4}$ & $r_{b_1\bar{n}_4}$ & $r_{b_2\bar{n}_3}$ & Configuration \\
				\midrule
				\multirow{8}{*}{$0(0^+)$} 
				& -- & $11009$ & $-297$ & 95\% & 5\%  & 0.90 & 0.90 & 0.55 & 0.84 & 0.93 & 0.93 & C. \\
				& -- & $11118$ & $-188$ & 28\% & 72\% & 0.85 & 0.85 & 1.02 & 1.25 & 1.20 & 1.20 & C. \\
				& -- & $11181$ & $-125$ & 64\% & 36\% & 0.90 & 0.90 & 0.96 & 1.09 & 1.14 & 1.14 & C. \\
				& $11195$ & -- & & & & & & & & & & \\
				& -- & $11208$ & $-98$  & 30\% & 70\% & 0.89 & 0.89 & 1.26 & 1.36 & 1.36 & 1.36 & C. \\
				& -- & $11271$ & $-35$  & 50\% & 50\% & 0.93 & 0.93 & 1.23 & 1.43 & 1.39 & 1.39 & C. \\
				& -- & $11295$ & $-11$  & 55\% & 45\% & 0.97 & 0.97 & 1.13 & 1.40 & 1.33 & 1.33 & C. \\
				& -- & $11304$ & $-3$   & 38\% & 62\% & 0.97 & 0.97 & 1.57 & 1.78 & 1.70 & 1.70 & C. \\
				\midrule
				\multirow{13}{*}{$0(1^+)$}
				& $10491$ & $10491$ & $-153$ & 97\% & 3\%  & 0.68 & 0.67 & 0.34 & 0.78 & 0.70 & 0.71 & C. \\
				& $10642$ & $10642$ & $-2$   & 33\% & 67\% & 0.66 & 0.63 & 1.84 & 2.01 & 1.92 & 1.93 & M. \\
				& $10700-1i$ & $10700-1i$ & & 44\% & 56\% & 0.68 & 0.67 & 1.90 & 2.05 & 1.99 & 1.99 & M. \\
				& $11025-1i$ & $11023-0.8i$ &      & 97\% & 3\%  & 1.08 & 1.07 & 0.35 & 0.83 & 1.08 & 1.08 & C. \\
				& -- & $11091-0.2i$ &      & 54\% & 46\% & 0.95 & 0.75 & 0.90 & 1.24 & 1.07 & 1.11 & C. \\
				& -- & $11141+0i$   &      & 29\% & 71\% & 0.89 & 0.82 & 1.05 & 1.35 & 1.21 & 1.26 & C. \\
				& $11152-0.5i$ & -- & & & & & & & & & & \\
				& -- & $11171+0i$   &      & 21\% & 79\% & 0.74 & 0.96 & 1.11 & 1.51 & 1.34 & 1.28 & C. \\
				& -- & $11203-0.5i$ &      & 42\% & 58\% & 1.04 & 0.69 & 1.18 & 1.41 & 1.31 & 1.29 & C. \\
				& -- & $11238-0.7i$ &      & 46\% & 54\% & 0.89 & 0.92 & 1.21 & 1.39 & 1.30 & 1.31 & C. \\
				& $11249-0.3i$ & -- & & & & & & & & & & \\
				& $11291-1i$ & -- & & & & & & & & & & \\
				& $11315-1i$ & -- & & & & & & & & & & \\
				\midrule
				\multirow{5}{*}{$0(2^+)$}
				& -- & $11181$ & $-210$ & 19\% & 81\% & 0.85 & 0.85 & 1.05 & 1.33 & 1.25 & 1.25 & C. \\
				& -- & $11267$ & $-124$ & 23\% & 77\% & 0.90 & 0.90 & 1.28 & 1.43 & 1.40 & 1.40 & C. \\
				& -- & $11338$ & $-53$  & 74\% & 26\% & 1.03 & 1.03 & 0.77 & 1.41 & 1.21 & 1.21 & C. \\
				& -- & $11366$ & $-25$  & 37\% & 63\% & 0.99 & 0.99 & 1.29 & 1.44 & 1.41 & 1.41 & C. \\
				& $11370$ & -- & & & & & & & & & & \\
			\end{tabular}
		\end{ruledtabular}
	\end{table*}
	
	\begin{table*}[htbp]
		\caption{Numerical results for the resonance poles of the $I=1$ $bb\bar{n}\bar{n}$ system. The complex energies $E_{\rm H}$ (with H-only configurations~\cite{Wu:2024zbx}) and $E_{\rm All}$ (with all Jacobi coordinates) are in units of MeV. The proportions of color configurations ($\chi_{\bar{3}_c\otimes 3_c}$ and $\chi_{6_c\otimes\bar{6}_c}$) are given in percentages. The root-mean-square radii of the relative distances between quarks are in units of fm. The quarks are denoted as $b_1$, $b_2$, and $\bar{n}_3, \bar{n}_4 = \bar{u}/\bar{d}$. The last column denotes the spatial configuration of the states, where M. and C. represent molecular and compact configurations, respectively, and M.$^*$ denotes a molecular candidate whose intra-cluster sizes differ from those of the corresponding 1S or 2S mesons.}
		\label{tab:bbnn_poles_I1}
		\begin{ruledtabular}
			\begin{tabular}{cccccccccccc}
				$I(J^P)$ & $E_{\rm H}$ & $E_{\rm All}$ & $\chi_{\bar{3}_c\otimes3_c}$ & $\chi_{6_c\otimes\bar{6}_c}$ & $r_{b_1\bar{n}_3}$ & $r_{b_2\bar{n}_4}$ & $r_{b_1b_2}$ & $r_{\bar{n}_3\bar{n}_4}$ & $r_{b_1\bar{n}_4}$ & $r_{b_2\bar{n}_3}$ & Configuration \\
				\midrule
				\multirow{8}{*}{$1(0^+)$}
				& -- & $11170-0.7i$ & 47\% & 53\% & 0.90 & 0.90 & 0.93 & 1.39 & 1.19 & 1.19 & C. \\
				& -- & $11194-1i$   & 63\% & 37\% & 1.02 & 1.02 & 0.77 & 1.27 & 1.21 & 1.21 & C. \\
				& $11200-1i$ & -- & & & & & & & & & \\
				& -- & $11230-2i$   & 62\% & 38\% & 0.98 & 0.98 & 0.83 & 1.35 & 1.21 & 1.21 & C. \\
				& -- & $11253-0.4i$ & 41\% & 59\% & 0.90 & 0.90 & 1.25 & 1.72 & 1.47 & 1.47 & C. \\
				& $11262-4i$ & -- & & & & & & & & & \\
				& -- & $11288-1i$   & 40\% & 60\% & 0.95 & 0.95 & 1.25 & 1.65 & 1.46 & 1.46 & C. \\
				& -- & $11338-8i$   & 46\% & 54\% & 0.99 & 0.99 & 1.10 & 1.56 & 1.38 & 1.38 & C. \\
				\midrule
				\multirow{12}{*}{$1(1^+)$}
				& $10685-7i$ & $10685-7i$   & 43\% & 57\% & 0.72 & 0.70 & 0.70 & 1.14 & 0.97 & 0.99 & C. \\
				& -- & $11167-0.2i$ & 47\% & 53\% & 1.01 & 0.73 & 0.91 & 1.30 & 1.18 & 1.19 & C. \\
				& -- & $11200-0.2i$ & 64\% & 36\% & 1.05 & 0.98 & 0.69 & 1.14 & 1.13 & 1.15 & C. \\
				& -- & $11201-0.1i$ & 74\% & 26\% & 0.90 & 1.04 & 0.66 & 1.11 & 1.05 & 1.12 & C. \\
				& $11210-0.1i$ & -- & & & & & & & & & \\
				& -- & $11237-1i$   & 65\% & 35\% & 0.93 & 0.99 & 0.84 & 1.36 & 1.23 & 1.19 & C. \\
				& -- & $11244-0.7i$ & 46\% & 54\% & 1.06 & 0.80 & 1.15 & 1.47 & 1.35 & 1.42 & C. \\
				& -- & $11272+0i$   & 48\% & 52\% & 0.95 & 0.92 & 1.11 & 1.45 & 1.37 & 1.33 & C. \\
				& $11286-5i$ & -- & & & & & & & & & \\
				& -- & $11295-1.3i$ & 54\% & 46\% & 0.84 & 1.05 & 1.15 & 1.63 & 1.46 & 1.41 & C. \\
				& -- & $11306-0.5i$ & 67\% & 33\% & 1.03 & 0.95 & 0.93 & 1.43 & 1.23 & 1.30 & C. \\
				& $11312$ & -- & & & & & & & & & \\
				\midrule
				\multirow{7}{*}{$1(2^+)$}
				& $10715-2i$ & $10715-2i$   & 87\% & 13\% & 0.74 & 0.74 & 0.81 & 1.16 & 1.02 & 1.02 & C. \\
				& -- & $11217-0.8i$ & 82\% & 18\% & 1.07 & 1.07 & 0.51 & 1.06 & 1.10 & 1.10 & C. \\
				& $11227-0.1i$ & -- & & & & & & & & & \\
				& -- & $11256-2i$   & 70\% & 30\% & 0.97 & 0.97 & 0.76 & 1.31 & 1.16 & 1.16 & C. \\
				& $11305-5i$ & -- & & & & & & & & & \\
				& -- & $11316-2i$   & 45\% & 55\% & 0.93 & 0.93 & 1.29 & 1.73 & 1.52 & 1.52 & C. \\
				& -- & $11408-4i$   & 59\% & 41\% & 1.04 & 1.04 & 1.32 & 1.59 & 1.48 & 1.48 & C. \\
			\end{tabular}
		\end{ruledtabular}
	\end{table*}
	
	We first focus on the $I(J^P)=0(1^+)$ $bb\bar{n}\bar{n}$ system. We obtain a deeply bound state at $10491$ MeV with a binding energy of $153$ MeV, and a shallow bound state at $10642$ MeV with a binding energy of $2$ MeV, which is consistent with previous investigations on doubly heavy tetraquarks~\cite{Wu:2024zbx}. For the deeply bound state, the $\chi_{\bar{3}_c\otimes3_c}$ color configuration is $97\%$ dominant. The strong attractive interaction between the two bottom quarks contributes to its deep binding energy and fosters a compact Helium-like structure. In contrast, the higher bound state at $10642$ MeV exhibits a molecular configuration, with $r_{b_1\bar{n}_3}$ ($0.66$~fm) and $r_{b_2\bar{n}_4}$ ($0.63$~fm) close to the sizes of the $\bar{B}^*$ ($0.66$~fm) and $\bar{B}$ ($0.63$~fm) mesons, respectively. It could be regarded as the bottom analog of the $D^*D$ molecular bound state $T_{cc}(3875)^+$ and may decay radiatively to $\bar{B}\bar{B}\gamma$. Above the physical thresholds, we also obtain six isoscalar resonant states with $J^P=1^+$, which are generally identified as compact tetraquarks.

\begin{figure*}[htbp]
		\centering
		\includegraphics[width=0.95\textwidth]{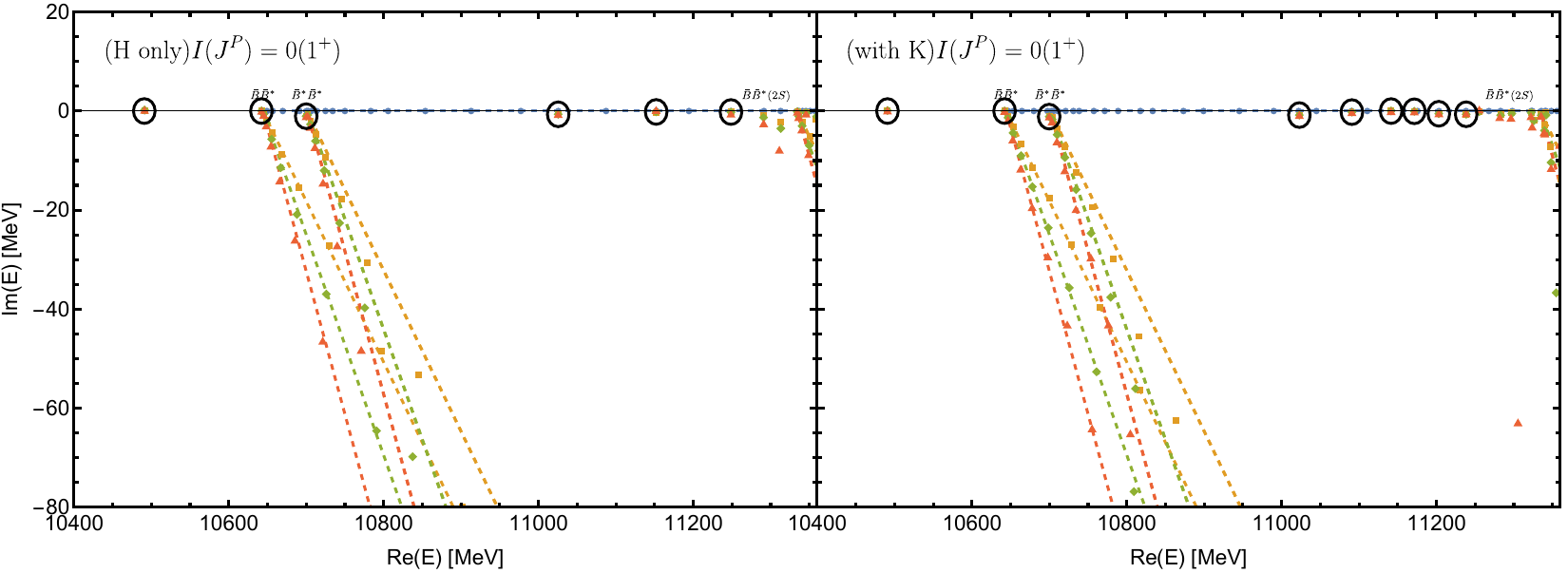}
		\caption{\label{fig:bbnn_compare} Comparison of the complex energy spectra for the $bb\bar{n}\bar{n}$ $I(J^P)=0(1^+)$ system. The left panel shows the results obtained using exclusively H-type configurations~\cite{Wu:2024zbx}, while the right panel incorporates all H- and K-type configurations. The stationary poles are marked by black circles.}
	\end{figure*}

	To further demonstrate the impact of the K-type Jacobi coordinates in the $bb\bar{n}\bar{n}$ system, we compare the results obtained using the restricted H-only basis~\cite{Wu:2024zbx} with those from the comprehensive all-Jacobi framework for the isoscalar $1^+$ channel. The comparative results are illustrated in Fig.~\ref{fig:bbnn_compare} and detailed in Table~\ref{tab:bbnn_compare}. A broader comparison of the complex energies between the H-only and all-Jacobi calculations across all channels can be found in the aforementioned Tables~\ref{tab:bbnn_poles_I0} and \ref{tab:bbnn_poles_I1}.

	\begin{table*}[htbp]
		\caption{A comparative summary of the calculated resonance poles and spatial root-mean-square radii for the $bb\bar{n}\bar{n}$ state with $I(J^P)=0(1^+)$. We contrast the results obtained using the traditional H-only Jacobi coordinates with those derived from the all-Jacobi configurations. The binding energy $\Delta E$ and complex eigenenergy $E = M - i\Gamma/2$ are given in units of MeV, while the spatial sizes are expressed in fm. The constituent quarks are labeled as $b_{1,2}$ and $\bar{n}_{3,4}$.}
		\label{tab:bbnn_compare}
		\resizebox{\textwidth}{!}{
			\begin{tabular}{cccccccccc|cccccccccc}
				\hline\hline
				\multicolumn{10}{c|}{\textbf{H-only framework}} & \multicolumn{10}{c}{\textbf{All-Jacobi framework}} \\
				\hline
				$E$ & $\Delta E$ & $\chi_{\bar{3}\otimes3}$ & $\chi_{6\otimes\bar{6}}$ & $r_{b_1\bar{n}_3}$ & $r_{b_2\bar{n}_4}$ & $r_{b_1b_2}$ & $r_{\bar{n}_3\bar{n}_4}$ & $r_{b_1\bar{n}_4}$ & $r_{b_2\bar{n}_3}$ & $E$ & $\Delta E$ & $\chi_{\bar{3}\otimes3}$ & $\chi_{6\otimes\bar{6}}$ & $r_{b_1\bar{n}_3}$ & $r_{b_2\bar{n}_4}$ & $r_{b_1b_2}$ & $r_{\bar{n}_3\bar{n}_4}$ & $r_{b_1\bar{n}_4}$ & $r_{b_2\bar{n}_3}$ \\
				\hline
				$10491$ & $-153$ & 97\% & 3\% & 0.68 & 0.67 & 0.33 & 0.78 & 0.70 & 0.71 & $10491$ & $-153$ & 97\% & 3\% & 0.68 & 0.67 & 0.34 & 0.78 & 0.70 & 0.71 \\
				$10642$ & $-1$ & 33\% & 67\% & 0.66 & 0.63 & 1.98 & 2.15 & 2.06 & 2.07 & $10642$ & $-2$ & 33\% & 67\% & 0.66 & 0.63 & 1.84 & 2.01 & 1.92 & 1.93 \\
				$10700-1i$ & & 44\% & 56\% & 0.67 & 0.67 & 1.88 & 2.02 & 1.96 & 1.96 & $10700-1i$ & & 44\% & 56\% & 0.68 & 0.67 & 1.90 & 2.05 & 1.99 & 1.99 \\
				$11025-1i$ & & 98\% & 2\% & 1.08 & 1.07 & 0.33 & 0.83 & 1.08 & 1.08 & $11023-0.8i$ & & 97\% & 3\% & 1.08 & 1.07 & 0.35 & 0.83 & 1.08 & 1.08 \\
				& & & & & & & & & & $11091-0.2i$ & & 54\% & 46\% & 0.95 & 0.75 & 0.90 & 1.24 & 1.07 & 1.11 \\
				& & & & & & & & & & $11141+0i$ & & 29\% & 71\% & 0.89 & 0.82 & 1.05 & 1.35 & 1.21 & 1.26 \\
				$11152-0.5i$ & & 60\% & 40\% & 0.89 & 0.88 & 0.96 & 0.97 & 0.77 & 1.07 & & & & & & & & & & \\
				& & & & & & & & & & $11171+0i$ & & 21\% & 79\% & 0.74 & 0.96 & 1.11 & 1.51 & 1.34 & 1.28 \\
				& & & & & & & & & & $11203-0.5i$ & & 42\% & 58\% & 1.04 & 0.69 & 1.18 & 1.41 & 1.31 & 1.29 \\
				$11249-0.3i$ & & 40\% & 60\% & 0.91 & 0.89 & 1.12 & 1.14 & 0.75 & 1.49 & $11238-0.7i$ & & 46\% & 54\% & 0.89 & 0.92 & 1.21 & 1.39 & 1.30 & 1.31 \\
				\hline\hline
			\end{tabular}
		}
	\end{table*}

	For the $0(0^+)$ and $0(2^+)$ channels, we find multiple bound states. As previously discussed for the $cc\bar{n}\bar{n}$ system, due to the identical symmetry constraints and the exclusion of $P$-wave couplings in our current $S$-wave framework, these isoscalar continuum manifest as ``pseudo-bound states''. Their true physical nature and pole positions might be altered upon the explicit inclusion of $P$-wave dynamics in future studies.
	
	Consistent with the $cc\bar{n}\bar{n}$ system, no bound states are found in the isovector $bb\bar{n}\bar{n}$ sector due to the absence of the ``good antidiquark'' $[\bar{n}\bar{n}]_{I=0}^{3_c}$  configuration~\cite{Deng:2022cld,Jaffe:2004ph}. Nevertheless, we obtain several narrow resonant states with $J^P=0^+$, $1^+$, and $2^+$ in the continuum. As listed in Table~\ref{tab:bbnn_poles_I1}, these isovector resonances are generally identified as compact tetraquark states. Among them, the two lowest-lying states, namely the $1(1^+)$ state at $10685$ MeV and the $1(2^+)$ state at $10715$ MeV, may be more accessible for experimental observation. We recommend future experimental searches for these states in the $\bar{B}^{(*)}\bar{B}^{(*)}$ decay channels.

\subsection{$QQ\bar{Q}\bar{Q}$}

\begin{figure*}[htbp]
	\centering
	\includegraphics[width=0.95\textwidth]{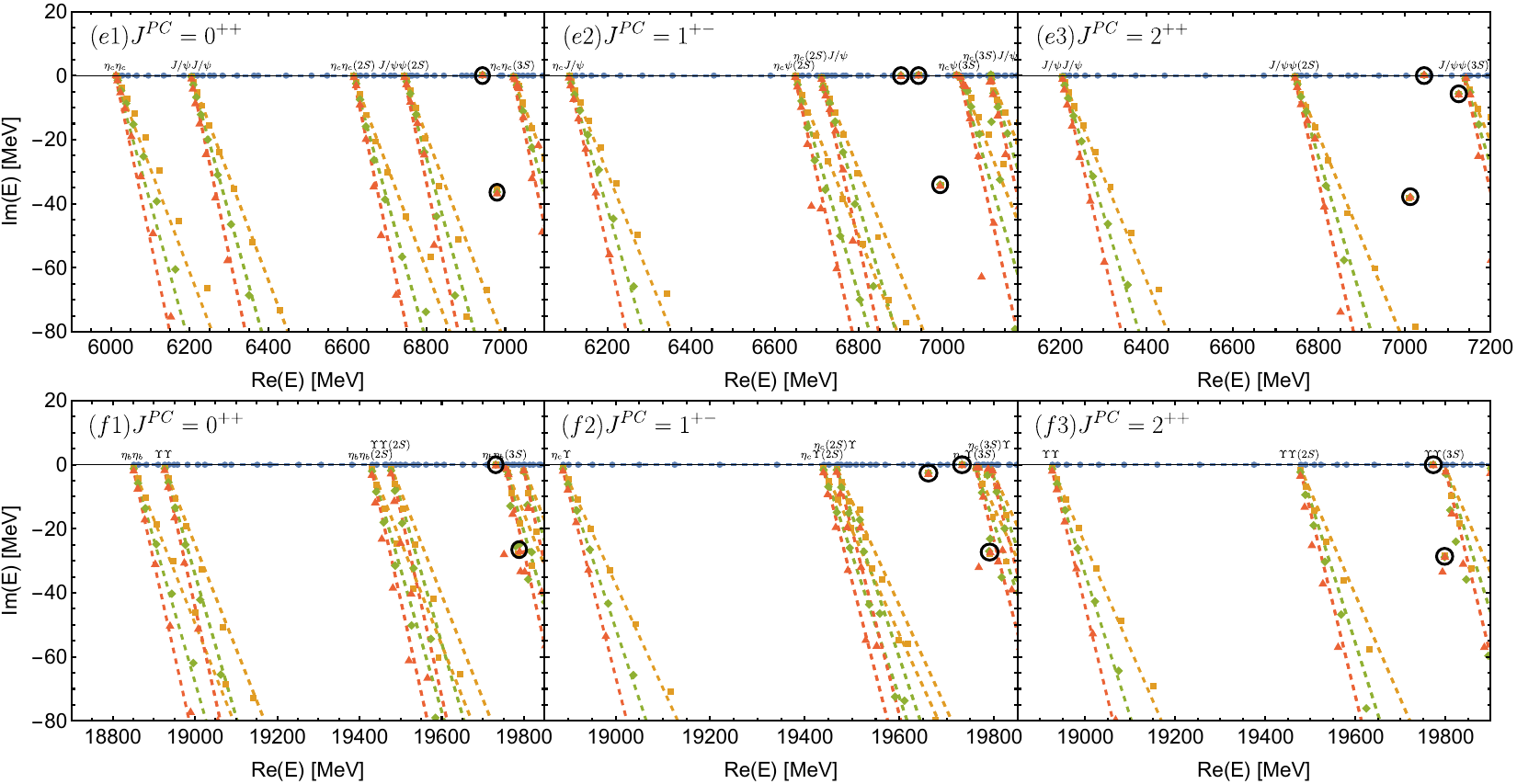}
	\caption{\label{fig:QQQQ_PlotGrid} Complex energy eigenvalues of the $cc\bar{c}\bar{c}$ and $bb\bar{b}\bar{b}$ states with normal C-parities in the AL1 potential with varying $\theta$ in the CSM. The dashed lines represent the continuum lines rotating along $\mathrm{Arg}(E)=-2\theta$. The resonant states do not shift as $\theta$ changes and are marked out by the corresponding circles.}
\end{figure*}

In this section, we investigate the fully heavy $cc\bar{c}\bar{c}$ and $bb\bar{b}\bar{b}$ tetraquarks in the $S$-wave. We restrict our focus to the normal C-parity states ($J^{PC}=0^{++}, 1^{+-}$, and $2^{++}$), which naturally couple to the $S$-wave thresholds of ground-state heavy quarkonia. Exotic C-parity configurations ($0^{+-}, 1^{++}, 2^{+-}$, etc.) are explicitly excluded, as authentically capturing their resonance dynamics and decay widths necessitates the inclusion of higher partial waves~\cite{Wu:2024ocq}. The calculated complex energy spectra are presented in Fig.~\ref{fig:QQQQ_PlotGrid}. Tables~\ref{tab:cccc_poles} and \ref{tab:bbbb_poles} summarize the complex poles $E = M - i\Gamma/2$, the color configuration ratios, and the root-mean-square radii for the charm and bottom sectors, respectively.

\begin{table*}[htbp]
	\caption{Numerical results for the bound state and resonance poles of the $cc\bar{c}\bar{c}$ system with normal C-parities. The complex energies $E_{\rm H}$ (with H-only configurations~\cite{Wu:2024euj}) and $E_{\rm All}$ (with all Jacobi coordinates) are in units of MeV. The proportions of color configurations ($\chi_{\bar{3}_c\otimes 3_c}$ and $\chi_{6_c\otimes\bar{6}_c}$) are given in percentages. The root-mean-square radii of the relative distances between quarks are in units of fm.  The ``?" indicates that the rms radii results are numerically unstable. The last column denotes the spatial configuration of the states, where M. and C. represent molecular and compact configurations, respectively, and M.$^*$ denotes a molecular candidate whose intra-cluster sizes differ from those of the corresponding 1S or 2S mesons.}
	\label{tab:cccc_poles}
	\begin{ruledtabular}
		\begin{tabular}{cccccccccccc}
			$J^{PC}$ & $E_{\rm H}$ & $E_{\rm All}$ & $\chi_{\bar{3}_c\otimes3_c}$ & $\chi_{6_c\otimes\bar{6}_c}$ & $r_{c_1\bar{c}_3}$ & $r_{c_2\bar{c}_4}$ & $r_{c_1c_2}$ & $r_{\bar{c}_3\bar{c}_4}$ & $r_{c_1\bar{c}_4}$ & $r_{c_2\bar{c}_3}$ & Config \\
			\midrule
			\multirow{2}{*}{$0^{++}$} 
			& -- & $6942+0i$ & 35\% & 65\% & 0.72 & 0.72 & 1.09 & 1.09 & 1.09 & 1.09 & C. \\
			& $6980-35i$ & $6979-36i$ & 88\% & 12\% & 0.78 & 0.78 & 0.58 & 0.58 & 0.76 & 0.76 & C. \\
			\midrule
			\multirow{4}{*}{$1^{+-}$}
			& -- & $6901+0i$ & 43\% & 57\% & 0.69 & 0.69 & 1.05 & 1.05 & 1.07 & 1.07 & C. \\
			& $6921-0.5i$ & -- & & & & & & & & & \\
			& -- & $6943+0i$ & 35\% & 65\% & 0.95 & 0.38 & 1.09 & 1.09 & 1.10 & 1.10 & C. \\
			& $6995-35i$ & $6995-34i$ & 87\% & 13\% & 0.77 & 0.77 & 0.63 & 0.63 & 0.80 & 0.80 & C. \\
			\midrule
			\multirow{3}{*}{$2^{++}$}
			& $7013-38i$ & $7013-38i$ & 92\% & 8\% & 0.77 & 0.77 & ? & ? & ? & ? & ? \\
			& -- & $7045+0i$ & 39\% & 61\% & 0.74 & 0.74 & 1.09 & 1.09 & 1.09 & 1.09 & C. \\
			& $7127-6i$ & $7126-6i$ & 77\% & 23\% & 0.90 & 0.90 & 0.78 & 0.78 & 0.97 & 0.97 & C. \\
		\end{tabular}
	\end{ruledtabular}
\end{table*}

For the $cc\bar{c}\bar{c}$ system, we obtain compact resonances at $E=6979-36i$ MeV ($0^{++}$) and $E=7013-38i$ MeV ($2^{++}$), both dominated by the $\bar{3}_c\otimes 3_c$ color configuration with small spatial radii. These poles provide theoretical candidates for the $X(6900)$ structure under different possible $J^{PC}$ assignments. Additionally, we extract a broad $1^{+-}$ resonance at $E=6995-34i$ MeV, and a higher $2^{++}$ resonance at $E=7126-6i$ MeV that serves as a candidate for the recently reported $X(7200)$ state.

\begin{table*}[htbp]
	\caption{Numerical results for the bound state and resonance poles of the $bb\bar{b}\bar{b}$ system with normal C-parities. The complex energies $E_{\rm H}$ (with H-only configurations~\cite{Wu:2024euj}) and $E_{\rm All}$ (with all Jacobi coordinates) are in units of MeV. The proportions of color configurations ($\chi_{\bar{3}_c\otimes 3_c}$ and $\chi_{6_c\otimes\bar{6}_c}$) are given in percentages. The root-mean-square radii of the relative distances between quarks are in units of fm. The ``?" indicates that the rms radii results are numerically unstable. The last column denotes the spatial configuration of the states, where M. and C. represent molecular and compact configurations, respectively, and M$^*$. denotes a molecular candidate whose intra-cluster sizes differ from those of the corresponding 1S or 2S mesons.}
	\label{tab:bbbb_poles}
	\begin{ruledtabular}
		\begin{tabular}{cccccccccccc}
			$J^{PC}$ & $E_{\rm H}$ & $E_{\rm All}$ & $\chi_{\bar{3}_c\otimes3_c}$ & $\chi_{6_c\otimes\bar{6}_c}$ & $r_{b_1\bar{b}_3}$ & $r_{b_2\bar{b}_4}$ & $r_{b_1b_2}$ & $r_{\bar{b}_3\bar{b}_4}$ & $r_{b_1\bar{b}_4}$ & $r_{b_2\bar{b}_3}$ & Config \\
			\midrule
			\multirow{2}{*}{$0^{++}$} 
			& -- & $19731+0i$ & 33\% & 67\% & 0.47 & 0.47 & 0.72 & 0.72 & 0.72 & 0.72 & C. \\
			& $19788-26i$ & $19788-27i$ & 86\% & 14\% & 0.51 & 0.51 & ? & ? & ? & ? & ? \\
			\midrule
			\multirow{3}{*}{$1^{+-}$}
			& -- & $19661-3i$ & 43\% & 57\% & 0.44 & 0.44 & 0.69 & 0.69 & 0.70 & 0.70 & C. \\
			& -- & $19732+0i$ & 34\% & 66\% & 0.63 & 0.22 & 0.72 & 0.72 & 0.72 & 0.72 & C. \\
			& $19791-26i$ & $19791-27i$ & 87\% & 13\% & 0.51 & 0.52 & 0.50 & 0.50 & 0.52 & 0.52 & C. \\
			\midrule
			\multirow{2}{*}{$2^{++}$}
			& -- & $19773+0i$ & 36\% & 64\% & 0.48 & 0.48 & 0.72 & 0.72 & 0.72 & 0.72 & C. \\
			& $19798-27i$ & $19798-29i$ & 87\% & 13\% & 0.52 & 0.52 & 0.67 & 0.67 & 0.77 & 0.77 & C. \\
		\end{tabular}
	\end{ruledtabular}
\end{table*}

\begin{figure*}[htbp]
		\centering
		\includegraphics[width=0.95\textwidth]{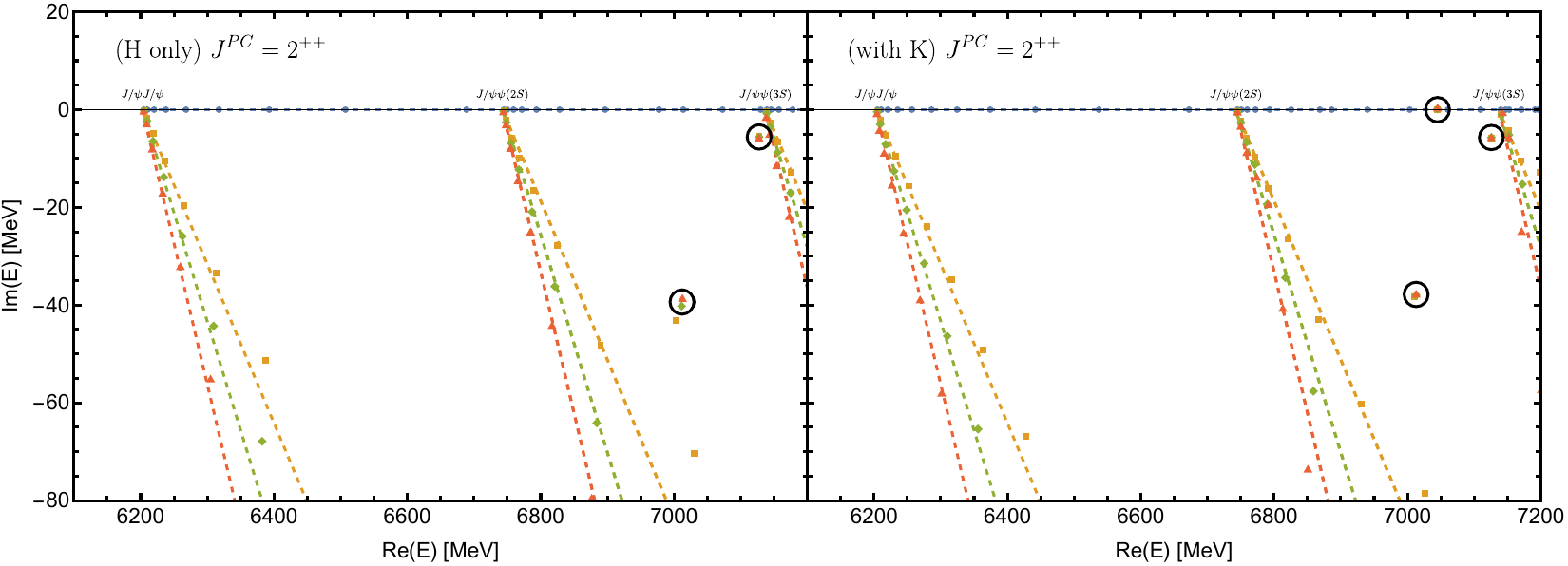}
		\caption{\label{fig:cccc_compare} Comparison of the complex energy spectra for the $cc\bar{c}\bar{c}$ $J^{PC}=2^{++}$ system. The left panel shows the results obtained using exclusively H-type configurations, while the right panel incorporates all H- and K-type configurations. The stationary poles are marked by black circles.}
	\end{figure*}

To evaluate the effects of the K-type Jacobi coordinates in the fully heavy sector, we compare the results of the $cc\bar{c}\bar{c}$ $J^{PC}=2^{++}$ system calculated using the traditional H-only basis with those using the all-Jacobi basis. The comparative results are presented in Table~\ref{tab:cccc_compare} and Fig.~\ref{fig:cccc_compare}. A broader comparison of the complex energies between the H-only and all-Jacobi calculations across all channels can be found in the aforementioned Tables~\ref{tab:cccc_poles} and \ref{tab:bbbb_poles}. Upon expanding the basis, the lowest $2^{++}$ pole only shifts slightly from $E = 7012 - 39i$~MeV to $E = 7013 - 38i$~MeV, and the higher $X(7200)$ candidate shifts from $E = 7128 - 6i$~MeV to $E = 7126 - 6i$~MeV. The color configurations and the rms radii remain consistent between the two calculations. However, the expanded basis incorporating all Jacobi configurations reveals an additional state located at $7045$~MeV, which is missing in the H-only framework.

\begin{table*}[htbp]
	\caption{A comparative summary of the calculated resonance poles and spatial root-mean-square radii for the $cc\bar{c}\bar{c}$ state with $J^{PC}=2^{++}$. We contrast the results obtained using the traditional H-only Jacobi coordinates with those derived from the all-Jacobi configurations. The complex eigenenergy $E = M - i\Gamma/2$ is given in units of MeV, while the spatial sizes are expressed in fm. The ``?" indicates that the rms radii results are numerically unstable. The constituent quarks are labeled as $c_{1,2}$ and $\bar{c}_{3,4}$.}
	\label{tab:cccc_compare}
	\resizebox{\textwidth}{!}{
		\begin{tabular}{cccccccccc|cccccccccc}
			\hline\hline
			\multicolumn{10}{c|}{\textbf{H-only framework}} & \multicolumn{10}{c}{\textbf{All-Jacobi framework}} \\
			\hline
			$E$ & $\chi_{\bar{3}\otimes3}$ & $\chi_{6\otimes\bar{6}}$ & $r_{c_1\bar{c}_3}$ & $r_{c_2\bar{c}_4}$ & $r_{c_1c_2}$ & $r_{\bar{c}_3\bar{c}_4}$ & $r_{c_1\bar{c}_4}$ & $r_{c_2\bar{c}_3}$ & Config & $E$ & $\chi_{\bar{3}\otimes3}$ & $\chi_{6\otimes\bar{6}}$ & $r_{c_1\bar{c}_3}$ & $r_{c_2\bar{c}_4}$ & $r_{c_1c_2}$ & $r_{\bar{c}_3\bar{c}_4}$ & $r_{c_1\bar{c}_4}$ & $r_{c_2\bar{c}_3}$ & Config \\
			\hline
			$7012-39i$ & 95\% & 5\% & 0.78 & 0.78 & ? & ? & ? & ? & ? & $7013-38i$ & 92\% & 8\% & 0.77 & 0.77 & ? & ? & ? & ? & ? \\
			& & & & & & & & & & $7045+0i$ & 39\% & 61\% & 0.74 & 0.74 & 1.09 & 1.09 & 1.09 & 1.09 & C. \\
			$7128-6i$ & 77\% & 23\% & 0.90 & 0.90 & 0.78 & 0.78 & 0.98 & 0.98 & C. & $7126-6i$ & 77\% & 23\% & 0.90 & 0.90 & 0.78 & 0.78 & 0.97 & 0.97 & C. \\
			\hline\hline
		\end{tabular}
	}
\end{table*}

The fully bottom $bb\bar{b}\bar{b}$ system displays similar spectral features, though the energy and spatial scales are different. For the $0^{++}$ channel, we identify a compact state at $19731$ MeV, along with a higher resonance at $E=19788-27i$ MeV ($86\%$ dominated by the $\bar{3}_c\otimes 3_c$ configuration). In the $1^{+-}$ channel, besides two lower states at $E=19661-3i$ MeV and $19732+0i$ MeV, we find a resonance at $E=19791-27i$ MeV ($87\%$ dominated by the $\bar{3}_c\otimes 3_c$ configuration). For the $2^{++}$ channel, a compact state at $19773+0i$ MeV is extracted alongside a resonant pole at $E=19798-29i$ MeV ($87\%$ dominated by the $\bar{3}_c\otimes 3_c$ configuration).
	
Since the spin-dependent chromomagnetic interactions are inversely proportional to the heavy quark mass ($\sim 1/m_Q$), tetraquark states differing only in heavy quark spin orientations are expected to form nearly degenerate multiplets. This heavy quark spin symmetry (HQSS) is clearly reflected in our calculated mass spectra for both the $cc\bar{c}\bar{c}$ and $bb\bar{b}\bar{b}$ systems. In the $cc\bar{c}\bar{c}$ sector, the $0^{++}$ state at $6979$ MeV, the $1^{+-}$ state at $6995$ MeV, and the $2^{++}$ state at $7013$ MeV share similar spatial sizes, comparable decay widths, and a dominant $\bar{3}_c\otimes 3_c$ color configuration. They form an HQSS triplet with mass splittings of 15--20 MeV. Similarly, the $bb\bar{b}\bar{b}$ resonances at $19788$ MeV ($0^{++}$), $19791$ MeV ($1^{+-}$), and $19798$ MeV ($2^{++}$) form a bottom spin triplet. They share comparable decay widths and a dominant $\bar{3}_c\otimes 3_c$ color configuration. The mass splittings in this bottom triplet are within $10$ MeV, which are much smaller than those in the charm sector, perfectly consistent with the expected $1/m_b < 1/m_c$ behavior. These results validate their assignments as compact spin partners within the fully heavy tetraquark family.

\subsection{$ss\bar{s}\bar{s}$}

\begin{figure*}[htbp]
	\centering
	\includegraphics[width=0.95\textwidth]{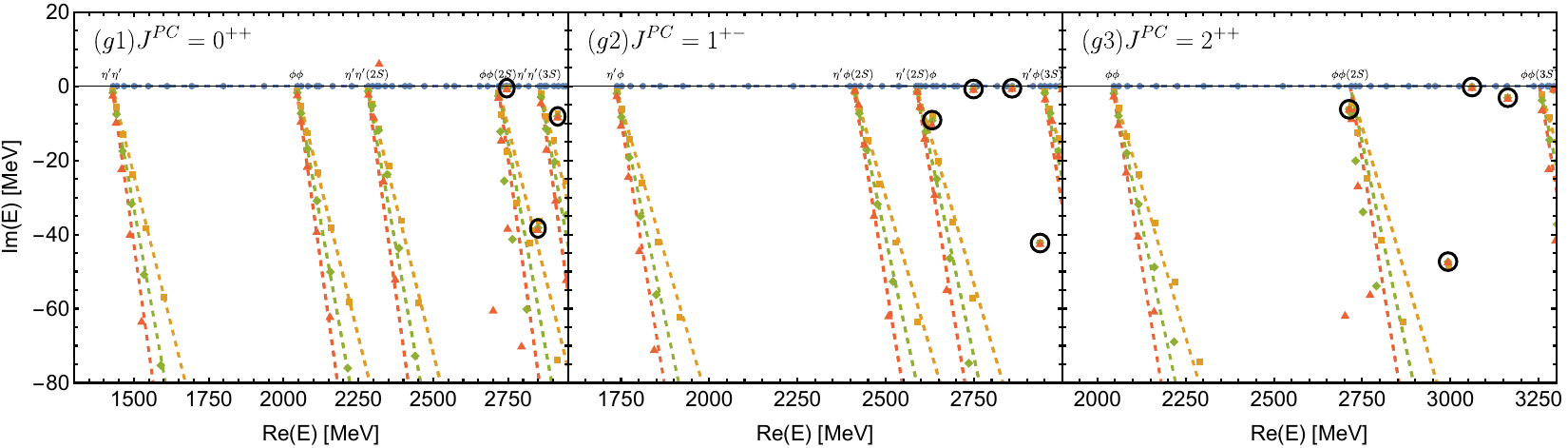}
	\caption{\label{fig:ssss_PlotGrid} Complex energy eigenvalues of the $ss\bar{s}\bar{s}$ states with normal C-parities in the AL1 potential with varying $\theta$ in the CSM. The dashed lines represent the continuum lines rotating along $\mathrm{Arg}(E)=-2\theta$. The resonant states do not shift as $\theta$ changes and are marked out by the black circles.}
\end{figure*}

In this section, we systematically investigate the $S$-wave fully strange tetraquark systems. We calculate the complex energy spectra for these systems under varying rotation angles $\theta$, as depicted in Fig.~\ref{fig:ssss_PlotGrid}. The numerical results of the extracted resonant states, including their complex eigenenergies $E=M-i\Gamma/2$, proportions of various color configurations, and root-mean-square radii, are summarized in Table~\ref{tab:ssss_poles}.

\begin{table*}[htbp]
	\caption{Numerical results for the bound state and resonance poles of the $ss\bar{s}\bar{s}$ system with normal C-parities. The complex energies $E_{\rm H}$ (with H-only configurations~\cite{Ma:2024vsi}) and $E_{\rm All}$ (with all Jacobi coordinates) are in units of MeV. The proportions of color configurations ($\chi_{\bar{3}_c\otimes 3_c}$ and $\chi_{6_c\otimes\bar{6}_c}$) are given in percentages. The root-mean-square radii of the relative distances between quarks are in units of fm.  The ``?" indicates that the rms radii results are numerically unstable. The last column denotes the spatial configuration of the states, where M. and C. represent molecular and compact configurations, respectively, and M.$^*$ denotes a molecular candidate whose intra-cluster sizes differ from those of the corresponding 1S or 2S mesons.}
	\label{tab:ssss_poles}
	\begin{ruledtabular}
		\begin{tabular}{cccccccccccc}
			$J^{PC}$ & $E_{\rm H}$ & $E_{\rm All}$ & $\chi_{\bar{3}_c\otimes3_c}$ & $\chi_{6_c\otimes\bar{6}_c}$ & $r_{s_1\bar{s}_3}$ & $r_{s_2\bar{s}_4}$ & $r_{s_1s_2}$ & $r_{\bar{s}_3\bar{s}_4}$ & $r_{s_1\bar{s}_4}$ & $r_{s_2\bar{s}_3}$ & Config \\
			\midrule
			\multirow{3}{*}{$0^{++}$}
			& -- & $2745-0.5i$ & 35\% & 65\% & 1.09 & 1.09 & 1.65 & 1.65 & 1.65 & 1.65 & C. \\
			& $2852-40i$ & $2849-38i$  & 83\% & 17\% & 1.19 & 1.19 & ?    & ?    & ?    & ?    & ? \\
			& $2917-9i$ & $2915-8i$   & 38\% & 62\% & 1.12 & 1.12 & 1.07 & 1.07 & 1.26 & 1.26 & C. \\
			\midrule
			\multirow{5}{*}{$1^{+-}$}
			& -- & $2632-9i$   & 43\% & 57\% & 0.88 & 1.05 & ?    & ?    & ?    & ?    & ?  \\
			& -- & $2749-0.8i$ & 35\% & 65\% & 1.44 & 0.56 & 1.66 & 1.66 & 1.67 & 1.67 & C. \\
			& $2819-3i$ & -- & & & & & & & & & \\
			& -- & $2858-0.6i$ & 43\% & 57\% & 1.10 & 1.07 & 1.59 & 1.59 & 1.65 & 1.65 & C. \\
			& $2940-46i$ & $2937-42i$  & 86\% & 14\% & 1.16 & 1.16 & 1.02 & 1.02 & 1.15 & 1.15 & C. \\
			\midrule
			\multirow{4}{*}{$2^{++}$}
			& $2714-6i$ & $2713-6i$   & 75\% & 25\% & 1.11 & 1.11 & ?    & ?    & ?    & ?    & ?  \\
			& $2993-48i$ & $2993-47i$  & 85\% & 15\% & 1.13 & 1.13 & 1.06 & 1.06 & 1.03 & 1.03 & C. \\
			& -- & $3061-0.2i$ & 42\% & 58\% & 1.15 & 1.15 & 1.64 & 1.64 & 1.68 & 1.68 & C. \\
			& $3164-3i$ & $3163-3i$   & 92\% & 8\%  & 1.47 & 1.47 & 0.93 & 0.93 & 1.50 & 1.50 & C. \\
		\end{tabular}
	\end{ruledtabular}
\end{table*}

For the scalar $0^{++}$ sector, three resonant states are obtained with the incorporation of all K-type Jacobi coordinates. The lowest state is located at $E=2745-0.5i$ MeV. As the energy increases, we identify a broader resonance at $E=2849-38i$ MeV, whose color wave function is highly dominated by the $\bar{3}_c\otimes3_c$ configuration with a proportion of $83\%$. Another higher resonant pole is extracted at $E=2915-8i$ MeV with an intermediate width.

For the $1^{+-}$ channel, we extract four complex resonance poles. A relatively low-mass state is observed at $E=2632-9i$ MeV. Furthermore, two narrow resonant states are identified at $E=2749-0.8i$ MeV and $E=2858-0.6i$ MeV. The other lies at $E=2937-42i$ MeV, which favors the $\bar{3}_c\otimes3_c$ color structure.

In the tensor $2^{++}$ channel, four resonant states are determined. The lowest pole is situated at $E=2713-6i$ MeV, where the $\bar{3}_c\otimes3_c$ color configuration accounts for a significant fraction of $75\%$. Above this state, we also find a broad resonance at $E=2993-47i$ MeV, along with two highly excited states at $E=3061-0.2i$ MeV and $E=3163-3i$ MeV.

Based on the calculated rms radii listed in Table~\ref{tab:ssss_poles}, the internal distances between different quarks for these resonant states are of the same order of magnitude. This feature indicates that the obtained fully strange resonances generally have compact tetraquark configurations rather than molecular structures.

According to their predicted mass range of $2.6-3.2$ GeV, these fully strange compact tetraquarks can readily decay into the $\phi\phi$ and $\eta^{(\prime)}\eta^{(\prime)}$ final states. Specifically, the $2^{++}$ state located at $2713$ MeV could serve as a target for future experimental searches. Its prominent decay channel naturally leads to a four-kaon final state ($K^+K^-K^+K^-$), which provides a clean and easily reconstructible signature at modern high-energy colliders.

\section{Summary and discussion}~\label{sec:summary}
	
We systematically investigated the $S$-wave $Qs\bar{n}\bar{n}$, $QQ\bar{n}\bar{n}$, $QQ\bar{Q}\bar{Q}$, and $ss\bar{s}\bar{s}$ tetraquarks within the AL1 constituent quark framework. By including the full set of H-type and K-type Jacobi coordinates with a stochastically optimized Gaussian expansion method and the complex scaling technique, we resolved both bound states and resonances from the scattering continua. \change{By comparing the pole positions obtained with and without the K-type configurations, we find that the H-type configurations are sufficient for the low-lying states below the $1S2S$ thresholds, while the K-type configurations become important for highly excited states, leading to substantial pole shifts and new resonances.}
	
Our results provide theoretical interpretations for several recent experimental observations. Specifically, we identified a compact state around 2.9 GeV matching the $T_{cs0}(2900)$, and a $D^*D$ molecular configuration consistent with the $T_{cc}(3875)^+$. The inclusion of all Jacobi coordinates is proved essential, uncovering isoscalar ``pseudo-bound states'' in the doubly heavy sector that are otherwise hidden in traditional H-type bases. Analogously, a significant portion of the near-zero-width resonances observed in our calculations are speculated to share a similar physical nature, likely corresponding to $P$-wave scattering states. Consequently, the credibility of these zero-width poles is relatively low, which warrants further investigation. For fully heavy systems, we proposed compact resonance candidates for the $X(6900)$ under different $J^{PC}$ assignments, alongside a candidate for the $X(7200)$. Furthermore, these predicted $QQ\bar{Q}\bar{Q}$ states naturally form distinct multiplets governed by heavy quark spin symmetry. Conversely, our calculations do not support the $\phi(2170)$ and $X(2370)$ as compact $S$-wave $ss\bar{s}\bar{s}$ tetraquarks, since no such resonances appear below 2.6 GeV. Instead, we propose a $2^{++}$ fully strange resonance near 2.7 GeV, which cleanly decays into the $K^+K^-K^+K^-$ channel, as a highly promising target for future searches. Ultimately, this work underscores the necessity of incorporating all spatial configurations in multiquark studies, offering concrete predictions for future experimental validation at high-energy colliders.

% \begin{appendix}

% \section{}~\label{app:}

% \end{appendix}

\begin{acknowledgements}

We are grateful to Hui-Min Yang, Yan-Ke Chen, and Wei-Lin Wu for the helpful discussions. This project was supported by the National Natural Science Foundation of China (Grant No. 12475137). Y. M. is supported by the Alexander von Humboldt Foundation. The computational resources were supported by High-performance Computing Platform of Peking University.

\end{acknowledgements}

\bibliography{Ref}

\begin{thebibliography}{94}%
\makeatletter
\providecommand \@ifxundefined [1]{%
 \@ifx{#1\undefined}
}%
\providecommand \@ifnum [1]{%
 \ifnum #1\expandafter \@firstoftwo
 \else \expandafter \@secondoftwo
 \fi
}%
\providecommand \@ifx [1]{%
 \ifx #1\expandafter \@firstoftwo
 \else \expandafter \@secondoftwo
 \fi
}%
\providecommand \natexlab [1]{#1}%
\providecommand \enquote  [1]{``#1''}%
\providecommand \bibnamefont  [1]{#1}%
\providecommand \bibfnamefont [1]{#1}%
\providecommand \citenamefont [1]{#1}%
\providecommand \href@noop [0]{\@secondoftwo}%
\providecommand \href [0]{\begingroup \@sanitize@url \@href}%
\providecommand \@href[1]{\@@startlink{#1}\@@href}%
\providecommand \@@href[1]{\endgroup#1\@@endlink}%
\providecommand \@sanitize@url [0]{\catcode `\\12\catcode `\$12\catcode
  `\&12\catcode `\#12\catcode `\^12\catcode `\_12\catcode `\%12\relax}%
\providecommand \@@startlink[1]{}%
\providecommand \@@endlink[0]{}%
\providecommand \url  [0]{\begingroup\@sanitize@url \@url }%
\providecommand \@url [1]{\endgroup\@href {#1}{\urlprefix }}%
\providecommand \urlprefix  [0]{URL }%
\providecommand \Eprint [0]{\href }%
\providecommand \doibase [0]{https://doi.org/}%
\providecommand \selectlanguage [0]{\@gobble}%
\providecommand \bibinfo  [0]{\@secondoftwo}%
\providecommand \bibfield  [0]{\@secondoftwo}%
\providecommand \translation [1]{[#1]}%
\providecommand \BibitemOpen [0]{}%
\providecommand \bibitemStop [0]{}%
\providecommand \bibitemNoStop [0]{.\EOS\space}%
\providecommand \EOS [0]{\spacefactor3000\relax}%
\providecommand \BibitemShut  [1]{\csname bibitem#1\endcsname}%
\let\auto@bib@innerbib\@empty
%</preamble>
\bibitem [{\citenamefont {Choi}\ \emph {et~al.}(2003)\citenamefont {Choi} \emph
  {et~al.}}]{Belle:2003nnu}%
  \BibitemOpen
  \bibfield  {author} {\bibinfo {author} {\bibfnamefont {S.~K.}\ \bibnamefont
  {Choi}} \emph {et~al.} (\bibinfo {collaboration} {Belle}),\ }\bibfield
  {title} {\bibinfo {title} {{Observation of a narrow charmonium-like state in
  exclusive $B^\pm \to K^\pm \pi^+ \pi^- J/\psi$ decays}},\ }\href
  {https://doi.org/10.1103/PhysRevLett.91.262001} {\bibfield  {journal}
  {\bibinfo  {journal} {Phys. Rev. Lett.}\ }\textbf {\bibinfo {volume} {91}},\
  \bibinfo {pages} {262001} (\bibinfo {year} {2003})},\ \Eprint
  {https://arxiv.org/abs/hep-ex/0309032} {arXiv:hep-ex/0309032} \BibitemShut
  {NoStop}%
\bibitem [{\citenamefont {Jaffe}(1977)}]{Jaffe:1976ig}%
  \BibitemOpen
  \bibfield  {author} {\bibinfo {author} {\bibfnamefont {R.~L.}\ \bibnamefont
  {Jaffe}},\ }\bibfield  {title} {\bibinfo {title} {{Multi-Quark Hadrons. 1.
  The Phenomenology of (2 Quark 2 anti-Quark) Mesons}},\ }\href
  {https://doi.org/10.1103/PhysRevD.15.267} {\bibfield  {journal} {\bibinfo
  {journal} {Phys. Rev. D}\ }\textbf {\bibinfo {volume} {15}},\ \bibinfo
  {pages} {267} (\bibinfo {year} {1977})}\BibitemShut {NoStop}%
\bibitem [{\citenamefont {Jaffe}\ and\ \citenamefont
  {Johnson}(1976)}]{Jaffe:1975fd}%
  \BibitemOpen
  \bibfield  {author} {\bibinfo {author} {\bibfnamefont {R.~L.}\ \bibnamefont
  {Jaffe}}\ and\ \bibinfo {author} {\bibfnamefont {K.}~\bibnamefont
  {Johnson}},\ }\bibfield  {title} {\bibinfo {title} {{Unconventional States of
  Confined Quarks and Gluons}},\ }\href
  {https://doi.org/10.1016/0370-2693(76)90423-8} {\bibfield  {journal}
  {\bibinfo  {journal} {Phys. Lett. B}\ }\textbf {\bibinfo {volume} {60}},\
  \bibinfo {pages} {201} (\bibinfo {year} {1976})}\BibitemShut {NoStop}%
\bibitem [{\citenamefont {Fritzsch}\ \emph {et~al.}(1973)\citenamefont
  {Fritzsch}, \citenamefont {Gell-Mann},\ and\ \citenamefont
  {Leutwyler}}]{Fritzsch:1973pi}%
  \BibitemOpen
  \bibfield  {author} {\bibinfo {author} {\bibfnamefont {H.}~\bibnamefont
  {Fritzsch}}, \bibinfo {author} {\bibfnamefont {M.}~\bibnamefont
  {Gell-Mann}},\ and\ \bibinfo {author} {\bibfnamefont {H.}~\bibnamefont
  {Leutwyler}},\ }\bibfield  {title} {\bibinfo {title} {{Advantages of the
  Color Octet Gluon Picture}},\ }\href
  {https://doi.org/10.1016/0370-2693(73)90625-4} {\bibfield  {journal}
  {\bibinfo  {journal} {Phys. Lett. B}\ }\textbf {\bibinfo {volume} {47}},\
  \bibinfo {pages} {365} (\bibinfo {year} {1973})}\BibitemShut {NoStop}%
\bibitem [{\citenamefont {Chen}\ \emph {et~al.}(2016)\citenamefont {Chen},
  \citenamefont {Chen}, \citenamefont {Liu},\ and\ \citenamefont
  {Zhu}}]{Chen:2016qju}%
  \BibitemOpen
  \bibfield  {author} {\bibinfo {author} {\bibfnamefont {H.-X.}\ \bibnamefont
  {Chen}}, \bibinfo {author} {\bibfnamefont {W.}~\bibnamefont {Chen}}, \bibinfo
  {author} {\bibfnamefont {X.}~\bibnamefont {Liu}},\ and\ \bibinfo {author}
  {\bibfnamefont {S.-L.}\ \bibnamefont {Zhu}},\ }\bibfield  {title} {\bibinfo
  {title} {The hidden-charm pentaquark and tetraquark states},\ }\href
  {https://doi.org/10.1016/j.physrep.2016.05.004} {\bibfield  {journal}
  {\bibinfo  {journal} {Phys. Rept.}\ }\textbf {\bibinfo {volume} {639}},\
  \bibinfo {pages} {1} (\bibinfo {year} {2016})},\ \Eprint
  {https://arxiv.org/abs/1601.02092} {arXiv:1601.02092 [hep-ph]} \BibitemShut
  {NoStop}%
\bibitem [{\citenamefont {Hosaka}\ \emph {et~al.}(2016)\citenamefont {Hosaka},
  \citenamefont {Iijima}, \citenamefont {Miyabayashi}, \citenamefont {Sakai},\
  and\ \citenamefont {Yasui}}]{Hosaka:2016pey}%
  \BibitemOpen
  \bibfield  {author} {\bibinfo {author} {\bibfnamefont {A.}~\bibnamefont
  {Hosaka}}, \bibinfo {author} {\bibfnamefont {T.}~\bibnamefont {Iijima}},
  \bibinfo {author} {\bibfnamefont {K.}~\bibnamefont {Miyabayashi}}, \bibinfo
  {author} {\bibfnamefont {Y.}~\bibnamefont {Sakai}},\ and\ \bibinfo {author}
  {\bibfnamefont {S.}~\bibnamefont {Yasui}},\ }\bibfield  {title} {\bibinfo
  {title} {{Exotic hadrons with heavy flavors: X, Y, Z, and related states}},\
  }\href {https://doi.org/10.1093/ptep/ptw045} {\bibfield  {journal} {\bibinfo
  {journal} {PTEP}\ }\textbf {\bibinfo {volume} {2016}},\ \bibinfo {pages}
  {062C01} (\bibinfo {year} {2016})},\ \Eprint
  {https://arxiv.org/abs/1603.09229} {arXiv:1603.09229 [hep-ph]} \BibitemShut
  {NoStop}%
\bibitem [{\citenamefont {Esposito}\ \emph {et~al.}(2017)\citenamefont
  {Esposito}, \citenamefont {Pilloni},\ and\ \citenamefont
  {Polosa}}]{Esposito:2016noz}%
  \BibitemOpen
  \bibfield  {author} {\bibinfo {author} {\bibfnamefont {A.}~\bibnamefont
  {Esposito}}, \bibinfo {author} {\bibfnamefont {A.}~\bibnamefont {Pilloni}},\
  and\ \bibinfo {author} {\bibfnamefont {A.~D.}\ \bibnamefont {Polosa}},\
  }\bibfield  {title} {\bibinfo {title} {{Multiquark Resonances}},\ }\href
  {https://doi.org/10.1016/j.physrep.2016.11.002} {\bibfield  {journal}
  {\bibinfo  {journal} {Phys. Rept.}\ }\textbf {\bibinfo {volume} {668}},\
  \bibinfo {pages} {1} (\bibinfo {year} {2017})},\ \Eprint
  {https://arxiv.org/abs/1611.07920} {arXiv:1611.07920 [hep-ph]} \BibitemShut
  {NoStop}%
\bibitem [{\citenamefont {Ali}\ \emph {et~al.}(2017)\citenamefont {Ali},
  \citenamefont {Lange},\ and\ \citenamefont {Stone}}]{Ali:2017jda}%
  \BibitemOpen
  \bibfield  {author} {\bibinfo {author} {\bibfnamefont {A.}~\bibnamefont
  {Ali}}, \bibinfo {author} {\bibfnamefont {J.~S.}\ \bibnamefont {Lange}},\
  and\ \bibinfo {author} {\bibfnamefont {S.}~\bibnamefont {Stone}},\ }\bibfield
   {title} {\bibinfo {title} {{Exotics: Heavy Pentaquarks and Tetraquarks}},\
  }\href {https://doi.org/10.1016/j.ppnp.2017.08.003} {\bibfield  {journal}
  {\bibinfo  {journal} {Prog. Part. Nucl. Phys.}\ }\textbf {\bibinfo {volume}
  {97}},\ \bibinfo {pages} {123} (\bibinfo {year} {2017})},\ \Eprint
  {https://arxiv.org/abs/1706.00610} {arXiv:1706.00610 [hep-ph]} \BibitemShut
  {NoStop}%
\bibitem [{\citenamefont {Lebed}\ \emph {et~al.}(2017)\citenamefont {Lebed},
  \citenamefont {Mitchell},\ and\ \citenamefont {Swanson}}]{Lebed:2016hpi}%
  \BibitemOpen
  \bibfield  {author} {\bibinfo {author} {\bibfnamefont {R.~F.}\ \bibnamefont
  {Lebed}}, \bibinfo {author} {\bibfnamefont {R.~E.}\ \bibnamefont
  {Mitchell}},\ and\ \bibinfo {author} {\bibfnamefont {E.~S.}\ \bibnamefont
  {Swanson}},\ }\bibfield  {title} {\bibinfo {title} {{Heavy-Quark QCD
  Exotica}},\ }\href {https://doi.org/10.1016/j.ppnp.2016.11.003} {\bibfield
  {journal} {\bibinfo  {journal} {Prog. Part. Nucl. Phys.}\ }\textbf {\bibinfo
  {volume} {93}},\ \bibinfo {pages} {143} (\bibinfo {year} {2017})},\ \Eprint
  {https://arxiv.org/abs/1610.04528} {arXiv:1610.04528 [hep-ph]} \BibitemShut
  {NoStop}%
\bibitem [{\citenamefont {Guo}\ \emph {et~al.}(2018)\citenamefont {Guo},
  \citenamefont {Hanhart}, \citenamefont {Mei{\ss}ner}, \citenamefont {Wang},
  \citenamefont {Zhao},\ and\ \citenamefont {Zou}}]{Guo:2017jvc}%
  \BibitemOpen
  \bibfield  {author} {\bibinfo {author} {\bibfnamefont {F.-K.}\ \bibnamefont
  {Guo}}, \bibinfo {author} {\bibfnamefont {C.}~\bibnamefont {Hanhart}},
  \bibinfo {author} {\bibfnamefont {U.-G.}\ \bibnamefont {Mei{\ss}ner}},
  \bibinfo {author} {\bibfnamefont {Q.}~\bibnamefont {Wang}}, \bibinfo {author}
  {\bibfnamefont {Q.}~\bibnamefont {Zhao}},\ and\ \bibinfo {author}
  {\bibfnamefont {B.-S.}\ \bibnamefont {Zou}},\ }\bibfield  {title} {\bibinfo
  {title} {Hadronic molecules},\ }\href
  {https://doi.org/10.1103/RevModPhys.90.015004} {\bibfield  {journal}
  {\bibinfo  {journal} {Rev. Mod. Phys.}\ }\textbf {\bibinfo {volume} {90}},\
  \bibinfo {pages} {015004} (\bibinfo {year} {2018})},\ \Eprint
  {https://arxiv.org/abs/1705.00141} {arXiv:1705.00141 [hep-ph]} \BibitemShut
  {NoStop}%
\bibitem [{\citenamefont {Liu}\ \emph {et~al.}(2019)\citenamefont {Liu},
  \citenamefont {Chen}, \citenamefont {Chen}, \citenamefont {Liu},\ and\
  \citenamefont {Zhu}}]{Liu:2019zoy}%
  \BibitemOpen
  \bibfield  {author} {\bibinfo {author} {\bibfnamefont {Y.-R.}\ \bibnamefont
  {Liu}}, \bibinfo {author} {\bibfnamefont {H.-X.}\ \bibnamefont {Chen}},
  \bibinfo {author} {\bibfnamefont {W.}~\bibnamefont {Chen}}, \bibinfo {author}
  {\bibfnamefont {X.}~\bibnamefont {Liu}},\ and\ \bibinfo {author}
  {\bibfnamefont {S.-L.}\ \bibnamefont {Zhu}},\ }\bibfield  {title} {\bibinfo
  {title} {Pentaquark and tetraquark states},\ }\href
  {https://doi.org/10.1016/j.ppnp.2019.04.003} {\bibfield  {journal} {\bibinfo
  {journal} {Prog. Part. Nucl. Phys.}\ }\textbf {\bibinfo {volume} {107}},\
  \bibinfo {pages} {237} (\bibinfo {year} {2019})},\ \Eprint
  {https://arxiv.org/abs/1903.11976} {arXiv:1903.11976 [hep-ph]} \BibitemShut
  {NoStop}%
\bibitem [{\citenamefont {Brambilla}\ \emph {et~al.}(2020)\citenamefont
  {Brambilla}, \citenamefont {Eidelman}, \citenamefont {Hanhart}, \citenamefont
  {Nefediev}, \citenamefont {Shen}, \citenamefont {Thomas}, \citenamefont
  {Vairo},\ and\ \citenamefont {Yuan}}]{Brambilla:2019esw}%
  \BibitemOpen
  \bibfield  {author} {\bibinfo {author} {\bibfnamefont {N.}~\bibnamefont
  {Brambilla}}, \bibinfo {author} {\bibfnamefont {S.}~\bibnamefont {Eidelman}},
  \bibinfo {author} {\bibfnamefont {C.}~\bibnamefont {Hanhart}}, \bibinfo
  {author} {\bibfnamefont {A.}~\bibnamefont {Nefediev}}, \bibinfo {author}
  {\bibfnamefont {C.-P.}\ \bibnamefont {Shen}}, \bibinfo {author}
  {\bibfnamefont {C.~E.}\ \bibnamefont {Thomas}}, \bibinfo {author}
  {\bibfnamefont {A.}~\bibnamefont {Vairo}},\ and\ \bibinfo {author}
  {\bibfnamefont {C.-Z.}\ \bibnamefont {Yuan}},\ }\bibfield  {title} {\bibinfo
  {title} {{The $XYZ$ states: experimental and theoretical status and
  perspectives}},\ }\href {https://doi.org/10.1016/j.physrep.2020.05.001}
  {\bibfield  {journal} {\bibinfo  {journal} {Phys. Rept.}\ }\textbf {\bibinfo
  {volume} {873}},\ \bibinfo {pages} {1} (\bibinfo {year} {2020})},\ \Eprint
  {https://arxiv.org/abs/1907.07583} {arXiv:1907.07583 [hep-ex]} \BibitemShut
  {NoStop}%
\bibitem [{\citenamefont {Meng}\ \emph {et~al.}(2023)\citenamefont {Meng},
  \citenamefont {Wang}, \citenamefont {Wang},\ and\ \citenamefont
  {Zhu}}]{Meng:2022ozq}%
  \BibitemOpen
  \bibfield  {author} {\bibinfo {author} {\bibfnamefont {L.}~\bibnamefont
  {Meng}}, \bibinfo {author} {\bibfnamefont {B.}~\bibnamefont {Wang}}, \bibinfo
  {author} {\bibfnamefont {G.-J.}\ \bibnamefont {Wang}},\ and\ \bibinfo
  {author} {\bibfnamefont {S.-L.}\ \bibnamefont {Zhu}},\ }\bibfield  {title}
  {\bibinfo {title} {Chiral effective field theory for heavy hadrons and its
  applications},\ }\href {https://doi.org/10.1016/j.physrep.2023.01.003}
  {\bibfield  {journal} {\bibinfo  {journal} {Phys. Rept.}\ }\textbf {\bibinfo
  {volume} {1044}},\ \bibinfo {pages} {1} (\bibinfo {year} {2023})},\ \Eprint
  {https://arxiv.org/abs/2204.08716} {arXiv:2204.08716 [hep-ph]} \BibitemShut
  {NoStop}%
\bibitem [{\citenamefont {Chen}\ \emph {et~al.}(2023)\citenamefont {Chen},
  \citenamefont {Chen}, \citenamefont {Liu}, \citenamefont {Liu},\ and\
  \citenamefont {Zhu}}]{Chen:2022asf}%
  \BibitemOpen
  \bibfield  {author} {\bibinfo {author} {\bibfnamefont {H.-X.}\ \bibnamefont
  {Chen}}, \bibinfo {author} {\bibfnamefont {W.}~\bibnamefont {Chen}}, \bibinfo
  {author} {\bibfnamefont {X.}~\bibnamefont {Liu}}, \bibinfo {author}
  {\bibfnamefont {Y.-R.}\ \bibnamefont {Liu}},\ and\ \bibinfo {author}
  {\bibfnamefont {S.-L.}\ \bibnamefont {Zhu}},\ }\bibfield  {title} {\bibinfo
  {title} {An updated review of the new hadron states},\ }\href
  {https://doi.org/10.1088/1361-6633/aca3b6} {\bibfield  {journal} {\bibinfo
  {journal} {Rept. Prog. Phys.}\ }\textbf {\bibinfo {volume} {86}},\ \bibinfo
  {pages} {026201} (\bibinfo {year} {2023})},\ \Eprint
  {https://arxiv.org/abs/2204.02649} {arXiv:2204.02649 [hep-ph]} \BibitemShut
  {NoStop}%
\bibitem [{\citenamefont {Mai}\ \emph {et~al.}(2023)\citenamefont {Mai},
  \citenamefont {Mei{\ss}ner},\ and\ \citenamefont {Urbach}}]{Mai:2022eur}%
  \BibitemOpen
  \bibfield  {author} {\bibinfo {author} {\bibfnamefont {M.}~\bibnamefont
  {Mai}}, \bibinfo {author} {\bibfnamefont {U.-G.}\ \bibnamefont
  {Mei{\ss}ner}},\ and\ \bibinfo {author} {\bibfnamefont {C.}~\bibnamefont
  {Urbach}},\ }\bibfield  {title} {\bibinfo {title} {{Towards a theory of
  hadron resonances}},\ }\href {https://doi.org/10.1016/j.physrep.2022.11.005}
  {\bibfield  {journal} {\bibinfo  {journal} {Phys. Rept.}\ }\textbf {\bibinfo
  {volume} {1001}},\ \bibinfo {pages} {1} (\bibinfo {year} {2023})},\ \Eprint
  {https://arxiv.org/abs/2206.01477} {arXiv:2206.01477 [hep-ph]} \BibitemShut
  {NoStop}%
\bibitem [{\citenamefont {Aaij}\ \emph
  {et~al.}(2020{\natexlab{a}})\citenamefont {Aaij} \emph
  {et~al.}}]{LHCb:2020bls}%
  \BibitemOpen
  \bibfield  {author} {\bibinfo {author} {\bibfnamefont {R.}~\bibnamefont
  {Aaij}} \emph {et~al.} (\bibinfo {collaboration} {LHCb}),\ }\bibfield
  {title} {\bibinfo {title} {{A model-independent study of resonant structure
  in $B^+\to D^+D^-K^+$ decays}},\ }\href
  {https://doi.org/10.1103/PhysRevLett.125.242001} {\bibfield  {journal}
  {\bibinfo  {journal} {Phys. Rev. Lett.}\ }\textbf {\bibinfo {volume} {125}},\
  \bibinfo {pages} {242001} (\bibinfo {year} {2020}{\natexlab{a}})},\ \Eprint
  {https://arxiv.org/abs/2009.00025} {arXiv:2009.00025 [hep-ex]} \BibitemShut
  {NoStop}%
\bibitem [{\citenamefont {Aaij}\ \emph
  {et~al.}(2020{\natexlab{b}})\citenamefont {Aaij} \emph
  {et~al.}}]{LHCb:2020pxc}%
  \BibitemOpen
  \bibfield  {author} {\bibinfo {author} {\bibfnamefont {R.}~\bibnamefont
  {Aaij}} \emph {et~al.} (\bibinfo {collaboration} {LHCb}),\ }\bibfield
  {title} {\bibinfo {title} {{Amplitude analysis of the $B^+\to D^+D^-K^+$
  decay}},\ }\href {https://doi.org/10.1103/PhysRevD.102.112003} {\bibfield
  {journal} {\bibinfo  {journal} {Phys. Rev. D}\ }\textbf {\bibinfo {volume}
  {102}},\ \bibinfo {pages} {112003} (\bibinfo {year} {2020}{\natexlab{b}})},\
  \Eprint {https://arxiv.org/abs/2009.00026} {arXiv:2009.00026 [hep-ex]}
  \BibitemShut {NoStop}%
\bibitem [{\citenamefont {Molina}\ \emph {et~al.}(2010)\citenamefont {Molina},
  \citenamefont {Branz},\ and\ \citenamefont {Oset}}]{Molina:2010tx}%
  \BibitemOpen
  \bibfield  {author} {\bibinfo {author} {\bibfnamefont {R.}~\bibnamefont
  {Molina}}, \bibinfo {author} {\bibfnamefont {T.}~\bibnamefont {Branz}},\ and\
  \bibinfo {author} {\bibfnamefont {E.}~\bibnamefont {Oset}},\ }\bibfield
  {title} {\bibinfo {title} {A new interpretation for the $d^*_s(2573)$ and the
  prediction of novel exotic charmed mesons},\ }\href
  {https://doi.org/10.1103/PhysRevD.82.014010} {\bibfield  {journal} {\bibinfo
  {journal} {Phys. Rev. D}\ }\textbf {\bibinfo {volume} {82}},\ \bibinfo
  {pages} {014010} (\bibinfo {year} {2010})},\ \Eprint
  {https://arxiv.org/abs/1005.0335} {arXiv:1005.0335 [hep-ph]} \BibitemShut
  {NoStop}%
\bibitem [{\citenamefont {Chen}\ \emph {et~al.}(2020)\citenamefont {Chen},
  \citenamefont {Chen}, \citenamefont {Dong},\ and\ \citenamefont
  {Su}}]{Chen:2020aos}%
  \BibitemOpen
  \bibfield  {author} {\bibinfo {author} {\bibfnamefont {H.-X.}\ \bibnamefont
  {Chen}}, \bibinfo {author} {\bibfnamefont {W.}~\bibnamefont {Chen}}, \bibinfo
  {author} {\bibfnamefont {R.-R.}\ \bibnamefont {Dong}},\ and\ \bibinfo
  {author} {\bibfnamefont {N.}~\bibnamefont {Su}},\ }\bibfield  {title}
  {\bibinfo {title} {$x_0(2900)$ and $x_1(2900)$: Hadronic molecules or compact
  tetraquarks},\ }\href {https://doi.org/10.1088/0256-307X/37/10/101201}
  {\bibfield  {journal} {\bibinfo  {journal} {Chin. Phys. Lett.}\ }\textbf
  {\bibinfo {volume} {37}},\ \bibinfo {pages} {101201} (\bibinfo {year}
  {2020})},\ \Eprint {https://arxiv.org/abs/2008.07516} {arXiv:2008.07516
  [hep-ph]} \BibitemShut {NoStop}%
\bibitem [{\citenamefont {He}\ and\ \citenamefont
  {Chen}(2021{\natexlab{a}})}]{He:2020btl}%
  \BibitemOpen
  \bibfield  {author} {\bibinfo {author} {\bibfnamefont {J.}~\bibnamefont
  {He}}\ and\ \bibinfo {author} {\bibfnamefont {D.-Y.}\ \bibnamefont {Chen}},\
  }\bibfield  {title} {\bibinfo {title} {{Molecular picture for $X_0(2900)$ and
  $X_1(2900)$}},\ }\href {https://doi.org/10.1088/1674-1137/abeda8} {\bibfield
  {journal} {\bibinfo  {journal} {Chin. Phys. C}\ }\textbf {\bibinfo {volume}
  {45}},\ \bibinfo {pages} {063102} (\bibinfo {year} {2021}{\natexlab{a}})},\
  \Eprint {https://arxiv.org/abs/2008.07782} {arXiv:2008.07782 [hep-ph]}
  \BibitemShut {NoStop}%
\bibitem [{\citenamefont {Liu}\ \emph {et~al.}(2020{\natexlab{a}})\citenamefont
  {Liu}, \citenamefont {Xie},\ and\ \citenamefont {Geng}}]{Liu:2020nil}%
  \BibitemOpen
  \bibfield  {author} {\bibinfo {author} {\bibfnamefont {M.-Z.}\ \bibnamefont
  {Liu}}, \bibinfo {author} {\bibfnamefont {J.-J.}\ \bibnamefont {Xie}},\ and\
  \bibinfo {author} {\bibfnamefont {L.-S.}\ \bibnamefont {Geng}},\ }\bibfield
  {title} {\bibinfo {title} {{$X_0(2866)$ as a $D^*\bar{K}^*$ molecular
  state}},\ }\href {https://doi.org/10.1103/PhysRevD.102.091502} {\bibfield
  {journal} {\bibinfo  {journal} {Phys. Rev. D}\ }\textbf {\bibinfo {volume}
  {102}},\ \bibinfo {pages} {091502} (\bibinfo {year} {2020}{\natexlab{a}})},\
  \Eprint {https://arxiv.org/abs/2008.07389} {arXiv:2008.07389 [hep-ph]}
  \BibitemShut {NoStop}%
\bibitem [{\citenamefont {Hu}\ \emph {et~al.}(2021)\citenamefont {Hu},
  \citenamefont {Lao}, \citenamefont {Ling},\ and\ \citenamefont
  {Wang}}]{Hu:2020mxp}%
  \BibitemOpen
  \bibfield  {author} {\bibinfo {author} {\bibfnamefont {M.-W.}\ \bibnamefont
  {Hu}}, \bibinfo {author} {\bibfnamefont {X.-Y.}\ \bibnamefont {Lao}},
  \bibinfo {author} {\bibfnamefont {P.}~\bibnamefont {Ling}},\ and\ \bibinfo
  {author} {\bibfnamefont {Q.}~\bibnamefont {Wang}},\ }\bibfield  {title}
  {\bibinfo {title} {$x_0(2900)$ and its heavy quark spin partners in molecular
  picture},\ }\href {https://doi.org/10.1088/1674-1137/abcfaa} {\bibfield
  {journal} {\bibinfo  {journal} {Chin. Phys. C}\ }\textbf {\bibinfo {volume}
  {45}},\ \bibinfo {pages} {021003} (\bibinfo {year} {2021})},\ \Eprint
  {https://arxiv.org/abs/2008.06894} {arXiv:2008.06894 [hep-ph]} \BibitemShut
  {NoStop}%
\bibitem [{\citenamefont {Agaev}\ \emph {et~al.}(2021)\citenamefont {Agaev},
  \citenamefont {Azizi},\ and\ \citenamefont {Sundu}}]{Agaev:2020nrc}%
  \BibitemOpen
  \bibfield  {author} {\bibinfo {author} {\bibfnamefont {S.~S.}\ \bibnamefont
  {Agaev}}, \bibinfo {author} {\bibfnamefont {K.}~\bibnamefont {Azizi}},\ and\
  \bibinfo {author} {\bibfnamefont {H.}~\bibnamefont {Sundu}},\ }\bibfield
  {title} {\bibinfo {title} {New scalar resonance $x_0(2900)$ as a molecule:
  mass and width},\ }\href {https://doi.org/10.1088/1361-6471/ac0b31}
  {\bibfield  {journal} {\bibinfo  {journal} {J. Phys. G}\ }\textbf {\bibinfo
  {volume} {48}},\ \bibinfo {pages} {085012} (\bibinfo {year} {2021})},\
  \Eprint {https://arxiv.org/abs/2008.13027} {arXiv:2008.13027 [hep-ph]}
  \BibitemShut {NoStop}%
\bibitem [{\citenamefont {Wang}\ and\ \citenamefont
  {Zhu}(2022)}]{Wang:2021lwy}%
  \BibitemOpen
  \bibfield  {author} {\bibinfo {author} {\bibfnamefont {B.}~\bibnamefont
  {Wang}}\ and\ \bibinfo {author} {\bibfnamefont {S.-L.}\ \bibnamefont {Zhu}},\
  }\bibfield  {title} {\bibinfo {title} {How to understand the $x(2900)$?},\
  }\href {https://doi.org/10.1140/epjc/s10052-022-10396-9} {\bibfield
  {journal} {\bibinfo  {journal} {Eur. Phys. J. C}\ }\textbf {\bibinfo {volume}
  {82}},\ \bibinfo {pages} {419} (\bibinfo {year} {2022})},\ \Eprint
  {https://arxiv.org/abs/2107.09275} {arXiv:2107.09275 [hep-ph]} \BibitemShut
  {NoStop}%
\bibitem [{\citenamefont {Wang}\ \emph
  {et~al.}(2024{\natexlab{a}})\citenamefont {Wang}, \citenamefont {Chen},
  \citenamefont {Meng},\ and\ \citenamefont {Zhu}}]{Wang:2023hpp}%
  \BibitemOpen
  \bibfield  {author} {\bibinfo {author} {\bibfnamefont {B.}~\bibnamefont
  {Wang}}, \bibinfo {author} {\bibfnamefont {K.}~\bibnamefont {Chen}}, \bibinfo
  {author} {\bibfnamefont {L.}~\bibnamefont {Meng}},\ and\ \bibinfo {author}
  {\bibfnamefont {S.-L.}\ \bibnamefont {Zhu}},\ }\bibfield  {title} {\bibinfo
  {title} {{Spectrum of the molecular tetraquarks: Unraveling the Tcs0(2900)
  and Tcs{\textasciimacron}0a(2900)}},\ }\href
  {https://doi.org/10.1103/PhysRevD.109.034027} {\bibfield  {journal} {\bibinfo
   {journal} {Phys. Rev. D}\ }\textbf {\bibinfo {volume} {109}},\ \bibinfo
  {pages} {034027} (\bibinfo {year} {2024}{\natexlab{a}})},\ \Eprint
  {https://arxiv.org/abs/2309.02191} {arXiv:2309.02191 [hep-ph]} \BibitemShut
  {NoStop}%
\bibitem [{\citenamefont {Karliner}\ and\ \citenamefont
  {Rosner}(2020)}]{Karliner:2020vsi}%
  \BibitemOpen
  \bibfield  {author} {\bibinfo {author} {\bibfnamefont {M.}~\bibnamefont
  {Karliner}}\ and\ \bibinfo {author} {\bibfnamefont {J.~L.}\ \bibnamefont
  {Rosner}},\ }\bibfield  {title} {\bibinfo {title} {First exotic hadron with
  open heavy flavor: $cs\bar u \bar d$ tetraquark},\ }\href
  {https://doi.org/10.1103/PhysRevD.102.094016} {\bibfield  {journal} {\bibinfo
   {journal} {Phys. Rev. D}\ }\textbf {\bibinfo {volume} {102}},\ \bibinfo
  {pages} {094016} (\bibinfo {year} {2020})},\ \Eprint
  {https://arxiv.org/abs/2008.05993} {arXiv:2008.05993 [hep-ph]} \BibitemShut
  {NoStop}%
\bibitem [{\citenamefont {He}\ and\ \citenamefont
  {Chen}(2021{\natexlab{b}})}]{He:2020jna}%
  \BibitemOpen
  \bibfield  {author} {\bibinfo {author} {\bibfnamefont {J.}~\bibnamefont
  {He}}\ and\ \bibinfo {author} {\bibfnamefont {D.-Y.}\ \bibnamefont {Chen}},\
  }\bibfield  {title} {\bibinfo {title} {Molecular picture for $x_0(2900)$ and
  $x_1(2900)$},\ }\href {https://doi.org/10.1088/1674-1137/abeda8} {\bibfield
  {journal} {\bibinfo  {journal} {Chin. Phys. C}\ }\textbf {\bibinfo {volume}
  {45}},\ \bibinfo {pages} {063102} (\bibinfo {year} {2021}{\natexlab{b}})},\
  \Eprint {https://arxiv.org/abs/2008.07782} {arXiv:2008.07782 [hep-ph]}
  \BibitemShut {NoStop}%
\bibitem [{\citenamefont {Wang}\ \emph
  {et~al.}(2021{\natexlab{a}})\citenamefont {Wang}, \citenamefont {Meng},
  \citenamefont {Xiao}, \citenamefont {Oka},\ and\ \citenamefont
  {Zhu}}]{Wang:2020xyc}%
  \BibitemOpen
  \bibfield  {author} {\bibinfo {author} {\bibfnamefont {G.-J.}\ \bibnamefont
  {Wang}}, \bibinfo {author} {\bibfnamefont {L.}~\bibnamefont {Meng}}, \bibinfo
  {author} {\bibfnamefont {L.-Y.}\ \bibnamefont {Xiao}}, \bibinfo {author}
  {\bibfnamefont {M.}~\bibnamefont {Oka}},\ and\ \bibinfo {author}
  {\bibfnamefont {S.-L.}\ \bibnamefont {Zhu}},\ }\bibfield  {title} {\bibinfo
  {title} {Mass spectrum and strong decays of tetraquark $c\bar s \bar q q$
  states},\ }\href {https://doi.org/10.1140/epjc/s10052-021-08978-0} {\bibfield
   {journal} {\bibinfo  {journal} {Eur. Phys. J. C}\ }\textbf {\bibinfo
  {volume} {81}},\ \bibinfo {pages} {188} (\bibinfo {year}
  {2021}{\natexlab{a}})},\ \Eprint {https://arxiv.org/abs/2010.09395}
  {arXiv:2010.09395 [hep-ph]} \BibitemShut {NoStop}%
\bibitem [{\citenamefont {Zhang}(2021)}]{Zhang:2020oze}%
  \BibitemOpen
  \bibfield  {author} {\bibinfo {author} {\bibfnamefont {J.-R.}\ \bibnamefont
  {Zhang}},\ }\bibfield  {title} {\bibinfo {title} {{Open-charm tetraquark
  candidate: Note on $X_0$(2900)}},\ }\href
  {https://doi.org/10.1103/PhysRevD.103.054019} {\bibfield  {journal} {\bibinfo
   {journal} {Phys. Rev. D}\ }\textbf {\bibinfo {volume} {103}},\ \bibinfo
  {pages} {054019} (\bibinfo {year} {2021})},\ \Eprint
  {https://arxiv.org/abs/2008.07295} {arXiv:2008.07295 [hep-ph]} \BibitemShut
  {NoStop}%
\bibitem [{\citenamefont {Wang}(2020)}]{Wang:2020prk}%
  \BibitemOpen
  \bibfield  {author} {\bibinfo {author} {\bibfnamefont {Z.-G.}\ \bibnamefont
  {Wang}},\ }\bibfield  {title} {\bibinfo {title} {Analysis of the $x_0(2900)$
  as the scalar tetraquark state via the qcd sum rules},\ }\href
  {https://doi.org/10.1142/S0217751X20501870} {\bibfield  {journal} {\bibinfo
  {journal} {Int. J. Mod. Phys. A}\ }\textbf {\bibinfo {volume} {35}},\
  \bibinfo {pages} {2050187} (\bibinfo {year} {2020})},\ \Eprint
  {https://arxiv.org/abs/2008.07833} {arXiv:2008.07833 [hep-ph]} \BibitemShut
  {NoStop}%
\bibitem [{\citenamefont {L{\"u}}\ \emph
  {et~al.}(2020{\natexlab{a}})\citenamefont {L{\"u}}, \citenamefont {Chen},\
  and\ \citenamefont {Dong}}]{Lu:2020qmp}%
  \BibitemOpen
  \bibfield  {author} {\bibinfo {author} {\bibfnamefont {Q.-F.}\ \bibnamefont
  {L{\"u}}}, \bibinfo {author} {\bibfnamefont {D.-Y.}\ \bibnamefont {Chen}},\
  and\ \bibinfo {author} {\bibfnamefont {Y.-B.}\ \bibnamefont {Dong}},\
  }\bibfield  {title} {\bibinfo {title} {Open charm and bottom tetraquarks in
  an extended relativized quark model},\ }\href
  {https://doi.org/10.1103/PhysRevD.102.074021} {\bibfield  {journal} {\bibinfo
   {journal} {Phys. Rev. D}\ }\textbf {\bibinfo {volume} {102}},\ \bibinfo
  {pages} {074021} (\bibinfo {year} {2020}{\natexlab{a}})},\ \Eprint
  {https://arxiv.org/abs/2008.07340} {arXiv:2008.07340 [hep-ph]} \BibitemShut
  {NoStop}%
\bibitem [{\citenamefont {Tan}\ and\ \citenamefont {Ping}(2021)}]{Tan:2020cpu}%
  \BibitemOpen
  \bibfield  {author} {\bibinfo {author} {\bibfnamefont {Y.}~\bibnamefont
  {Tan}}\ and\ \bibinfo {author} {\bibfnamefont {J.}~\bibnamefont {Ping}},\
  }\bibfield  {title} {\bibinfo {title} {{X(2900) in a chiral quark model}},\
  }\href {https://doi.org/10.1088/1674-1137/ac0ba4} {\bibfield  {journal}
  {\bibinfo  {journal} {Chin. Phys. C}\ }\textbf {\bibinfo {volume} {45}},\
  \bibinfo {pages} {093104} (\bibinfo {year} {2021})},\ \Eprint
  {https://arxiv.org/abs/2010.04045} {arXiv:2010.04045 [hep-ph]} \BibitemShut
  {NoStop}%
\bibitem [{\citenamefont {Albuquerque}\ \emph {et~al.}(2021)\citenamefont
  {Albuquerque}, \citenamefont {Narison}, \citenamefont {Rabetiarivony},\ and\
  \citenamefont {Randriamanatrika}}]{Albuquerque:2020ugi}%
  \BibitemOpen
  \bibfield  {author} {\bibinfo {author} {\bibfnamefont {R.~M.}\ \bibnamefont
  {Albuquerque}}, \bibinfo {author} {\bibfnamefont {S.}~\bibnamefont
  {Narison}}, \bibinfo {author} {\bibfnamefont {D.}~\bibnamefont
  {Rabetiarivony}},\ and\ \bibinfo {author} {\bibfnamefont {G.}~\bibnamefont
  {Randriamanatrika}},\ }\bibfield  {title} {\bibinfo {title} {$x_0,1(2900)$
  and $(d^-k^+)$ invariant mass from qcd laplace sum rules at nlo},\ }\href
  {https://doi.org/10.1016/j.nuclphysa.2020.122113} {\bibfield  {journal}
  {\bibinfo  {journal} {Nucl. Phys. A}\ }\textbf {\bibinfo {volume} {1007}},\
  \bibinfo {pages} {122113} (\bibinfo {year} {2021})},\ \Eprint
  {https://arxiv.org/abs/2008.13463} {arXiv:2008.13463 [hep-ph]} \BibitemShut
  {NoStop}%
\bibitem [{\citenamefont {Yang}\ \emph {et~al.}(2021)\citenamefont {Yang},
  \citenamefont {Ping},\ and\ \citenamefont {Segovia}}]{Yang:2021izl}%
  \BibitemOpen
  \bibfield  {author} {\bibinfo {author} {\bibfnamefont {G.}~\bibnamefont
  {Yang}}, \bibinfo {author} {\bibfnamefont {J.}~\bibnamefont {Ping}},\ and\
  \bibinfo {author} {\bibfnamefont {J.}~\bibnamefont {Segovia}},\ }\bibfield
  {title} {\bibinfo {title} {$sq\bar q \bar q$ ($q=u,d; q=c,b$) tetraquarks in
  the chiral quark model},\ }\href
  {https://doi.org/10.1103/PhysRevD.103.074011} {\bibfield  {journal} {\bibinfo
   {journal} {Phys. Rev. D}\ }\textbf {\bibinfo {volume} {103}},\ \bibinfo
  {pages} {074011} (\bibinfo {year} {2021})},\ \Eprint
  {https://arxiv.org/abs/2101.04933} {arXiv:2101.04933 [hep-ph]} \BibitemShut
  {NoStop}%
\bibitem [{\citenamefont {Agaev}\ \emph {et~al.}(2022)\citenamefont {Agaev},
  \citenamefont {Azizi},\ and\ \citenamefont {Sundu}}]{Agaev:2022eeh}%
  \BibitemOpen
  \bibfield  {author} {\bibinfo {author} {\bibfnamefont {S.~S.}\ \bibnamefont
  {Agaev}}, \bibinfo {author} {\bibfnamefont {K.}~\bibnamefont {Azizi}},\ and\
  \bibinfo {author} {\bibfnamefont {H.}~\bibnamefont {Sundu}},\ }\bibfield
  {title} {\bibinfo {title} {{Is the resonance X0(2900) a ground-state or
  radially excited scalar tetraquark
  [ud][c{\textasciimacron}s{\textasciimacron}]?}},\ }\href
  {https://doi.org/10.1103/PhysRevD.106.014019} {\bibfield  {journal} {\bibinfo
   {journal} {Phys. Rev. D}\ }\textbf {\bibinfo {volume} {106}},\ \bibinfo
  {pages} {014019} (\bibinfo {year} {2022})},\ \Eprint
  {https://arxiv.org/abs/2204.08498} {arXiv:2204.08498 [hep-ph]} \BibitemShut
  {NoStop}%
\bibitem [{\citenamefont {Liu}\ \emph {et~al.}(2023)\citenamefont {Liu},
  \citenamefont {Ni}, \citenamefont {Zhong},\ and\ \citenamefont
  {Zhao}}]{Liu:2022hbk}%
  \BibitemOpen
  \bibfield  {author} {\bibinfo {author} {\bibfnamefont {F.-X.}\ \bibnamefont
  {Liu}}, \bibinfo {author} {\bibfnamefont {R.-H.}\ \bibnamefont {Ni}},
  \bibinfo {author} {\bibfnamefont {X.-H.}\ \bibnamefont {Zhong}},\ and\
  \bibinfo {author} {\bibfnamefont {Q.}~\bibnamefont {Zhao}},\ }\bibfield
  {title} {\bibinfo {title} {{Charmed-strange tetraquarks and their decays in
  the potential quark model}},\ }\href
  {https://doi.org/10.1103/PhysRevD.107.096020} {\bibfield  {journal} {\bibinfo
   {journal} {Phys. Rev. D}\ }\textbf {\bibinfo {volume} {107}},\ \bibinfo
  {pages} {096020} (\bibinfo {year} {2023})},\ \Eprint
  {https://arxiv.org/abs/2211.01711} {arXiv:2211.01711 [hep-ph]} \BibitemShut
  {NoStop}%
\bibitem [{\citenamefont {Liu}\ \emph {et~al.}(2020{\natexlab{b}})\citenamefont
  {Liu}, \citenamefont {Xie},\ and\ \citenamefont {Geng}}]{Liu:2020orv}%
  \BibitemOpen
  \bibfield  {author} {\bibinfo {author} {\bibfnamefont {M.-Z.}\ \bibnamefont
  {Liu}}, \bibinfo {author} {\bibfnamefont {J.-J.}\ \bibnamefont {Xie}},\ and\
  \bibinfo {author} {\bibfnamefont {L.-S.}\ \bibnamefont {Geng}},\ }\bibfield
  {title} {\bibinfo {title} {$x_0(2866)$ as a $d^* \bar k^*$ molecular state},\
  }\href {https://doi.org/10.1103/PhysRevD.102.091502} {\bibfield  {journal}
  {\bibinfo  {journal} {Phys. Rev. D}\ }\textbf {\bibinfo {volume} {102}},\
  \bibinfo {pages} {091502} (\bibinfo {year} {2020}{\natexlab{b}})},\ \Eprint
  {https://arxiv.org/abs/2008.07389} {arXiv:2008.07389 [hep-ph]} \BibitemShut
  {NoStop}%
\bibitem [{\citenamefont {Burns}\ and\ \citenamefont
  {Swanson}(2021)}]{Burns:2020epm}%
  \BibitemOpen
  \bibfield  {author} {\bibinfo {author} {\bibfnamefont {T.~J.}\ \bibnamefont
  {Burns}}\ and\ \bibinfo {author} {\bibfnamefont {E.~S.}\ \bibnamefont
  {Swanson}},\ }\bibfield  {title} {\bibinfo {title} {Kinematical cusp and
  resonance interpretations of the $x(2900)$},\ }\href
  {https://doi.org/10.1016/j.physletb.2020.136057} {\bibfield  {journal}
  {\bibinfo  {journal} {Phys. Lett. B}\ }\textbf {\bibinfo {volume} {813}},\
  \bibinfo {pages} {136057} (\bibinfo {year} {2021})},\ \Eprint
  {https://arxiv.org/abs/2008.12838} {arXiv:2008.12838 [hep-ph]} \BibitemShut
  {NoStop}%
\bibitem [{\citenamefont {Chen}\ \emph {et~al.}(2024)\citenamefont {Chen},
  \citenamefont {Wu}, \citenamefont {Meng},\ and\ \citenamefont
  {Zhu}}]{Chen:2023syh}%
  \BibitemOpen
  \bibfield  {author} {\bibinfo {author} {\bibfnamefont {Y.-K.}\ \bibnamefont
  {Chen}}, \bibinfo {author} {\bibfnamefont {W.-L.}\ \bibnamefont {Wu}},
  \bibinfo {author} {\bibfnamefont {L.}~\bibnamefont {Meng}},\ and\ \bibinfo
  {author} {\bibfnamefont {S.-L.}\ \bibnamefont {Zhu}},\ }\bibfield  {title}
  {\bibinfo {title} {{Unified description of the
  Qsq\textasciimacron{}q\textasciimacron{} molecular bound states, molecular
  resonances, and compact tetraquark states in the quark potential model}},\
  }\href {https://doi.org/10.1103/PhysRevD.109.014010} {\bibfield  {journal}
  {\bibinfo  {journal} {Phys. Rev. D}\ }\textbf {\bibinfo {volume} {109}},\
  \bibinfo {pages} {014010} (\bibinfo {year} {2024})},\ \Eprint
  {https://arxiv.org/abs/2310.14597} {arXiv:2310.14597 [hep-ph]} \BibitemShut
  {NoStop}%
\bibitem [{\citenamefont {Aaij}\ \emph
  {et~al.}(2022{\natexlab{a}})\citenamefont {Aaij} \emph
  {et~al.}}]{LHCb:2021vvq}%
  \BibitemOpen
  \bibfield  {author} {\bibinfo {author} {\bibfnamefont {R.}~\bibnamefont
  {Aaij}} \emph {et~al.} (\bibinfo {collaboration} {LHCb}),\ }\bibfield
  {title} {\bibinfo {title} {{Observation of an exotic narrow doubly charmed
  tetraquark}},\ }\href {https://doi.org/10.1038/s41567-022-01614-y} {\bibfield
   {journal} {\bibinfo  {journal} {Nature Phys.}\ }\textbf {\bibinfo {volume}
  {18}},\ \bibinfo {pages} {751} (\bibinfo {year} {2022}{\natexlab{a}})},\
  \Eprint {https://arxiv.org/abs/2109.01038} {arXiv:2109.01038 [hep-ex]}
  \BibitemShut {NoStop}%
\bibitem [{\citenamefont {Aaij}\ \emph
  {et~al.}(2022{\natexlab{b}})\citenamefont {Aaij} \emph
  {et~al.}}]{LHCb:2021auc}%
  \BibitemOpen
  \bibfield  {author} {\bibinfo {author} {\bibfnamefont {R.}~\bibnamefont
  {Aaij}} \emph {et~al.} (\bibinfo {collaboration} {LHCb}),\ }\bibfield
  {title} {\bibinfo {title} {{Study of the doubly charmed tetraquark
  $T_{cc}^{+}$}},\ }\href {https://doi.org/10.1038/s41467-022-30206-w}
  {\bibfield  {journal} {\bibinfo  {journal} {Nature Commun.}\ }\textbf
  {\bibinfo {volume} {13}},\ \bibinfo {pages} {3351} (\bibinfo {year}
  {2022}{\natexlab{b}})},\ \Eprint {https://arxiv.org/abs/2109.01056}
  {arXiv:2109.01056 [hep-ex]} \BibitemShut {NoStop}%
\bibitem [{\citenamefont {Pepin}\ \emph {et~al.}(1997)\citenamefont {Pepin},
  \citenamefont {Stancu}, \citenamefont {Genovese},\ and\ \citenamefont
  {Richard}}]{Pepin:1996id}%
  \BibitemOpen
  \bibfield  {author} {\bibinfo {author} {\bibfnamefont {S.}~\bibnamefont
  {Pepin}}, \bibinfo {author} {\bibfnamefont {F.}~\bibnamefont {Stancu}},
  \bibinfo {author} {\bibfnamefont {M.}~\bibnamefont {Genovese}},\ and\
  \bibinfo {author} {\bibfnamefont {J.~M.}\ \bibnamefont {Richard}},\
  }\bibfield  {title} {\bibinfo {title} {{Tetraquarks with color blind forces
  in chiral quark models}},\ }\href
  {https://doi.org/10.1016/S0370-2693(96)01597-3} {\bibfield  {journal}
  {\bibinfo  {journal} {Phys. Lett. B}\ }\textbf {\bibinfo {volume} {393}},\
  \bibinfo {pages} {119} (\bibinfo {year} {1997})},\ \Eprint
  {https://arxiv.org/abs/hep-ph/9609348} {arXiv:hep-ph/9609348} \BibitemShut
  {NoStop}%
\bibitem [{\citenamefont {Gelman}\ and\ \citenamefont
  {Nussinov}(2003)}]{Gelman:2002wf}%
  \BibitemOpen
  \bibfield  {author} {\bibinfo {author} {\bibfnamefont {B.~A.}\ \bibnamefont
  {Gelman}}\ and\ \bibinfo {author} {\bibfnamefont {S.}~\bibnamefont
  {Nussinov}},\ }\bibfield  {title} {\bibinfo {title} {{Does a narrow
  tetraquark cc anti-u anti-d state exist?}},\ }\href
  {https://doi.org/10.1016/S0370-2693(02)03069-1} {\bibfield  {journal}
  {\bibinfo  {journal} {Phys. Lett. B}\ }\textbf {\bibinfo {volume} {551}},\
  \bibinfo {pages} {296} (\bibinfo {year} {2003})},\ \Eprint
  {https://arxiv.org/abs/hep-ph/0209095} {arXiv:hep-ph/0209095} \BibitemShut
  {NoStop}%
\bibitem [{\citenamefont {Vijande}\ \emph {et~al.}(2004)\citenamefont
  {Vijande}, \citenamefont {Fernandez}, \citenamefont {Valcarce},\ and\
  \citenamefont {Silvestre-Brac}}]{Vijande:2003ki}%
  \BibitemOpen
  \bibfield  {author} {\bibinfo {author} {\bibfnamefont {J.}~\bibnamefont
  {Vijande}}, \bibinfo {author} {\bibfnamefont {F.}~\bibnamefont {Fernandez}},
  \bibinfo {author} {\bibfnamefont {A.}~\bibnamefont {Valcarce}},\ and\
  \bibinfo {author} {\bibfnamefont {B.}~\bibnamefont {Silvestre-Brac}},\
  }\bibfield  {title} {\bibinfo {title} {{Tetraquarks in a chiral constituent
  quark model}},\ }\href {https://doi.org/10.1140/epja/i2003-10128-9}
  {\bibfield  {journal} {\bibinfo  {journal} {Eur. Phys. J. A}\ }\textbf
  {\bibinfo {volume} {19}},\ \bibinfo {pages} {383} (\bibinfo {year} {2004})},\
  \Eprint {https://arxiv.org/abs/hep-ph/0310007} {arXiv:hep-ph/0310007}
  \BibitemShut {NoStop}%
\bibitem [{\citenamefont {Ebert}\ \emph {et~al.}(2007)\citenamefont {Ebert},
  \citenamefont {Faustov}, \citenamefont {Galkin},\ and\ \citenamefont
  {Lucha}}]{Ebert:2007rn}%
  \BibitemOpen
  \bibfield  {author} {\bibinfo {author} {\bibfnamefont {D.}~\bibnamefont
  {Ebert}}, \bibinfo {author} {\bibfnamefont {R.~N.}\ \bibnamefont {Faustov}},
  \bibinfo {author} {\bibfnamefont {V.~O.}\ \bibnamefont {Galkin}},\ and\
  \bibinfo {author} {\bibfnamefont {W.}~\bibnamefont {Lucha}},\ }\bibfield
  {title} {\bibinfo {title} {{Masses of tetraquarks with two heavy quarks in
  the relativistic quark model}},\ }\href
  {https://doi.org/10.1103/PhysRevD.76.114015} {\bibfield  {journal} {\bibinfo
  {journal} {Phys. Rev. D}\ }\textbf {\bibinfo {volume} {76}},\ \bibinfo
  {pages} {114015} (\bibinfo {year} {2007})},\ \Eprint
  {https://arxiv.org/abs/0706.3853} {arXiv:0706.3853 [hep-ph]} \BibitemShut
  {NoStop}%
\bibitem [{\citenamefont {Du}\ \emph {et~al.}(2013)\citenamefont {Du},
  \citenamefont {Chen}, \citenamefont {Chen},\ and\ \citenamefont
  {Zhu}}]{Du:2012wp}%
  \BibitemOpen
  \bibfield  {author} {\bibinfo {author} {\bibfnamefont {M.-L.}\ \bibnamefont
  {Du}}, \bibinfo {author} {\bibfnamefont {W.}~\bibnamefont {Chen}}, \bibinfo
  {author} {\bibfnamefont {X.-L.}\ \bibnamefont {Chen}},\ and\ \bibinfo
  {author} {\bibfnamefont {S.-L.}\ \bibnamefont {Zhu}},\ }\bibfield  {title}
  {\bibinfo {title} {{Exotic $QQ\bar{q}\bar{q}$, $QQ\bar{q}\bar{s}$ and
  $QQ\bar{s}\bar{s}$ states}},\ }\href
  {https://doi.org/10.1103/PhysRevD.87.014003} {\bibfield  {journal} {\bibinfo
  {journal} {Phys. Rev. D}\ }\textbf {\bibinfo {volume} {87}},\ \bibinfo
  {pages} {014003} (\bibinfo {year} {2013})},\ \Eprint
  {https://arxiv.org/abs/1209.5134} {arXiv:1209.5134 [hep-ph]} \BibitemShut
  {NoStop}%
\bibitem [{\citenamefont {Karliner}\ \emph {et~al.}(2018)\citenamefont
  {Karliner}, \citenamefont {Rosner},\ and\ \citenamefont
  {Skwarnicki}}]{Karliner:2017qhf}%
  \BibitemOpen
  \bibfield  {author} {\bibinfo {author} {\bibfnamefont {M.}~\bibnamefont
  {Karliner}}, \bibinfo {author} {\bibfnamefont {J.~L.}\ \bibnamefont
  {Rosner}},\ and\ \bibinfo {author} {\bibfnamefont {T.}~\bibnamefont
  {Skwarnicki}},\ }\bibfield  {title} {\bibinfo {title} {{Multiquark States}},\
  }\href {https://doi.org/10.1146/annurev-nucl-101917-020902} {\bibfield
  {journal} {\bibinfo  {journal} {Ann. Rev. Nucl. Part. Sci.}\ }\textbf
  {\bibinfo {volume} {68}},\ \bibinfo {pages} {17} (\bibinfo {year} {2018})},\
  \Eprint {https://arxiv.org/abs/1711.10626} {arXiv:1711.10626 [hep-ph]}
  \BibitemShut {NoStop}%
\bibitem [{\citenamefont {Eichten}\ and\ \citenamefont
  {Quigg}(2017)}]{Eichten:2017ffp}%
  \BibitemOpen
  \bibfield  {author} {\bibinfo {author} {\bibfnamefont {E.~J.}\ \bibnamefont
  {Eichten}}\ and\ \bibinfo {author} {\bibfnamefont {C.}~\bibnamefont
  {Quigg}},\ }\bibfield  {title} {\bibinfo {title} {{Heavy-quark symmetry
  implies stable heavy tetraquark mesons $Q_iQ_j \bar q_k \bar q_l$}},\ }\href
  {https://doi.org/10.1103/PhysRevLett.119.202002} {\bibfield  {journal}
  {\bibinfo  {journal} {Phys. Rev. Lett.}\ }\textbf {\bibinfo {volume} {119}},\
  \bibinfo {pages} {202002} (\bibinfo {year} {2017})},\ \Eprint
  {https://arxiv.org/abs/1707.09575} {arXiv:1707.09575 [hep-ph]} \BibitemShut
  {NoStop}%
\bibitem [{\citenamefont {Yang}\ \emph {et~al.}(2020)\citenamefont {Yang},
  \citenamefont {Ping},\ and\ \citenamefont {Segovia}}]{Yang:2019itm}%
  \BibitemOpen
  \bibfield  {author} {\bibinfo {author} {\bibfnamefont {G.}~\bibnamefont
  {Yang}}, \bibinfo {author} {\bibfnamefont {J.}~\bibnamefont {Ping}},\ and\
  \bibinfo {author} {\bibfnamefont {J.}~\bibnamefont {Segovia}},\ }\bibfield
  {title} {\bibinfo {title} {{Doubly-heavy tetraquarks}},\ }\href
  {https://doi.org/10.1103/PhysRevD.101.014001} {\bibfield  {journal} {\bibinfo
   {journal} {Phys. Rev. D}\ }\textbf {\bibinfo {volume} {101}},\ \bibinfo
  {pages} {014001} (\bibinfo {year} {2020})},\ \Eprint
  {https://arxiv.org/abs/1911.00215} {arXiv:1911.00215 [hep-ph]} \BibitemShut
  {NoStop}%
\bibitem [{\citenamefont {Albaladejo}(2022)}]{Albaladejo:2021vln}%
  \BibitemOpen
  \bibfield  {author} {\bibinfo {author} {\bibfnamefont {M.}~\bibnamefont
  {Albaladejo}},\ }\bibfield  {title} {\bibinfo {title} {{Tcc+ coupled channel
  analysis and predictions}},\ }\href
  {https://doi.org/10.1016/j.physletb.2022.137052} {\bibfield  {journal}
  {\bibinfo  {journal} {Phys. Lett. B}\ }\textbf {\bibinfo {volume} {829}},\
  \bibinfo {pages} {137052} (\bibinfo {year} {2022})},\ \Eprint
  {https://arxiv.org/abs/2110.02944} {arXiv:2110.02944 [hep-ph]} \BibitemShut
  {NoStop}%
\bibitem [{\citenamefont {Meng}\ \emph {et~al.}(2025)\citenamefont {Meng},
  \citenamefont {Wang},\ and\ \citenamefont {Oka}}]{Meng:2024yhu}%
  \BibitemOpen
  \bibfield  {author} {\bibinfo {author} {\bibfnamefont {Q.}~\bibnamefont
  {Meng}}, \bibinfo {author} {\bibfnamefont {G.-J.}\ \bibnamefont {Wang}},\
  and\ \bibinfo {author} {\bibfnamefont {M.}~\bibnamefont {Oka}},\ }\bibfield
  {title} {\bibinfo {title} {{Mass spectra of full-heavy and double-heavy
  tetraquark states in the conventional quark model}},\ }\href
  {https://doi.org/10.1103/PhysRevD.111.014018} {\bibfield  {journal} {\bibinfo
   {journal} {Phys. Rev. D}\ }\textbf {\bibinfo {volume} {111}},\ \bibinfo
  {pages} {014018} (\bibinfo {year} {2025})},\ \Eprint
  {https://arxiv.org/abs/2404.01238} {arXiv:2404.01238 [hep-ph]} \BibitemShut
  {NoStop}%
\bibitem [{\citenamefont {Whyte}\ \emph {et~al.}(2025)\citenamefont {Whyte},
  \citenamefont {Wilson},\ and\ \citenamefont {Thomas}}]{Whyte:2024ihh}%
  \BibitemOpen
  \bibfield  {author} {\bibinfo {author} {\bibfnamefont {T.}~\bibnamefont
  {Whyte}}, \bibinfo {author} {\bibfnamefont {D.~J.}\ \bibnamefont {Wilson}},\
  and\ \bibinfo {author} {\bibfnamefont {C.~E.}\ \bibnamefont {Thomas}}
  (\bibinfo {collaboration} {Hadron Spectrum}),\ }\bibfield  {title} {\bibinfo
  {title} {{Near-threshold states in coupled DD*-D*D* scattering from lattice
  QCD}},\ }\href {https://doi.org/10.1103/PhysRevD.111.034511} {\bibfield
  {journal} {\bibinfo  {journal} {Phys. Rev. D}\ }\textbf {\bibinfo {volume}
  {111}},\ \bibinfo {pages} {034511} (\bibinfo {year} {2025})},\ \Eprint
  {https://arxiv.org/abs/2405.15741} {arXiv:2405.15741 [hep-lat]} \BibitemShut
  {NoStop}%
\bibitem [{\citenamefont {Wu}\ \emph {et~al.}(2024{\natexlab{a}})\citenamefont
  {Wu}, \citenamefont {Ma}, \citenamefont {Chen}, \citenamefont {Meng},\ and\
  \citenamefont {Zhu}}]{Wu:2024zbx}%
  \BibitemOpen
  \bibfield  {author} {\bibinfo {author} {\bibfnamefont {W.-L.}\ \bibnamefont
  {Wu}}, \bibinfo {author} {\bibfnamefont {Y.}~\bibnamefont {Ma}}, \bibinfo
  {author} {\bibfnamefont {Y.-K.}\ \bibnamefont {Chen}}, \bibinfo {author}
  {\bibfnamefont {L.}~\bibnamefont {Meng}},\ and\ \bibinfo {author}
  {\bibfnamefont {S.-L.}\ \bibnamefont {Zhu}},\ }\bibfield  {title} {\bibinfo
  {title} {{Doubly heavy tetraquark bound and resonant states}},\ }\href
  {https://doi.org/10.1103/PhysRevD.110.094041} {\bibfield  {journal} {\bibinfo
   {journal} {Phys. Rev. D}\ }\textbf {\bibinfo {volume} {110}},\ \bibinfo
  {pages} {094041} (\bibinfo {year} {2024}{\natexlab{a}})},\ \Eprint
  {https://arxiv.org/abs/2409.03373} {arXiv:2409.03373 [hep-ph]} \BibitemShut
  {NoStop}%
\bibitem [{\citenamefont {Aaij}\ \emph
  {et~al.}(2020{\natexlab{c}})\citenamefont {Aaij} \emph
  {et~al.}}]{LHCb:2020bwg}%
  \BibitemOpen
  \bibfield  {author} {\bibinfo {author} {\bibfnamefont {R.}~\bibnamefont
  {Aaij}} \emph {et~al.} (\bibinfo {collaboration} {LHCb}),\ }\bibfield
  {title} {\bibinfo {title} {{Observation of structure in the $J /\psi$ -pair
  mass spectrum}},\ }\href {https://doi.org/10.1016/j.scib.2020.08.032}
  {\bibfield  {journal} {\bibinfo  {journal} {Sci. Bull.}\ }\textbf {\bibinfo
  {volume} {65}},\ \bibinfo {pages} {1983} (\bibinfo {year}
  {2020}{\natexlab{c}})},\ \Eprint {https://arxiv.org/abs/2006.16957}
  {arXiv:2006.16957 [hep-ex]} \BibitemShut {NoStop}%
\bibitem [{\citenamefont {Hayrapetyan}\ \emph {et~al.}(2024)\citenamefont
  {Hayrapetyan} \emph {et~al.}}]{CMS:2023owd}%
  \BibitemOpen
  \bibfield  {author} {\bibinfo {author} {\bibfnamefont {A.}~\bibnamefont
  {Hayrapetyan}} \emph {et~al.} (\bibinfo {collaboration} {CMS}),\ }\bibfield
  {title} {\bibinfo {title} {{New Structures in the
  J/{\ensuremath{\psi}}J/{\ensuremath{\psi}} Mass Spectrum in Proton-Proton
  Collisions at s=13{\,}{\,}TeV}},\ }\href
  {https://doi.org/10.1103/PhysRevLett.132.111901} {\bibfield  {journal}
  {\bibinfo  {journal} {Phys. Rev. Lett.}\ }\textbf {\bibinfo {volume} {132}},\
  \bibinfo {pages} {111901} (\bibinfo {year} {2024})},\ \Eprint
  {https://arxiv.org/abs/2306.07164} {arXiv:2306.07164 [hep-ex]} \BibitemShut
  {NoStop}%
\bibitem [{\citenamefont {Aad}\ \emph {et~al.}(2023)\citenamefont {Aad} \emph
  {et~al.}}]{ATLAS:2023bft}%
  \BibitemOpen
  \bibfield  {author} {\bibinfo {author} {\bibfnamefont {G.}~\bibnamefont
  {Aad}} \emph {et~al.} (\bibinfo {collaboration} {ATLAS}),\ }\bibfield
  {title} {\bibinfo {title} {{Observation of an Excess of Dicharmonium Events
  in the Four-Muon Final State with the ATLAS Detector}},\ }\href
  {https://doi.org/10.1103/PhysRevLett.131.151902} {\bibfield  {journal}
  {\bibinfo  {journal} {Phys. Rev. Lett.}\ }\textbf {\bibinfo {volume} {131}},\
  \bibinfo {pages} {151902} (\bibinfo {year} {2023})},\ \Eprint
  {https://arxiv.org/abs/2304.08962} {arXiv:2304.08962 [hep-ex]} \BibitemShut
  {NoStop}%
\bibitem [{\citenamefont {Aubert}\ \emph {et~al.}(2006)\citenamefont {Aubert}
  \emph {et~al.}}]{BaBar:2006gsq}%
  \BibitemOpen
  \bibfield  {author} {\bibinfo {author} {\bibfnamefont {B.}~\bibnamefont
  {Aubert}} \emph {et~al.} (\bibinfo {collaboration} {BaBar}),\ }\bibfield
  {title} {\bibinfo {title} {{A Structure at 2175-MeV in $e^{+} e^{-} \to \phi$
  f0(980) Observed via Initial-State Radiation}},\ }\href
  {https://doi.org/10.1103/PhysRevD.74.091103} {\bibfield  {journal} {\bibinfo
  {journal} {Phys. Rev. D}\ }\textbf {\bibinfo {volume} {74}},\ \bibinfo
  {pages} {091103} (\bibinfo {year} {2006})},\ \Eprint
  {https://arxiv.org/abs/hep-ex/0610018} {arXiv:hep-ex/0610018} \BibitemShut
  {NoStop}%
\bibitem [{\citenamefont {Aubert}\ \emph {et~al.}(2007)\citenamefont {Aubert}
  \emph {et~al.}}]{BaBar:2007ptr}%
  \BibitemOpen
  \bibfield  {author} {\bibinfo {author} {\bibfnamefont {B.}~\bibnamefont
  {Aubert}} \emph {et~al.} (\bibinfo {collaboration} {BaBar}),\ }\bibfield
  {title} {\bibinfo {title} {{The e+ e- ---{\ensuremath{>}} K+ K- pi+ pi-, K+
  K- pi0 pi0 and K+ K- K+ K- cross-sections measured with initial-state
  radiation}},\ }\href {https://doi.org/10.1103/PhysRevD.76.012008} {\bibfield
  {journal} {\bibinfo  {journal} {Phys. Rev. D}\ }\textbf {\bibinfo {volume}
  {76}},\ \bibinfo {pages} {012008} (\bibinfo {year} {2007})},\ \Eprint
  {https://arxiv.org/abs/0704.0630} {arXiv:0704.0630 [hep-ex]} \BibitemShut
  {NoStop}%
\bibitem [{\citenamefont {Aubert}\ \emph {et~al.}(2008)\citenamefont {Aubert}
  \emph {et~al.}}]{BaBar:2007ceh}%
  \BibitemOpen
  \bibfield  {author} {\bibinfo {author} {\bibfnamefont {B.}~\bibnamefont
  {Aubert}} \emph {et~al.} (\bibinfo {collaboration} {BaBar}),\ }\bibfield
  {title} {\bibinfo {title} {{Measurements of $e^{+} e^{-} \to K^{+} K^{-}
  \eta$, $K^{+} K^{-} \pi^0$ and $K^0_{s} K^\pm \pi^\mp$ cross- sections using
  initial state radiation events}},\ }\href
  {https://doi.org/10.1103/PhysRevD.77.092002} {\bibfield  {journal} {\bibinfo
  {journal} {Phys. Rev. D}\ }\textbf {\bibinfo {volume} {77}},\ \bibinfo
  {pages} {092002} (\bibinfo {year} {2008})},\ \Eprint
  {https://arxiv.org/abs/0710.4451} {arXiv:0710.4451 [hep-ex]} \BibitemShut
  {NoStop}%
\bibitem [{\citenamefont {Lees}\ \emph {et~al.}(2012)\citenamefont {Lees} \emph
  {et~al.}}]{BaBar:2011btv}%
  \BibitemOpen
  \bibfield  {author} {\bibinfo {author} {\bibfnamefont {J.~P.}\ \bibnamefont
  {Lees}} \emph {et~al.} (\bibinfo {collaboration} {BaBar}),\ }\bibfield
  {title} {\bibinfo {title} {{Cross Sections for the Reactions e+e-
  --{\ensuremath{>}} K+ K- pi+pi-, K+ K- pi0pi0, and K+ K- K+ K- Measured Using
  Initial-State Radiation Events}},\ }\href
  {https://doi.org/10.1103/PhysRevD.86.012008} {\bibfield  {journal} {\bibinfo
  {journal} {Phys. Rev. D}\ }\textbf {\bibinfo {volume} {86}},\ \bibinfo
  {pages} {012008} (\bibinfo {year} {2012})},\ \Eprint
  {https://arxiv.org/abs/1103.3001} {arXiv:1103.3001 [hep-ex]} \BibitemShut
  {NoStop}%
\bibitem [{\citenamefont {Ablikim}\ \emph {et~al.}(2011)\citenamefont {Ablikim}
  \emph {et~al.}}]{BESIII:2010gmv}%
  \BibitemOpen
  \bibfield  {author} {\bibinfo {author} {\bibfnamefont {M.}~\bibnamefont
  {Ablikim}} \emph {et~al.} (\bibinfo {collaboration} {BESIII}),\ }\bibfield
  {title} {\bibinfo {title} {{Confirmation of the $X(1835)$ and observation of
  the resonances $X(2120)$ and $X(2370)$ in $J/\psi\to \gamma
  \pi^+\pi^-\eta^\prime$}},\ }\href
  {https://doi.org/10.1103/PhysRevLett.106.072002} {\bibfield  {journal}
  {\bibinfo  {journal} {Phys. Rev. Lett.}\ }\textbf {\bibinfo {volume} {106}},\
  \bibinfo {pages} {072002} (\bibinfo {year} {2011})},\ \Eprint
  {https://arxiv.org/abs/1012.3510} {arXiv:1012.3510 [hep-ex]} \BibitemShut
  {NoStop}%
\bibitem [{\citenamefont {Ablikim}\ \emph {et~al.}(2020)\citenamefont {Ablikim}
  \emph {et~al.}}]{BESIII:2019wkp}%
  \BibitemOpen
  \bibfield  {author} {\bibinfo {author} {\bibfnamefont {M.}~\bibnamefont
  {Ablikim}} \emph {et~al.} (\bibinfo {collaboration} {BESIII}),\ }\bibfield
  {title} {\bibinfo {title} {{Observation of $X(2370)$ and search for X(2120)
  in $J/\psi \rightarrow \gamma K{\bar{K}} \eta '$}},\ }\href
  {https://doi.org/10.1140/epjc/s10052-020-8078-4} {\bibfield  {journal}
  {\bibinfo  {journal} {Eur. Phys. J. C}\ }\textbf {\bibinfo {volume} {80}},\
  \bibinfo {pages} {746} (\bibinfo {year} {2020})},\ \Eprint
  {https://arxiv.org/abs/1912.11253} {arXiv:1912.11253 [hep-ex]} \BibitemShut
  {NoStop}%
\bibitem [{\citenamefont {liu}\ \emph {et~al.}(2024)\citenamefont {liu},
  \citenamefont {Liu}, \citenamefont {Zhong},\ and\ \citenamefont
  {Zhao}}]{liu:2020eha}%
  \BibitemOpen
  \bibfield  {author} {\bibinfo {author} {\bibfnamefont {M.-S.}\ \bibnamefont
  {liu}}, \bibinfo {author} {\bibfnamefont {F.-X.}\ \bibnamefont {Liu}},
  \bibinfo {author} {\bibfnamefont {X.-H.}\ \bibnamefont {Zhong}},\ and\
  \bibinfo {author} {\bibfnamefont {Q.}~\bibnamefont {Zhao}},\ }\bibfield
  {title} {\bibinfo {title} {{Fully heavy tetraquark states and their evidences
  in LHC observations}},\ }\href {https://doi.org/10.1103/PhysRevD.109.076017}
  {\bibfield  {journal} {\bibinfo  {journal} {Phys. Rev. D}\ }\textbf {\bibinfo
  {volume} {109}},\ \bibinfo {pages} {076017} (\bibinfo {year} {2024})},\
  \Eprint {https://arxiv.org/abs/2006.11952} {arXiv:2006.11952 [hep-ph]}
  \BibitemShut {NoStop}%
\bibitem [{\citenamefont {L{\"u}}\ \emph
  {et~al.}(2020{\natexlab{b}})\citenamefont {L{\"u}}, \citenamefont {Chen},\
  and\ \citenamefont {Dong}}]{Lu:2020cns}%
  \BibitemOpen
  \bibfield  {author} {\bibinfo {author} {\bibfnamefont {Q.-F.}\ \bibnamefont
  {L{\"u}}}, \bibinfo {author} {\bibfnamefont {D.-Y.}\ \bibnamefont {Chen}},\
  and\ \bibinfo {author} {\bibfnamefont {Y.-B.}\ \bibnamefont {Dong}},\
  }\bibfield  {title} {\bibinfo {title} {{Masses of fully heavy tetraquarks $QQ
  {\bar{Q}} {\bar{Q}}$ in an extended relativized quark model}},\ }\href
  {https://doi.org/10.1140/epjc/s10052-020-08454-1} {\bibfield  {journal}
  {\bibinfo  {journal} {Eur. Phys. J. C}\ }\textbf {\bibinfo {volume} {80}},\
  \bibinfo {pages} {871} (\bibinfo {year} {2020}{\natexlab{b}})},\ \Eprint
  {https://arxiv.org/abs/2006.14445} {arXiv:2006.14445 [hep-ph]} \BibitemShut
  {NoStop}%
\bibitem [{\citenamefont {Dong}\ \emph {et~al.}(2020)\citenamefont {Dong},
  \citenamefont {Su}, \citenamefont {Chen}, \citenamefont {Cui},\ and\
  \citenamefont {Zhou}}]{Dong:2020okt}%
  \BibitemOpen
  \bibfield  {author} {\bibinfo {author} {\bibfnamefont {R.-R.}\ \bibnamefont
  {Dong}}, \bibinfo {author} {\bibfnamefont {N.}~\bibnamefont {Su}}, \bibinfo
  {author} {\bibfnamefont {H.-X.}\ \bibnamefont {Chen}}, \bibinfo {author}
  {\bibfnamefont {E.-L.}\ \bibnamefont {Cui}},\ and\ \bibinfo {author}
  {\bibfnamefont {Z.-Y.}\ \bibnamefont {Zhou}},\ }\bibfield  {title} {\bibinfo
  {title} {{QCD sum rule studies on the $s s \bar s \bar s$ tetraquark states
  of $J^{PC} = 0^{-+}$}},\ }\href
  {https://doi.org/10.1140/epjc/s10052-020-8340-9} {\bibfield  {journal}
  {\bibinfo  {journal} {Eur. Phys. J. C}\ }\textbf {\bibinfo {volume} {80}},\
  \bibinfo {pages} {749} (\bibinfo {year} {2020})},\ \Eprint
  {https://arxiv.org/abs/2003.07670} {arXiv:2003.07670 [hep-ph]} \BibitemShut
  {NoStop}%
\bibitem [{\citenamefont {Giron}\ and\ \citenamefont
  {Lebed}(2020)}]{Giron:2020wpx}%
  \BibitemOpen
  \bibfield  {author} {\bibinfo {author} {\bibfnamefont {J.~F.}\ \bibnamefont
  {Giron}}\ and\ \bibinfo {author} {\bibfnamefont {R.~F.}\ \bibnamefont
  {Lebed}},\ }\bibfield  {title} {\bibinfo {title} {{Simple spectrum of $c\bar
  c c\bar c$ states in the dynamical diquark model}},\ }\href
  {https://doi.org/10.1103/PhysRevD.102.074003} {\bibfield  {journal} {\bibinfo
   {journal} {Phys. Rev. D}\ }\textbf {\bibinfo {volume} {102}},\ \bibinfo
  {pages} {074003} (\bibinfo {year} {2020})},\ \Eprint
  {https://arxiv.org/abs/2008.01631} {arXiv:2008.01631 [hep-ph]} \BibitemShut
  {NoStop}%
\bibitem [{\citenamefont {Wang}\ \emph
  {et~al.}(2021{\natexlab{b}})\citenamefont {Wang}, \citenamefont {Chen},
  \citenamefont {Liu},\ and\ \citenamefont {Matsuki}}]{Wang:2020wrp}%
  \BibitemOpen
  \bibfield  {author} {\bibinfo {author} {\bibfnamefont {J.-Z.}\ \bibnamefont
  {Wang}}, \bibinfo {author} {\bibfnamefont {D.-Y.}\ \bibnamefont {Chen}},
  \bibinfo {author} {\bibfnamefont {X.}~\bibnamefont {Liu}},\ and\ \bibinfo
  {author} {\bibfnamefont {T.}~\bibnamefont {Matsuki}},\ }\bibfield  {title}
  {\bibinfo {title} {{Producing fully charm structures in the $J/\psi$ -pair
  invariant mass spectrum}},\ }\href
  {https://doi.org/10.1103/PhysRevD.103.L071503} {\bibfield  {journal}
  {\bibinfo  {journal} {Phys. Rev. D}\ }\textbf {\bibinfo {volume} {103}},\
  \bibinfo {pages} {071503} (\bibinfo {year} {2021}{\natexlab{b}})},\ \Eprint
  {https://arxiv.org/abs/2008.07430} {arXiv:2008.07430 [hep-ph]} \BibitemShut
  {NoStop}%
\bibitem [{\citenamefont {Dong}\ \emph {et~al.}(2021)\citenamefont {Dong},
  \citenamefont {Baru}, \citenamefont {Guo}, \citenamefont {Hanhart},\ and\
  \citenamefont {Nefediev}}]{Dong:2020nwy}%
  \BibitemOpen
  \bibfield  {author} {\bibinfo {author} {\bibfnamefont {X.-K.}\ \bibnamefont
  {Dong}}, \bibinfo {author} {\bibfnamefont {V.}~\bibnamefont {Baru}}, \bibinfo
  {author} {\bibfnamefont {F.-K.}\ \bibnamefont {Guo}}, \bibinfo {author}
  {\bibfnamefont {C.}~\bibnamefont {Hanhart}},\ and\ \bibinfo {author}
  {\bibfnamefont {A.}~\bibnamefont {Nefediev}},\ }\bibfield  {title} {\bibinfo
  {title} {{Coupled-Channel Interpretation of the LHCb
  Double-~$J/\psi$~Spectrum and Hints of a New State Near the~ $J/\psi
  J/\psi$~~Threshold}},\ }\href
  {https://doi.org/10.1103/PhysRevLett.127.119901} {\bibfield  {journal}
  {\bibinfo  {journal} {Phys. Rev. Lett.}\ }\textbf {\bibinfo {volume} {126}},\
  \bibinfo {pages} {132001} (\bibinfo {year} {2021})},\ \bibinfo {note}
  {[Erratum: Phys.Rev.Lett. 127, 119901 (2021)]},\ \Eprint
  {https://arxiv.org/abs/2009.07795} {arXiv:2009.07795 [hep-ph]} \BibitemShut
  {NoStop}%
\bibitem [{\citenamefont {Wang}(2007)}]{Wang:2006ri}%
  \BibitemOpen
  \bibfield  {author} {\bibinfo {author} {\bibfnamefont {Z.-G.}\ \bibnamefont
  {Wang}},\ }\bibfield  {title} {\bibinfo {title} {{Analysis of the Y(2175) as
  a tetraquark state with QCD sum rules}},\ }\href
  {https://doi.org/10.1016/j.nuclphysa.2007.04.012} {\bibfield  {journal}
  {\bibinfo  {journal} {Nucl. Phys. A}\ }\textbf {\bibinfo {volume} {791}},\
  \bibinfo {pages} {106} (\bibinfo {year} {2007})},\ \Eprint
  {https://arxiv.org/abs/hep-ph/0610171} {arXiv:hep-ph/0610171} \BibitemShut
  {NoStop}%
\bibitem [{\citenamefont {Liu}\ \emph {et~al.}(2021)\citenamefont {Liu},
  \citenamefont {Liu}, \citenamefont {Zhong},\ and\ \citenamefont
  {Zhao}}]{Liu:2020lpw}%
  \BibitemOpen
  \bibfield  {author} {\bibinfo {author} {\bibfnamefont {F.-X.}\ \bibnamefont
  {Liu}}, \bibinfo {author} {\bibfnamefont {M.-S.}\ \bibnamefont {Liu}},
  \bibinfo {author} {\bibfnamefont {X.-H.}\ \bibnamefont {Zhong}},\ and\
  \bibinfo {author} {\bibfnamefont {Q.}~\bibnamefont {Zhao}},\ }\bibfield
  {title} {\bibinfo {title} {{Fully-strange tetraquark $ss\bar{s}\bar{s}$
  spectrum and possible experimental evidence}},\ }\href
  {https://doi.org/10.1103/PhysRevD.103.016016} {\bibfield  {journal} {\bibinfo
   {journal} {Phys. Rev. D}\ }\textbf {\bibinfo {volume} {103}},\ \bibinfo
  {pages} {016016} (\bibinfo {year} {2021})},\ \Eprint
  {https://arxiv.org/abs/2008.01372} {arXiv:2008.01372 [hep-ph]} \BibitemShut
  {NoStop}%
\bibitem [{\citenamefont {Wu}\ \emph {et~al.}(2024{\natexlab{b}})\citenamefont
  {Wu}, \citenamefont {Chen}, \citenamefont {Meng},\ and\ \citenamefont
  {Zhu}}]{Wu:2024euj}%
  \BibitemOpen
  \bibfield  {author} {\bibinfo {author} {\bibfnamefont {W.-L.}\ \bibnamefont
  {Wu}}, \bibinfo {author} {\bibfnamefont {Y.-K.}\ \bibnamefont {Chen}},
  \bibinfo {author} {\bibfnamefont {L.}~\bibnamefont {Meng}},\ and\ \bibinfo
  {author} {\bibfnamefont {S.-L.}\ \bibnamefont {Zhu}},\ }\bibfield  {title}
  {\bibinfo {title} {{Benchmark calculations of fully heavy compact and
  molecular tetraquark states}},\ }\href
  {https://doi.org/10.1103/PhysRevD.109.054034} {\bibfield  {journal} {\bibinfo
   {journal} {Phys. Rev. D}\ }\textbf {\bibinfo {volume} {109}},\ \bibinfo
  {pages} {054034} (\bibinfo {year} {2024}{\natexlab{b}})},\ \Eprint
  {https://arxiv.org/abs/2401.14899} {arXiv:2401.14899 [hep-ph]} \BibitemShut
  {NoStop}%
\bibitem [{\citenamefont {Ma}\ \emph {et~al.}(2024)\citenamefont {Ma},
  \citenamefont {Wu}, \citenamefont {Meng}, \citenamefont {Chen},\ and\
  \citenamefont {Zhu}}]{Ma:2024vsi}%
  \BibitemOpen
  \bibfield  {author} {\bibinfo {author} {\bibfnamefont {Y.}~\bibnamefont
  {Ma}}, \bibinfo {author} {\bibfnamefont {W.-L.}\ \bibnamefont {Wu}}, \bibinfo
  {author} {\bibfnamefont {L.}~\bibnamefont {Meng}}, \bibinfo {author}
  {\bibfnamefont {Y.-K.}\ \bibnamefont {Chen}},\ and\ \bibinfo {author}
  {\bibfnamefont {S.-L.}\ \bibnamefont {Zhu}},\ }\bibfield  {title} {\bibinfo
  {title} {{Fully strange tetraquark resonant states as the cousins of
  X(6900)}},\ }\href {https://doi.org/10.1103/PhysRevD.110.074026} {\bibfield
  {journal} {\bibinfo  {journal} {Phys. Rev. D}\ }\textbf {\bibinfo {volume}
  {110}},\ \bibinfo {pages} {074026} (\bibinfo {year} {2024})},\ \Eprint
  {https://arxiv.org/abs/2408.00503} {arXiv:2408.00503 [hep-ph]} \BibitemShut
  {NoStop}%
\bibitem [{\citenamefont {Semay}\ and\ \citenamefont
  {Silvestre-Brac}(1994)}]{Semay:1994ht}%
  \BibitemOpen
  \bibfield  {author} {\bibinfo {author} {\bibfnamefont {C.}~\bibnamefont
  {Semay}}\ and\ \bibinfo {author} {\bibfnamefont {B.}~\bibnamefont
  {Silvestre-Brac}},\ }\bibfield  {title} {\bibinfo {title} {{Diquonia and
  potential models}},\ }\href {https://doi.org/10.1007/BF01413104} {\bibfield
  {journal} {\bibinfo  {journal} {Z. Phys. C}\ }\textbf {\bibinfo {volume}
  {61}},\ \bibinfo {pages} {271} (\bibinfo {year} {1994})}\BibitemShut
  {NoStop}%
\bibitem [{\citenamefont {Silvestre-Brac}(1996)}]{Silvestre-Brac:1996myf}%
  \BibitemOpen
  \bibfield  {author} {\bibinfo {author} {\bibfnamefont {B.}~\bibnamefont
  {Silvestre-Brac}},\ }\bibfield  {title} {\bibinfo {title} {{Spectrum and
  static properties of heavy baryons}},\ }\href
  {https://doi.org/10.1007/s006010050028} {\bibfield  {journal} {\bibinfo
  {journal} {Few Body Syst.}\ }\textbf {\bibinfo {volume} {20}},\ \bibinfo
  {pages} {1} (\bibinfo {year} {1996})}\BibitemShut {NoStop}%
\bibitem [{\citenamefont {Hiyama}\ \emph {et~al.}(2003)\citenamefont {Hiyama},
  \citenamefont {Kino},\ and\ \citenamefont {Kamimura}}]{Hiyama:2003cu}%
  \BibitemOpen
  \bibfield  {author} {\bibinfo {author} {\bibfnamefont {E.}~\bibnamefont
  {Hiyama}}, \bibinfo {author} {\bibfnamefont {Y.}~\bibnamefont {Kino}},\ and\
  \bibinfo {author} {\bibfnamefont {M.}~\bibnamefont {Kamimura}},\ }\bibfield
  {title} {\bibinfo {title} {{Gaussian expansion method for few-body
  systems}},\ }\href {https://doi.org/10.1016/S0146-6410(03)90015-9} {\bibfield
   {journal} {\bibinfo  {journal} {Prog. Part. Nucl. Phys.}\ }\textbf {\bibinfo
  {volume} {51}},\ \bibinfo {pages} {223} (\bibinfo {year} {2003})}\BibitemShut
  {NoStop}%
\bibitem [{\citenamefont {Aguilar}\ and\ \citenamefont
  {Combes}(1971)}]{Aguilar:1971ve}%
  \BibitemOpen
  \bibfield  {author} {\bibinfo {author} {\bibfnamefont {J.}~\bibnamefont
  {Aguilar}}\ and\ \bibinfo {author} {\bibfnamefont {J.~M.}\ \bibnamefont
  {Combes}},\ }\bibfield  {title} {\bibinfo {title} {{A class of analytic
  perturbations for one-body schroedinger hamiltonians}},\ }\href
  {https://doi.org/10.1007/BF01877510} {\bibfield  {journal} {\bibinfo
  {journal} {Commun. Math. Phys.}\ }\textbf {\bibinfo {volume} {22}},\ \bibinfo
  {pages} {269} (\bibinfo {year} {1971})}\BibitemShut {NoStop}%
\bibitem [{\citenamefont {Balslev}\ and\ \citenamefont
  {Combes}(1971)}]{Balslev:1971vb}%
  \BibitemOpen
  \bibfield  {author} {\bibinfo {author} {\bibfnamefont {E.}~\bibnamefont
  {Balslev}}\ and\ \bibinfo {author} {\bibfnamefont {J.~M.}\ \bibnamefont
  {Combes}},\ }\bibfield  {title} {\bibinfo {title} {{Spectral properties of
  many-body schroedinger operators with dilatation-analytic interactions}},\
  }\href {https://doi.org/10.1007/BF01877511} {\bibfield  {journal} {\bibinfo
  {journal} {Commun. Math. Phys.}\ }\textbf {\bibinfo {volume} {22}},\ \bibinfo
  {pages} {280} (\bibinfo {year} {1971})}\BibitemShut {NoStop}%
\bibitem [{\citenamefont {Aoyama}\ \emph {et~al.}(2006)\citenamefont {Aoyama},
  \citenamefont {Myo}, \citenamefont {Kat{\={o}}},\ and\ \citenamefont
  {Ikeda}}]{Aoyama:2006hrz}%
  \BibitemOpen
  \bibfield  {author} {\bibinfo {author} {\bibfnamefont {S.}~\bibnamefont
  {Aoyama}}, \bibinfo {author} {\bibfnamefont {T.}~\bibnamefont {Myo}},
  \bibinfo {author} {\bibfnamefont {K.}~\bibnamefont {Kat{\={o}}}},\ and\
  \bibinfo {author} {\bibfnamefont {K.}~\bibnamefont {Ikeda}},\ }\bibfield
  {title} {\bibinfo {title} {{The Complex Scaling Method for Many-Body
  Resonances and Its Applications to Three-Body Resonances}},\ }\href
  {https://doi.org/10.1143/ptp.116.1} {\bibfield  {journal} {\bibinfo
  {journal} {Prog. Theor. Phys.}\ }\textbf {\bibinfo {volume} {116}},\ \bibinfo
  {pages} {1} (\bibinfo {year} {2006})}\BibitemShut {NoStop}%
\bibitem [{\citenamefont {Workman}\ \emph {et~al.}(2022)\citenamefont {Workman}
  \emph {et~al.}}]{ParticleDataGroup:2022pth}%
  \BibitemOpen
  \bibfield  {author} {\bibinfo {author} {\bibfnamefont {R.~L.}\ \bibnamefont
  {Workman}} \emph {et~al.} (\bibinfo {collaboration} {Particle Data Group}),\
  }\bibfield  {title} {\bibinfo {title} {{Review of Particle Physics}},\ }\href
  {https://doi.org/10.1093/ptep/ptac097} {\bibfield  {journal} {\bibinfo
  {journal} {PTEP}\ }\textbf {\bibinfo {volume} {2022}},\ \bibinfo {pages}
  {083C01} (\bibinfo {year} {2022})}\BibitemShut {NoStop}%
\bibitem [{\citenamefont {Yang}\ \emph {et~al.}(2025)\citenamefont {Yang},
  \citenamefont {Ma}, \citenamefont {Wu},\ and\ \citenamefont
  {Zhu}}]{Yang:2025wqo}%
  \BibitemOpen
  \bibfield  {author} {\bibinfo {author} {\bibfnamefont {H.-M.}\ \bibnamefont
  {Yang}}, \bibinfo {author} {\bibfnamefont {Y.}~\bibnamefont {Ma}}, \bibinfo
  {author} {\bibfnamefont {W.-L.}\ \bibnamefont {Wu}},\ and\ \bibinfo {author}
  {\bibfnamefont {S.-L.}\ \bibnamefont {Zhu}},\ }\bibfield  {title} {\bibinfo
  {title} {{Triply heavy tetraquark states with different flavors}},\ }\href
  {https://doi.org/10.1103/PhysRevD.111.074040} {\bibfield  {journal} {\bibinfo
   {journal} {Phys. Rev. D}\ }\textbf {\bibinfo {volume} {111}},\ \bibinfo
  {pages} {074040} (\bibinfo {year} {2025})},\ \Eprint
  {https://arxiv.org/abs/2502.10798} {arXiv:2502.10798 [hep-ph]} \BibitemShut
  {NoStop}%
\bibitem [{\citenamefont {Wu}\ \emph {et~al.}(2024{\natexlab{c}})\citenamefont
  {Wu}, \citenamefont {Ma}, \citenamefont {Chen}, \citenamefont {Meng},\ and\
  \citenamefont {Zhu}}]{Wu:2024hrv}%
  \BibitemOpen
  \bibfield  {author} {\bibinfo {author} {\bibfnamefont {W.-L.}\ \bibnamefont
  {Wu}}, \bibinfo {author} {\bibfnamefont {Y.}~\bibnamefont {Ma}}, \bibinfo
  {author} {\bibfnamefont {Y.-K.}\ \bibnamefont {Chen}}, \bibinfo {author}
  {\bibfnamefont {L.}~\bibnamefont {Meng}},\ and\ \bibinfo {author}
  {\bibfnamefont {S.-L.}\ \bibnamefont {Zhu}},\ }\bibfield  {title} {\bibinfo
  {title} {{Fully heavy tetraquark resonant states with different flavors}},\
  }\href@noop {} {\  (\bibinfo {year} {2024}{\natexlab{c}})},\ \Eprint
  {https://arxiv.org/abs/2406.17824} {arXiv:2406.17824 [hep-ph]} \BibitemShut
  {NoStop}%
\bibitem [{\citenamefont {Wu}\ and\ \citenamefont {Zhu}(2025)}]{Wu:2024ocq}%
  \BibitemOpen
  \bibfield  {author} {\bibinfo {author} {\bibfnamefont {W.-L.}\ \bibnamefont
  {Wu}}\ and\ \bibinfo {author} {\bibfnamefont {S.-L.}\ \bibnamefont {Zhu}},\
  }\bibfield  {title} {\bibinfo {title} {{Fully charmed P-wave tetraquark
  resonant states in the quark model}},\ }\href
  {https://doi.org/10.1103/PhysRevD.111.034044} {\bibfield  {journal} {\bibinfo
   {journal} {Phys. Rev. D}\ }\textbf {\bibinfo {volume} {111}},\ \bibinfo
  {pages} {034044} (\bibinfo {year} {2025})},\ \Eprint
  {https://arxiv.org/abs/2411.17962} {arXiv:2411.17962 [hep-ph]} \BibitemShut
  {NoStop}%
\bibitem [{\citenamefont {Zheng}\ \emph {et~al.}(2026)\citenamefont {Zheng},
  \citenamefont {Ma},\ and\ \citenamefont {Zhu}}]{Zheng:2025uzy}%
  \BibitemOpen
  \bibfield  {author} {\bibinfo {author} {\bibfnamefont {X.-H.}\ \bibnamefont
  {Zheng}}, \bibinfo {author} {\bibfnamefont {Y.}~\bibnamefont {Ma}},\ and\
  \bibinfo {author} {\bibfnamefont {S.-L.}\ \bibnamefont {Zhu}},\ }\bibfield
  {title} {\bibinfo {title} {{Singly heavy tetraquark resonant states with
  multiple strange quarks}},\ }\href {https://doi.org/10.1103/2k7s-hm8b}
  {\bibfield  {journal} {\bibinfo  {journal} {Phys. Rev. D}\ }\textbf {\bibinfo
  {volume} {113}},\ \bibinfo {pages} {054027} (\bibinfo {year} {2026})},\
  \Eprint {https://arxiv.org/abs/2510.01505} {arXiv:2510.01505 [hep-ph]}
  \BibitemShut {NoStop}%
\bibitem [{\citenamefont {Kamada}\ \emph {et~al.}(2001)\citenamefont {Kamada}
  \emph {et~al.}}]{Kamada:2001tv}%
  \BibitemOpen
  \bibfield  {author} {\bibinfo {author} {\bibfnamefont {H.}~\bibnamefont
  {Kamada}} \emph {et~al.},\ }\bibfield  {title} {\bibinfo {title} {{Benchmark
  test calculation of a four nucleon bound state}},\ }\href
  {https://doi.org/10.1103/PhysRevC.64.044001} {\bibfield  {journal} {\bibinfo
  {journal} {Phys. Rev. C}\ }\textbf {\bibinfo {volume} {64}},\ \bibinfo
  {pages} {044001} (\bibinfo {year} {2001})},\ \Eprint
  {https://arxiv.org/abs/nucl-th/0104057} {arXiv:nucl-th/0104057} \BibitemShut
  {NoStop}%
\bibitem [{\citenamefont {Hiyama}\ \emph {et~al.}(2002)\citenamefont {Hiyama},
  \citenamefont {Kamimura}, \citenamefont {Motoba}, \citenamefont {Yamada},\
  and\ \citenamefont {Yamamoto}}]{Hiyama:2001zt}%
  \BibitemOpen
  \bibfield  {author} {\bibinfo {author} {\bibfnamefont {E.}~\bibnamefont
  {Hiyama}}, \bibinfo {author} {\bibfnamefont {M.}~\bibnamefont {Kamimura}},
  \bibinfo {author} {\bibfnamefont {T.}~\bibnamefont {Motoba}}, \bibinfo
  {author} {\bibfnamefont {T.}~\bibnamefont {Yamada}},\ and\ \bibinfo {author}
  {\bibfnamefont {Y.}~\bibnamefont {Yamamoto}},\ }\bibfield  {title} {\bibinfo
  {title} {{Lambda - Sigma conversion in He-4(Lambda) and He-4(Lambda) based on
  four body calculation}},\ }\href {https://doi.org/10.1103/PhysRevC.65.011301}
  {\bibfield  {journal} {\bibinfo  {journal} {Phys. Rev. C}\ }\textbf {\bibinfo
  {volume} {65}},\ \bibinfo {pages} {011301} (\bibinfo {year} {2002})},\
  \Eprint {https://arxiv.org/abs/nucl-th/0106070} {arXiv:nucl-th/0106070}
  \BibitemShut {NoStop}%
\bibitem [{\citenamefont {Dagum}\ and\ \citenamefont
  {Menon}(1998)}]{Dagum:1998rbq}%
  \BibitemOpen
  \bibfield  {author} {\bibinfo {author} {\bibfnamefont {L.}~\bibnamefont
  {Dagum}}\ and\ \bibinfo {author} {\bibfnamefont {R.}~\bibnamefont {Menon}},\
  }\bibfield  {title} {\bibinfo {title} {{OpenMP: an industry standard API for
  shared-memory programming}},\ }\href {https://doi.org/10.1109/99.660313}
  {\bibfield  {journal} {\bibinfo  {journal} {Comput. Sci. Eng.}\ }\textbf
  {\bibinfo {volume} {5}},\ \bibinfo {pages} {46} (\bibinfo {year}
  {1998})}\BibitemShut {NoStop}%
\bibitem [{\citenamefont {Wang}\ \emph {et~al.}(2013)\citenamefont {Wang},
  \citenamefont {Zhang}, \citenamefont {Zhang},\ and\ \citenamefont
  {Yi}}]{Wang:2013pot}%
  \BibitemOpen
  \bibfield  {author} {\bibinfo {author} {\bibfnamefont {Q.}~\bibnamefont
  {Wang}}, \bibinfo {author} {\bibfnamefont {X.}~\bibnamefont {Zhang}},
  \bibinfo {author} {\bibfnamefont {Y.}~\bibnamefont {Zhang}},\ and\ \bibinfo
  {author} {\bibfnamefont {Q.}~\bibnamefont {Yi}},\ }\bibfield  {title}
  {\bibinfo {title} {Augem: Automatically generate high performance dense
  linear algebra kernels on x86 cpus},\ }in\ \href
  {https://doi.org/10.1145/2503210.2503219} {\emph {\bibinfo {booktitle} {SC
  '13: Proceedings of the International Conference on High Performance
  Computing, Networking, Storage and Analysis}}}\ (\bibinfo {year} {2013})\
  pp.\ \bibinfo {pages} {1--12}\BibitemShut {NoStop}%
\bibitem [{\citenamefont {Lehoucq}\ \emph {et~al.}(1989)\citenamefont
  {Lehoucq}, \citenamefont {Sorensen},\ and\ \citenamefont
  {Yang}}]{Lehoucq:1989aug}%
  \BibitemOpen
  \bibfield  {author} {\bibinfo {author} {\bibfnamefont {R.~B.}\ \bibnamefont
  {Lehoucq}}, \bibinfo {author} {\bibfnamefont {D.}~\bibnamefont {Sorensen}},\
  and\ \bibinfo {author} {\bibfnamefont {C.}~\bibnamefont {Yang}},\ }\bibfield
  {title} {\bibinfo {title} {Arpack users’ guide: Solution of large-scale
  eigenvalue problems with implicitly restarted arnoldi methods, siam,
  philadelphia, 1998},\ }\href@noop {} {\bibfield  {journal} {\bibinfo
  {journal} {The software and this manual are available at URL http://www.
  caam. rice. edu/software/ARPACK}\ } (\bibinfo {year} {1989})}\BibitemShut
  {NoStop}%
\bibitem [{\citenamefont {Yu}(2022)}]{Yu:2017pmn}%
  \BibitemOpen
  \bibfield  {author} {\bibinfo {author} {\bibfnamefont {F.-S.}\ \bibnamefont
  {Yu}},\ }\bibfield  {title} {\bibinfo {title} {{Weak-decay searches for
  $Qs{\bar{u}}{\bar{d}}$ tetraquarks}},\ }\href
  {https://doi.org/10.1140/epjc/s10052-022-10567-8} {\bibfield  {journal}
  {\bibinfo  {journal} {Eur. Phys. J. C}\ }\textbf {\bibinfo {volume} {82}},\
  \bibinfo {pages} {641} (\bibinfo {year} {2022})},\ \Eprint
  {https://arxiv.org/abs/1709.02571} {arXiv:1709.02571 [hep-ph]} \BibitemShut
  {NoStop}%
\bibitem [{\citenamefont {Chen}\ \emph {et~al.}(2021)\citenamefont {Chen},
  \citenamefont {Han}, \citenamefont {Lü}, \citenamefont {Wang},\ and\
  \citenamefont {Yu}}]{Chen:2020eyu}%
  \BibitemOpen
  \bibfield  {author} {\bibinfo {author} {\bibfnamefont {Y.-K.}\ \bibnamefont
  {Chen}}, \bibinfo {author} {\bibfnamefont {J.-J.}\ \bibnamefont {Han}},
  \bibinfo {author} {\bibfnamefont {Q.-F.}\ \bibnamefont {Lü}}, \bibinfo
  {author} {\bibfnamefont {J.-P.}\ \bibnamefont {Wang}},\ and\ \bibinfo
  {author} {\bibfnamefont {F.-S.}\ \bibnamefont {Yu}},\ }\bibfield  {title}
  {\bibinfo {title} {Branching fractions of $b^- \to d^-x_{0,1}(2900)$ and
  their implications},\ }\href
  {https://doi.org/10.1140/epjc/s10052-021-08857-8} {\bibfield  {journal}
  {\bibinfo  {journal} {Eur. Phys. J. C}\ }\textbf {\bibinfo {volume} {81}},\
  \bibinfo {pages} {71} (\bibinfo {year} {2021})},\ \Eprint
  {https://arxiv.org/abs/2009.01182} {arXiv:2009.01182 [hep-ph]} \BibitemShut
  {NoStop}%
\bibitem [{\citenamefont {Wang}\ \emph
  {et~al.}(2024{\natexlab{b}})\citenamefont {Wang}, \citenamefont {Lin},
  \citenamefont {Wang}, \citenamefont {Meng},\ and\ \citenamefont
  {Zhu}}]{Wang:2024ukc}%
  \BibitemOpen
  \bibfield  {author} {\bibinfo {author} {\bibfnamefont {J.-Z.}\ \bibnamefont
  {Wang}}, \bibinfo {author} {\bibfnamefont {Z.-Y.}\ \bibnamefont {Lin}},
  \bibinfo {author} {\bibfnamefont {B.}~\bibnamefont {Wang}}, \bibinfo {author}
  {\bibfnamefont {L.}~\bibnamefont {Meng}},\ and\ \bibinfo {author}
  {\bibfnamefont {S.-L.}\ \bibnamefont {Zhu}},\ }\bibfield  {title} {\bibinfo
  {title} {{Double pole structures of X1(2900) as the P-wave
  D{\textasciimacron}*K* resonances}},\ }\href
  {https://doi.org/10.1103/PhysRevD.110.114003} {\bibfield  {journal} {\bibinfo
   {journal} {Phys. Rev. D}\ }\textbf {\bibinfo {volume} {110}},\ \bibinfo
  {pages} {114003} (\bibinfo {year} {2024}{\natexlab{b}})},\ \Eprint
  {https://arxiv.org/abs/2408.08965} {arXiv:2408.08965 [hep-ph]} \BibitemShut
  {NoStop}%
\bibitem [{\citenamefont {Qin}\ \emph {et~al.}(2021)\citenamefont {Qin},
  \citenamefont {Shen},\ and\ \citenamefont {Yu}}]{Qin:2020zlg}%
  \BibitemOpen
  \bibfield  {author} {\bibinfo {author} {\bibfnamefont {Q.}~\bibnamefont
  {Qin}}, \bibinfo {author} {\bibfnamefont {Y.-F.}\ \bibnamefont {Shen}},\ and\
  \bibinfo {author} {\bibfnamefont {F.-S.}\ \bibnamefont {Yu}},\ }\bibfield
  {title} {\bibinfo {title} {{Discovery potentials of double-charm
  tetraquarks}},\ }\href {https://doi.org/10.1088/1674-1137/ac1b97} {\bibfield
  {journal} {\bibinfo  {journal} {Chin. Phys. C}\ }\textbf {\bibinfo {volume}
  {45}},\ \bibinfo {pages} {103106} (\bibinfo {year} {2021})},\ \Eprint
  {https://arxiv.org/abs/2008.08026} {arXiv:2008.08026 [hep-ph]} \BibitemShut
  {NoStop}%
\bibitem [{\citenamefont {Deng}\ and\ \citenamefont
  {Zhu}(2022)}]{Deng:2022cld}%
  \BibitemOpen
  \bibfield  {author} {\bibinfo {author} {\bibfnamefont {C.-R.}\ \bibnamefont
  {Deng}}\ and\ \bibinfo {author} {\bibfnamefont {S.-L.}\ \bibnamefont {Zhu}},\
  }\bibfield  {title} {\bibinfo {title} {{Decoding the double heavy tetraquark
  state $T^+_{cc}$}},\ }\href {https://doi.org/10.1016/j.scib.2022.06.016}
  {\bibfield  {journal} {\bibinfo  {journal} {Sci. Bull.}\ }\textbf {\bibinfo
  {volume} {67}},\ \bibinfo {pages} {1522} (\bibinfo {year} {2022})},\ \Eprint
  {https://arxiv.org/abs/2204.11079} {arXiv:2204.11079 [hep-ph]} \BibitemShut
  {NoStop}%
\bibitem [{\citenamefont {Jaffe}(2005)}]{Jaffe:2004ph}%
  \BibitemOpen
  \bibfield  {author} {\bibinfo {author} {\bibfnamefont {R.~L.}\ \bibnamefont
  {Jaffe}},\ }\bibfield  {title} {\bibinfo {title} {{Exotica}},\ }\href
  {https://doi.org/10.1016/j.physrep.2004.11.005} {\bibfield  {journal}
  {\bibinfo  {journal} {Phys. Rept.}\ }\textbf {\bibinfo {volume} {409}},\
  \bibinfo {pages} {1} (\bibinfo {year} {2005})},\ \Eprint
  {https://arxiv.org/abs/hep-ph/0409065} {arXiv:hep-ph/0409065} \BibitemShut
  {NoStop}%
\end{thebibliography}%

\end{document}